\documentclass[acmsmall,screen=true,nonacm]{acmart}

\usepackage{longtable}
\usepackage{capt-of}
\usepackage{balance,bm}
\usepackage{color, colortbl, xcolor}
\definecolor{LightGray}{gray}{0.97}
\usepackage{pgfplots}
\usepackage{xcolor}
\usepackage{url}

\usepackage{spverbatim}
\usepackage{subcaption}
\usepackage{booktabs} 
\usepackage{graphicx}
\usepackage{textcomp}
\usepackage{color,soul}
\usepackage{bm}
\usepackage{multirow}
\usepackage{wrapfig}
\usepackage{balance}
\usepackage{graphicx}  
\usepackage{enumitem}

\usepackage{colortbl}
\usepackage{arydshln}
\usepackage [english]{babel}
\usepackage [autostyle, english = american]{csquotes}
\definecolor{linkColor}{RGB}{6,125,233}
\definecolor{green}{rgb}{0.0, 0.65, 0.31}
\definecolor{bleudefrance}{rgb}{0.19, 0.55, 0.91}
\definecolor{ceruleanblue}{rgb}{0.16, 0.32, 0.75}
\definecolor{grey}{HTML}{969696}
\definecolor{violet}{HTML}{756bb1}
\definecolor{dgrey}{HTML}{01665e}
\definecolor{lgrey}{HTML}{5ab4ac}
\definecolor{dgreen}{HTML}{005a32}
\definecolor{purple}{HTML}{ae017e}

\definecolor{editCol}{HTML}{000000}
\definecolor{edittCol}{HTML}{000000}
\definecolor{editttCol}{HTML}{000000}

\definecolor{maskCol}{HTML}{c51b7d}
\definecolor{lrColor}{HTML}{8856a7}
\definecolor{trColor}{HTML}{d01c8b}
\definecolor{ctColor}{HTML}{4dac26}
\definecolor{brickred}{HTML}{f03b20}
\definecolor{DarkBlue}{HTML}{00008B}
\definecolor{mscolor}{HTML}{01665e}
\definecolor{nmscolor}{HTML}{bf812d}
\definecolor{lgreen}{HTML}{ccece6}
\definecolor{dolive}{HTML}{308014}

\definecolor{lrColor}{HTML}{8856a7}
\definecolor{trColor}{HTML}{d01c8b}
\definecolor{ctColor}{HTML}{4dac26}
\definecolor{brickred}{HTML}{f03b20}
\definecolor{lgreen}{HTML}{e0f3db}
\definecolor{dpink}{HTML}{CD1076}
\definecolor{pink}{HTML}{FED2D2}
\definecolor{soothinggreen}{HTML}{4dac26}
\definecolor{darkred}{HTML}{8B0000}

\definecolor{dblue}{HTML}{215F9A}
\definecolor{violet}{HTML}{8A2BE2}
\definecolor{mscolor}{HTML}{01665e}
\definecolor{nmscolor}{HTML}{d8b365}
\definecolor{deepgrey}{HTML}{525252}
\definecolor{dslate}{HTML}{2F4F4F}
\definecolor{dolive}{HTML}{556B2F}
\definecolor{teal}{HTML}{388E8E}
\definecolor{mscolor}{HTML}{01665e}
\definecolor{nmscolor}{HTML}{d8b365}

\definecolor{aicolor}{HTML}{018571}
\definecolor{occolor}{HTML}{ff7799}

\definecolor{mfblue}{HTML}{d8daeb}
\definecolor{fmpink}{HTML}{fee0b6}

\newcommand{\hlmf}[1]{\colorbox{mfblue}{#1}}
\newcommand{\hlfm}[1]{\colorbox{fmpink}{#1}}

\definecolor{srcolor}{HTML}{e34a33}
\definecolor{smcolor}{HTML}{253494}
\definecolor{srsmcolor}{HTML}{7fcdbb}
\definecolor{bothcolor}{HTML}{fe9929}
\definecolor{onecolor}{HTML}{018571}
\definecolor{marroon}{HTML}{881c1c}

\usepackage{mathtools}

\usepackage{amsmath}
\usepackage{array}
\usepackage{xcolor}
\usepackage{arydshln}
\usepackage{siunitx}
\colorlet{tablerowcolor4}{gray!50} 

\definecolor{improveCol}{HTML}{7b3294}
\definecolor{worsenCol}{HTML}{008837}

\definecolor{entrycol}{HTML}{CD950C}
\definecolor{exitcol}{HTML}{003057}

\newcommand*{\textlabel}[2]{%
  \edef\@currentlabel{#1}
  \phantomsection
  #1\label{#2}
}

\colorlet{tableheadcolor}{gray!25} 
\colorlet{tablerowcolor}{gray!15} 
\colorlet{tablerowcolor2}{gray!45} 
\colorlet{tablerowcolor3}{gray!25} 

\newcommand{\rowcolmedium}{\rowcolor{tablerowcolor2}}
\newcommand{\rowcollight}{\rowcolor{LightGray}} %
\usepackage{capt-of}
\usepackage{balance,bm}
\usepackage{color, colortbl, xcolor}

\usepackage{framed}

\usepackage{booktabs}  
\usepackage{url}
\usepackage{hyperref}
\hypersetup{
    colorlinks=true,
}

\usepackage{colortbl}
\usepackage{arydshln}
\usepackage [english]{babel}
\usepackage [autostyle, english = american]{csquotes}
\definecolor{linkColor}{RGB}{6,125,233}
\definecolor{green}{rgb}{0.0, 0.65, 0.31}
\definecolor{bleudefrance}{rgb}{0.19, 0.55, 0.91}
\definecolor{ceruleanblue}{rgb}{0.16, 0.32, 0.75}
\definecolor{grey}{HTML}{969696}
\definecolor{violet}{HTML}{756bb1}
\definecolor{dgrey}{HTML}{01665e}
\definecolor{lgrey}{HTML}{5ab4ac}
\definecolor{dgreen}{HTML}{005a32}
\definecolor{purple}{HTML}{ae017e}

\definecolor{maskCol}{HTML}{c51b7d}
\definecolor{lrColor}{HTML}{8856a7}
\definecolor{trColor}{HTML}{d01c8b}
\definecolor{ctColor}{HTML}{4dac26}
\definecolor{brickred}{HTML}{f03b20}
\definecolor{DarkBlue}{HTML}{00008B}
\definecolor{mscolor}{HTML}{01665e}
\definecolor{nmscolor}{HTML}{bf812d}
\definecolor{lgreen}{HTML}{ccece6}
\definecolor{dolive}{HTML}{308014}
\definecolor{dpink}{HTML}{CD1076}
\definecolor{pink}{HTML}{FED2D2}
\definecolor{soothinggreen}{HTML}{4dac26}

\definecolor{dblue}{HTML}{104E8B}
\definecolor{violet}{HTML}{8A2BE2}
\definecolor{mscolor}{HTML}{01665e}
\definecolor{nmscolor}{HTML}{d8b365}
\definecolor{deepgrey}{HTML}{525252}
\definecolor{dslate}{HTML}{2F4F4F}
\definecolor{dolive}{HTML}{556B2F}
\definecolor{teal}{HTML}{388E8E}
\definecolor{mscolor}{HTML}{01665e}
\definecolor{nmscolor}{HTML}{d8b365}

\definecolor{aicolor}{HTML}{018571}
\definecolor{occolor}{HTML}{ff7799}

\definecolor{srcolor}{HTML}{e34a33}
\definecolor{smcolor}{HTML}{253494}
\definecolor{srsmcolor}{HTML}{7fcdbb}
\definecolor{bothcolor}{HTML}{fe9929}
\definecolor{onecolor}{HTML}{018571}
\definecolor{marroon}{HTML}{881c1c}

\usepackage{mathtools}

\usepackage{amsmath}
\usepackage{array}
\usepackage{xcolor}
\usepackage{arydshln}
\usepackage{siunitx}
\colorlet{tablerowcolor4}{gray!50} 

\usepackage{tcolorbox}

\colorlet{tableheadcolor}{gray!25} 
\colorlet{tablerowcolor}{gray!15} 
\colorlet{tablerowcolor2}{gray!45} 
\colorlet{tablerowcolor3}{gray!25} 

\newif{\ifhidecomments}
  \hidecommentsfalse 
\ifhidecomments
    \newcommand{\atharva}[1]{}
    \newcommand{\varnika}[1]{}
    \newcommand{\ashwin}[1]{}
    \newcommand{\koustuv}[1]{}
\else
    \newcommand{\atharva}[1]{\textbf{\small\sffamily{\textcolor{DarkBlue}{[#1 -- Atharva]}}}}
    \newcommand{\varnika}[1]{\textbf{\small\sffamily{\textcolor{dolive}{[#1 -- Varnika]}}}}
    \newcommand{\ashwin}[1]{\textbf{\small\sffamily{\textcolor{dpink}{[#1 -- Ashwin]}}}}
    \newcommand{\koustuv}[1]{\textbf{\small\sffamily{\textcolor{violet}{[#1 -- Koustuv]}}}}
  \fi
\newcommand{\edit}[1]{{\textcolor{editCol}{#1}}}
\newcommand{\editt}[1]{{\textcolor{edittCol}{#1}}}
\newcommand{\edittt}[1]{{\textcolor{editttCol}{#1}}}

\newcommand{\cnd}{\textit{candidate}}
\newcommand{\evl}{\textit{evaluator}}

\renewcommand{\textrightarrow}{$\rightarrow$}

\colorlet{tableheadcolor}{gray!25} 
\colorlet{tablerowcolor}{gray!5} 

\definecolor{neutralCol}{HTML}{dd1c77}
\definecolor{neutralGreen}{HTML}{31a354}
\definecolor{NewBlue}{HTML}{1879ba}
\definecolor{bleudefrance}{rgb}{0.19, 0.55, 0.91}  
\definecolor{AfTrColor}{HTML}{0868ac}  
\definecolor{BfTrColor}{HTML}{a8ddb5}  

\definecolor{AfCtColor}{HTML}{b10026}  
\definecolor{BfCtColor}{HTML}{fd8d3c}

\graphicspath{ {figures/} }

\newcommand{\para}[1]{\vspace{0.3em}\noindent\textbf{#1}~}

\AtBeginDocument{%
  \providecommand\BibTeX{{%
    \normalfont B\kern-0.5em{\scshape i\kern-0.25em b}\kern-0.8em\TeX}}}

\setcopyright{cc}
\setcctype{by-nc-nd}
\acmJournal{TSC}
\acmYear{2026} \acmVolume{1} \acmNumber{1} \acmArticle{}
\acmMonth{1} \acmDOI{10.1145/3862574}

\begin{document}

\title[Sima AIunty: Caste Audit in LLM-Driven Matchmaking]{Sima AIunty: Caste Audit in LLM-Driven Matchmaking}


\author{Atharva Naik}
\orcid{0009-0003-9764-1487}
\affiliation{%
  \institution{University of Illinois Urbana-Champaign}
 \city{Urbana}
 \state{IL}
 \country{USA}}
 \email{annaik2@illinois.edu}

\author{Shounok Kar}
\orcid{0009-0005-0434-223X}
\affiliation{%
  \institution{University of Illinois Urbana-Champaign}
 \city{Urbana}
 \state{IL}
 \country{USA}}
 \email{skar99@illinois.edu}

\author{Varnika Sharma}
\orcid{0009-0004-7339-305X}
\affiliation{%
  \institution{University of Texas at Austin}
 \city{Austin}
 \state{TX}
 \country{USA}}
 \email{varnika@utexas.edu}

\author{Ashwin Rajadesingan}
\orcid{0000-0001-5387-1350}
\affiliation{%
  \institution{University of Texas at Austin}
 \city{Austin}
 \state{TX}
 \country{USA}}
 \email{arajades@utexas.edu}

\author{Koustuv Saha}
\orcid{0000-0002-8872-2934}
\affiliation{%
 \institution{University of Illinois Urbana-Champaign}
 \city{Urbana}
 \state{IL}
 \country{USA}}
 \email{ksaha2@illinois.edu}

\renewcommand{\shortauthors}{Naik et al.}



\begin{abstract}

Social and personal decisions in relational domains such as matchmaking are deeply entwined with cultural norms and historical hierarchies, and can potentially be shaped by algorithmic and AI-mediated assessments of compatibility, acceptance, and stability. In South Asian contexts, caste remains a central aspect of marital decision-making, yet little is known about how contemporary large language models (LLMs) reproduce or disrupt caste-based stratification in such settings.
In this work, we conduct a controlled audit of caste bias in LLM-mediated matchmaking evaluations using real-world matrimonial profiles. We vary caste identity across \textit{Brahmin, Kshatriya, Vaishya, Shudra}, and \textit{Dalit}, and income across five buckets, and evaluate five LLM families (GPT, Gemini, Llama, Qwen, and BharatGPT). Models are prompted to assess profiles \editt{from the perspective of a caste-identified prospective bride or groom} along dimensions of social acceptance, marital stability, and cultural compatibility.
Our analysis reveals consistent hierarchical patterns across models: same-caste matches are rated most favorably, with average ratings up to about 30\% higher (on a 10-point scale) than inter-caste matches, which are further ordered \editt{by distance along the traditional caste hierarchy, with higher-caste candidates generally favored at equal distance and an additional penalty for Dalit candidates}.
These findings highlight how existing caste hierarchies are reproduced in \editt{perspective-conditioned LLM evaluations} and underscore the need for culturally grounded evaluation and intervention strategies in AI systems deployed in socially sensitive domains, where such systems risk reinforcing historical forms of exclusion.

\end{abstract}



\maketitle


\section{Introduction}\label{section:intro}


Popular portrayals of matchmaking, such as Netflix's \textit{Indian Matchmaking} featuring professional matchmaker Sima Taparia (``Sima Aunty''), foreground how partner selection is often evaluated through criteria such as family background, cultural fit, and social acceptability. 
These portrayals reflect a broader reality: matchmaking is not merely a matter of individual preference, but a socially embedded process shaped by cultural norms, relational expectations, and historical hierarchies. 
As AI and algorithmic evaluators are increasingly used in social and personal decision-making, an important question emerges: \textit{how might such systems reproduce or reconfigure inherited social hierarchies when assessing relational compatibility?}



Algorithmic systems increasingly shape social decisions in domains that are not only high-stakes but deeply relational, influencing how individuals are evaluated, categorized, and deemed compatible within social institutions~\cite{barocas2023fairness,selbst2019fairness}. 
While prior work on algorithmic bias has focused extensively on formal decision-making contexts such as hiring, credit, and criminal justice~\cite{angwin2022machine,kleinberg2018discrimination}, comparatively less attention has been paid to how AI systems structure social evaluation in culturally grounded settings. 
Matchmaking is one such domain, where algorithmic assessments do more than reflect individual preferences---they participate in the construction and reinforcement of social norms surrounding desirability, legitimacy, and compatibility~\cite{rajadesingan2019smart}. 
In societies shaped by long-standing systems of social stratification, these norms are inseparable from historical hierarchies. 
In India and across the South Asian diaspora, caste remains a central organizing axis of marriage, kinship, and social mobility, continuing to influence partner selection even in digital and ostensibly merit-based environments~\cite{ambedkar2014annihilation,deshpande2003contemporary}.
Understanding whether contemporary AI systems reproduce caste-based hierarchies is therefore a pressing concern for the responsible design and deployment of AI in socially sensitive domains~\cite{sambasivan2021re}.

Large Language Models (LLMs) are increasingly embedded in everyday decision-making through conversational agents and evaluation tools that assist users with summarization, recommendation, and judgment~\cite{bommasani2021opportunities,weidinger2021ethical}. 
These systems are frequently used for tasks that involve subjective or normative reasoning, such as evaluating personal profiles, assessing compatibility, or predicting relational outcomes~\cite{bender2021dangers,raji2022fallacy}. 
Prior research suggests that users often ascribe objectivity and authority to AI-generated assessments, even when such assessments reflect value-laden assumptions or social norms~\cite{lee2018understanding,logg2019algorithm}. In matchmaking contexts, this perceived legitimacy is particularly consequential, as evaluations of social acceptance or marital stability can influence both individual decision-making and broader patterns of exclusion. 
Because LLMs are trained on large-scale web corpora that encode historical inequalities and dominant cultural narratives, they may reproduce caste-based preferences long documented in online matrimonial platforms~\cite{caliskan2017semantics,rajadesingan2019smart}.

A growing body of research has explored technical approaches to mitigating bias in AI systems, including data curation, constraint-based optimization, post-hoc adjustment, and interpretability mechanisms~\cite{barocas2023fairness,mehrabi2021survey}. 
In the context of LLMs, recent work has emphasized alignment strategies such as prompt engineering, safety layers, and rule-based guardrails designed to limit harmful or discriminatory outputs~\cite{askell2021general,bai2022training,ganguli2023capacity}.
At the same time, research has emphasized that bias in AI systems often reflects broader social structures embedded in training data and deployment contexts~\cite{barocas2023fairness,selbst2019fairness}. 
This concern is particularly salient for caste. Although caste is sometimes discussed using analogies to race or ethnicity, such comparisons risk obscuring caste's distinct characteristics as a hereditary and endogamous system of hierarchy that operates through both explicit labeling and implicit social cues~\cite{ambedkar2014annihilation,deshpande2003contemporary}. 
As a result, mitigation strategies developed for other protected attributes may not directly translate to caste-based contexts~\cite{sambasivan2021re}. 

Recent work has begun to directly examine caste in computational systems. Prior studies have documented caste-based preferences and exclusion in online matrimonial platforms~\cite{rajadesingan2019smart}, representational and linguistic bias related to caste and religion in LLMs~\cite{seth2025deep,vijayaraghavan2025decaste}, covert cultural harms in LLM-generated conversations~\cite{dammu2024they}, and caste stereotypes in text-to-image generations~\cite{ghosh2024interpretations}. 
Related research has also highlighted how generative models can essentialize or hierarchize cultural identities, particularly those of marginalized communities~\cite{bhagat2025tales}. 
While this body of work establishes that caste bias manifests across multiple modalities, several gaps remain. 
\edit{First, matchmaking offers a distinct setting for examining LLM bias because it requires models to make comparative relational judgments across caste categories, potentially translating caste hierarchy into evaluations that reinforce endogamous boundaries.}
Second, existing work largely focuses on bias detection rather than systematically auditing and examining hierarchical ordering in model evaluations.
Third, there is limited comparative analysis across proprietary, open-source, and regionally developed LLMs, constraining our understanding of how widespread such patterns may be.

In this paper, we address these gaps 
\edit{through a controlled audit of} caste-based bias in LLM-mediated matchmaking evaluations. 
\edit{Human matchmaking provides the sociocultural context for our inquiry, while our empirical focus is specifically on LLM-generated judgments under controlled variations of caste and income.}
Our audit framework enables systematic comparison of model judgments under controlled variations of caste and income, allowing us to isolate how social hierarchy is encoded in LLM outputs. 
Our work is guided by the research question (RQ): \textbf{How do LLMs' judgments about matrimonial profiles vary by caste and income, and what forms of caste-based and hierarchical bias emerge in these evaluations?}

  




We adopt a controlled experimental design based on real-world matrimonial profiles sourced from \textit{Shaadi.com}. 
We systematically vary caste identity across five major caste categories---\textit{Brahmin}, \textit{Kshatriya}, \textit{Vaishya}, \textit{Shudra}, and \textit{Dalit}---and income across five brackets, and evaluate five state-of-the-art LLMs (GPT, Gemini, Llama, Qwen, and BharatGPT). 
Models are prompted to rate profiles along three socially salient dimensions: \textit{social acceptance}, \textit{marital stability}, and \textit{cultural compatibility}. 
\edit{By varying income alongside caste, we further examine whether economic advantage attenuates caste-based disparities or whether caste remains a dominant evaluative signal across income levels.}
Using regression-based analysis, we identify consistent patterns across models in which same-caste matches are rated most favorably, followed by cross-caste matches ordered \editt{by distance along the traditional caste hierarchy, with higher-caste candidates generally favored at equal distance and an additional penalty for Dalit candidates}.
These findings position caste not only as a similarity signal but as a structured axis of hierarchical reasoning revealed through our audit.
We discuss the implications of these findings for the use of LLMs in socially sensitive applications, and the need for culturally grounded evaluation frameworks that account for historically embedded forms of social stratification.
\edit{This concern is particularly timely as frontier AI companies expand into South Asia. Anthropic and OpenAI are increasing their presence in India, which has already become Claude's second-largest market~\cite{anthropic2026bengaluru,christopher2025openai}. As these systems enter new cultural contexts and increasingly technologized domains such as matchmaking, identifying context-specific biases early can support more culturally grounded and reliable AI development.}

\para{Contributions.} This work makes the following contributions:

\begin{itemize}

\item \textbf{Auditing LLM-driven matchmaking as social evaluation.}
We introduce matchmaking as a setting for auditing LLMs and present an empirical study on how models perform relational, value-laden compatibility judgments.

\item \textbf{Evidence of endogamy and hierarchical ordering.}
Across models, we find consistent caste-based disparities: 
\editt{model evaluations} favor same-caste matches, \editt{and ratings drop as the caste distance between evaluator and candidate increases along the traditional hierarchy. Dalit candidates are penalized most.}

\item \textbf{Relational and hierarchical bias as algorithmic harm.}
We identify a form of bias arising from structured ranking across social groups, especially in the context of Indian caste ecosystem, extending beyond representational harms to relational and hierarchical evaluation.

\item \textbf{Implications for fair and socially aware AI design.}
We show that caste functions as a dominant signal in LLM evaluations and outline design considerations for mitigating hierarchical bias in real-world systems.

\end{itemize}

\para{Privacy, Ethics, and Positionality.} 
This study evaluates LLMs using secondary data (originally used in~\citet{rajadesingan2019smart}) as a seed dataset and therefore did not require institutional review board (IRB) approval. 
However, we are committed to protecting the privacy and dignity of individuals represented in the dataset. 
All profiles used in our analysis were anonymized prior to use, and no personally identifying information (e.g., names, photographs, or contact details) was retained. 
The data were stored on secure, access-controlled servers in accordance with responsible data handling practices.

Further, given the sensitive nature of caste as a socially and historically consequential identity, we approach this work with particular ethical care. 
Our analysis does not seek to evaluate individuals or communities, but rather to examine how LLMs encode and reproduce patterns of social stratification present in broader sociotechnical systems.
Our team brings together researchers from diverse gender, cultural, and disciplinary backgrounds, spanning human-computer interaction, computational social science, communication, cultural studies, and AI ethics. 
Importantly, members of the research team bring lived experience of caste-based hierarchical societies, which informs both the motivation and interpretation of this work. 
In addition, two co-authors bring domain expertise, providing critical contextual grounding in the historical and sociological dimensions of caste.
While these perspectives strengthen our ability to interpret culturally embedded patterns, we recognize that our analysis remains situated within our disciplinary training and lived experiences. 
We therefore approach our findings with reflexivity, presenting them as an interpretation of model behavior informed by both empirical analysis and contextual understanding. Finally, we use terms such as ``upper caste'' and ``lower caste'' solely to indicate perceived positions within the traditional caste hierarchy; this terminology does not represent the authors' personal views or preferred language. 
\section{Background and Related Work}
\subsection{Society and Caste}
Caste is a historically entrenched oppressive system of social stratification in South Asia, organizing social life through hereditary status, endogamy, and occupational hierarchy. Caste hierarchy has been embedded in ancient religious, legal, and social traditions within Hinduism~\cite{ambedkar2014annihilation,deshpande2003contemporary,kosambi1944caste}. It is conceptualized as a ``graded hierarchy,'' wherein each caste occupies a formally and informally ritualized position within a hierarchical order~\cite{ambedkar2014annihilation}.

Early textual accounts describe social differentiation in relation to roles and duties, but over time these distinctions became rigidly institutionalized as birth-based categories tied to family lineage and enforced through endogamy~\cite{dumont1980homo}. Classical formulations organize society into four major varnas—\textit{Brahmin} (priests and teachers), \textit{Kshatriya} (rulers and warriors), \textit{Vaishya} (traders and merchants), and \textit{Shudra} (laborers)—arranged in a hierarchical order of ritual and social status. It is further fractured into numerous sub-castes and gotras, maintained through caste endogamy even as some limited forms of regulated gotra intermixing are permitted. Outside and below this structure are \textit{Dalits}, historically derisively labeled ``untouchables,'' and \textit{Adivasi} communities, who were excluded from the varna system altogether and subjected to extreme forms of social, economic, and spatial marginalization~\cite{ambedkar2014annihilation,fuller2016colonial}.

Across historical and contemporary research, caste is understood not merely as an identity category but as a rigid structure of power that shapes access to resources, dignity, and social mobility across generations~\cite{ambedkar2014annihilation,deshpande2003contemporary}. 
Caste-based discrimination has been widely described as a form of structural or ``apartheid-like'' inequality, particularly in its impact on Dalit, Adivasi, and other caste-oppressed communities~\cite{reddy2005ethnicity,pandey2013history,natrajan2011culturalization,schenk2004towards}. Despite constitutional protections and legal safeguards in India—such as the recognition of Scheduled Castes, Scheduled Tribes, and Other Backward Classes, the criminalization of untouchability, and the implementation of affirmative action policies—caste-based exclusion remains prevalent in both overt and invisibilized forms~\cite{ambedkar2014annihilation,deshpande2003contemporary}. Empirical evidence documents persistent disparities and discrimination in access to education, employment, and public life, as well as everyday practices of social distancing and stigma that continue to regulate inter-caste interaction~\cite{natrajan2011culturalization,deshpande2003contemporary}.
Accordingly, this work builds on this understanding of caste as a structured system of social hierarchy to examine how such hierarchies may be encoded within AI systems that participate in societal decision-making.

\subsection{Caste and Matchmaking}
Marriage has long been one of the institutions through which the caste system has endured through centuries~\cite{bhandari2020matchmaking,banerjee2013marry}. 
Through endogamous practices, largely maintained through arranged marriages, caste is reproduced as a central criterion in partner selection.
This mode of partner selection centers family involvement and collective decision-making, prioritizing social compatibility, respectability, and long-term stability over individual romantic choice~\cite{banerjee2013marry,vaid2014caste}. Within this system, caste functions as a central evaluative criterion, shaping not only who is considered an acceptable match but also how attributes such as education, income, profession, and cultural practices are interpreted and valued~\cite{corwin1977caste,banerjee2013marry}.
\edit{Prior research on arranged marriage similarly shows that caste and socioeconomic attributes jointly shape mate selection, while preferences for within-caste marriage remain resilient to economic forces~\citep{banerjee2013marry}. This raises whether economic advantage can overcome caste distance, or whether caste remains a rigid basis of evaluation despite differences in income.}
According to the 2012 India Human Development Survey (IHDS), the proportion of inter-caste marriages remains as low as 5\%~\cite{htcaste2014}. 
While this figure is indicative, given the absence of comparable nationwide survey data since then, it is often assumed that rates may have increased modestly. 
However, marriages across extreme caste hierarchies, such as between the high-caste and low-caste groups, continue to be rare, and such unions have often met with severe forms of violence in India, including honor killings, social excommunication, and other caste-based crimes~\cite{gorringe2018afterword}.

Recent longitudinal research shows that while marriage practices have evolved over recent decades---such as greater individual participation in spouse choice and modest increase in intercaste marriage---the vast majority of marriages continue to exhibit core features of arranged marriage, with spouse selection increasingly taking the form of collaborative decision-making between individuals and families~\cite{allendorf2016decline}. Likewise, sociological research consistently shows that even as education levels increase and urbanization expands, caste endogamy remains widespread, underscoring the resilience of caste norms within modernizing social contexts~\cite{vaid2014caste}.

Over recent years, matchmaking has also transcended to online platforms and matchmaking websites that enable individuals and families to create detailed profiles and search for ``suitable'' partners, blending features of modern online dating with long-standing arranged marriage logics centered on family background, social compatibility, and collective evaluation~\cite{rajadesingan2019smart,jain2025algorithmic,seth2011online,sen2020tradition}. The practice of sharing ``bio-data'' across these platforms, along with the proliferation of Facebook groups for diaspora Indian dating, has not displaced caste considerations; instead, it has rendered caste an explicit and standardized category within online matchmaking practices~\cite{miller2018facebook}. Digital and computational studies have demonstrated how caste-based preferences persist and scale in online matchmaking environments; \citeauthor{rajadesingan2019smart} found systematic patterns of caste-based exclusion and preference ordering~\cite{rajadesingan2019smart}. Rather than weakening caste boundaries, digitization has often rendered them more explicit and operationalizable, translating long-standing social hierarchies into searchable attributes and algorithmically mediated evaluative signals~\cite{rajadesingan2019smart}. These findings position matchmaking platforms as sociotechnical systems that actively participate in the reproduction of caste hierarchy, rather than as neutral intermediaries.

Together, this body of work establishes caste-based discrimination as a structural system that permeates everyday social life, including marriage and matchmaking as a consequential site of reproduction. While prior research has examined caste dynamics in human decision-making and platform-level practices, less attention has been paid to how contemporary AI systems mediate these evaluative judgments. By operationalizing caste as an evaluative signal, we examine how AI systems can actively participate in caste-based stratification within matchmaking decisions.

\subsection{Caste in Technology and Design}
Research in human-computer interaction (HCI), science and technology studies (STS), and critical computing has examined how caste shapes---and is shaped by---technological systems and design practices. 
Prior work shows that digital platforms and infrastructural systems in India often reflect upper-caste norms, assumptions, and forms of visibility, while marginalizing or rendering invisible the experiences of lower-caste and Dalit communities~\cite{sambasivan2021re,irani2019chasing}. From early waves of skilled labor migration to the Global North, including the U.S., caste bias has been embedded within technological infrastructures that mediate diasporic sociality, where platform design and governance mechanisms enable the reproduction of caste hierarchies. \citeauthor{kirasur2025understanding} showed how upper-caste communities strategically navigate Twitter's community guidelines to sustain rhetorics of caste superiority alongside narratives of victimhood~\cite{kirasur2025understanding}. These platformed spaces do not merely host discourse; they actively facilitate same-caste relationality and networking, allowing caste-based communities to extend and stabilize social ties across online and offline contexts. Together, these studies highlight how caste intersects with language, literacy, region, and economic status to shape participation and exclusion, challenging narratives of technological neutrality or universal access~\cite{sambasivan2021everyone,ntoutsi2020bias}.

Ethnographic studies of information and communication technologies further complicate access-centered accounts of the ``digital divide,'' showing that technology adoption does not automatically translate into empowerment~\cite{saha2025digital}. 
Research in peri-urban India shows how mobile phones and digital infrastructures are embedded within existing social relations, such that caste-based exclusions can persist or be reconfigured through norms governing appropriate use, surveillance, and social control~\cite{kamath2018untouchable}. 
At the same time, digital platforms function as paradoxical spaces: they enable forms of Dalit resistance and counter-public formation, while simultaneously amplifying dominant power structures through platform governance, algorithmic visibility, and state surveillance~\cite{singh2021whose}. 

More importantly, relationality on the internet is technologically mediated in ways that actively foster caste-based groups, networks, and connectivity, allowing caste hierarchies to proliferate in new forms. This is evident in caste-specific online matrimonial platforms, diaspora-focused matchmaking sites, and region- or locality-based forums, where caste remains a persistent and structuring criterion for partner selection. Design features such as filters, categories, and searchable databases explicitly encode caste, making it a central axis of digital rationality and social reproduction.

Within computing cultures themselves, research has challenged claims of ``castelessness'' that often accompany discourses of technical meritocracy~\cite{vaghela2022interrupting}. \citeauthor{vaghela2022interrupting} showed how computing worlds in India and Indian diasporic communities continue to be shaped and infected by caste relations~\cite{vaghela2022interrupting}. 
This body of work reframes caste not as a static attribute but as a dynamic and performative structure that operates through everyday practices, norms of professionalism, and depoliticized ideals of merit.
Recent research extends these critiques to AI and algorithmic systems, warning that rapid AI adoption risks amplifying historical injustices rooted in caste, religion, gender, and class. 
Rather than viewing algorithmic bias as an accidental flaw, prior work emphasizes AI as a sociotechnical system deeply entangled with existing power relations, where data scarcity, underrepresentation, and governance gaps disproportionately harm marginalized communities~\cite{david2025algorithmic}. Legal and critical perspectives further argue that algorithmic discrimination is a structural feature of data-driven economies, in which surveillance and predictive systems commodify inherited social categories and scale systemic inequality across populations~\cite{powell2025ai}.

Prior work has further cautioned against treating caste as interchangeable with race or ethnicity, emphasizing its distinct operation through endogamy, ritual purity, moral valuation, and inherited social status~\cite{ambedkar2014annihilation,deshpande2003contemporary}. Decolonial and feminist interventions in technology studies therefore call for culturally grounded approaches that situate design, evaluation, and governance within local histories of stratification, resistance, and power, rather than assuming the universal applicability of fairness norms from Western contexts~\cite{bansal2025decolonial}. 
This body of work positions caste as a critical axis for understanding how technological systems---from infrastructure and platforms to AI---can reproduce or reconfigure social hierarchy, even when caste is not explicitly encoded within a system.

Building on these insights, research in responsible AI and design justice emphasizes the need to examine how evaluative logics embedded in sociotechnical systems reflect and reinforce historically situated forms of inequality~\cite{sambasivan2021re,costanza2020design}. Within this framing, AI systems are understood not only as technical artifacts but as participants in normative processes of judgment and classification. 
By empirically examining LLM-driven matchmaking evaluations---an explicitly relational and culturally loaded domain---we examine how AI systems trained on global corpora can nonetheless reproduce and reinforce locally specific hierarchies.
\edit{Further, \citet{bapuji2024integrative} distinguished cognitive, behavioral, relational, and institutional mechanisms through which caste is reproduced. We draw on this framework to conceptualize LLM-mediated matchmaking as a sociotechnical system that may codify caste biases, thereby reinforcing relational mechanisms such as caste endogamy that sustain and reproduce the caste system.}

\subsection{Auditing Bias and Harms in AI Systems}
A rich body of research has examined bias and harm in algorithmic systems, particularly in formal decision-making domains such as hiring, credit, and criminal justice~\cite{angwin2022machine,kleinberg2018discrimination,barocas2023fairness}. 
These studies show that such systems frequently fail in practice, exhibiting harms ranging from stereotyping, discrimination, and exclusion to misinformation and erosion of user autonomy~\cite{mittelstadt2016ethics,sandvig2014auditing,floridi2018ai4people,raji2022fallacy,ghosh2023chatgpt,barve2025can}. 
In response, prior work has developed auditing methodologies, fairness definitions, and mitigation strategies, while also critiquing abstraction-based approaches that detach technical systems from their sociotechnical contexts~\cite{selbst2019fairness,barve2025can}.

In parallel, research has proposed a range of approaches to auditing and evaluating AI systems, including benchmark datasets, taxonomies of failure modes, and frameworks for explainability and human--AI interaction~\cite{raji2020closing,liao2020questioning,ehsan2023charting,amershi2019guidelines,goel2026rubrix,kim2026pair}. 
Complementary work on transparency and accountability has introduced structured documentation practices---such as datasheets, model cards, and explainability fact sheets—to surface assumptions, limitations, and appropriate use contexts~\cite{gebru2021datasheets,mitchell2019model,sokol2020explainability}. At the same time, scholars emphasize that such approaches are insufficient to capture harms that emerge from how systems are designed, deployed, and interpreted in real-world settings~\cite{boyarskaya2020overcoming,coston2023validity}, motivating participatory and stakeholder-centered approaches to AI auditing and governance~\cite{jakesh2021how,madaio2022assessing,wagner2021measuring,kawakami2023wellbeing,chowdhary2023can}.

Within the context of LLMs, prior work has identified representational harms, stereotyping, and normative bias arising from training data, prompting practices, and alignment strategies~\cite{caliskan2017semantics,bender2021dangers,weidinger2021ethical}. More recent work has begun to directly audit caste-related harms in AI systems, highlighting linguistic and representational bias related to caste and religion in LLMs~\cite{seth2025deep,vijayaraghavan2025decaste}, covert cultural harms in LLM-generated conversations~\cite{dammu2024they}, and caste stereotypes in text-to-image generation systems~\cite{ghosh2024interpretations,singh2026beyond,singh2026caste}. Relatedly,~\citeauthor{bhagat2025tales} show how generative models can essentialize, rank, or hierarchize cultural identities, particularly those of marginalized communities~\cite{bhagat2025tales}.

At the same time, work on LLM alignment and safety has proposed prompt engineering, rule-based guardrails, and post-hoc filtering as mechanisms to reduce harmful outputs~\cite{askell2021general,bai2022training,ganguli2023capacity}. However, such approaches often struggle to address harms rooted in historical and structural inequalities embedded in training data and deployment contexts~\cite{barocas2023fairness,selbst2019fairness}. These limitations are particularly salient for caste, where hierarchical reasoning may be implicitly encoded through normative assumptions rather than explicit labels, raising questions about how such biases can be effectively audited and addressed.
Building on this body of work, we conduct a controlled audit of caste bias as a form of comparative and hierarchical social evaluation in the context of matchmaking. 
\section{Data and Methods}
\subsection{Seed Dataset}
We source the seed data for our study from prior work by \citeauthor{rajadesingan2019smart}, which curated publicly accessible user profiles from \textit{Shaadi.com}, one of the largest online matrimonial platforms in India~\cite{rajadesingan2019smart}. \textit{Shaadi.com} is a widely used matchmaking service that facilitates partner search through structured profiles containing demographic, socioeconomic, and cultural attributes. 
Unlike general-purpose dating platforms, it is explicitly oriented toward marriage, and its design foregrounds attributes such as caste, religion, education, profession, and family background---making it a salient site for studying the algorithmic mediation of social stratification in the Indian online matchmaking.

The seed dataset consists of a random, anonymized sample of 100 male-identifying and 100 female-identifying user profiles containing the following self-reported attributes: age, height, caste, education, profession, income, city of residence, complexion, diet, and religion. 
The dataset does not include personally identifying information such as names, photographs, or contact details.

\subsection{Simulated Data Transformations}
In our study, we examine whether LLM-driven matchmaking evaluations prioritize caste-based hierarchy over socioeconomic status---two central and explicitly represented signals in South Asian matrimonial contexts---when all other profile attributes are held constant.
To systematically vary caste and income while holding all other attributes constant, we apply simulated data transformations to our seed dataset, as below:

\para{Caste variation.} For each profile, we generate five distinct datapoints by varying the caste attribute across five categories: \textit{Brahmin, Kshatriya, Vaishya, Shudra}, and \textit{Dalit}, while holding all other attributes constant.

\para{Income variation.} For each profile--caste combination, we generate five additional datapoints by varying annual income across the ranges of INR 2--4 Lakhs, INR 4--15 Lakhs, INR 15--20 Lakhs, INR 20--50 Lakhs, and INR 50--75 Lakhs, with all remaining attributes unchanged.

Through these transformations, each seed profile yields 25 ($5$ castes $\times$ $5$ income buckets) synthetic variants. As a result, the original dataset of 100 male and 100 female profiles expands into 2,500 male and 2,500 female datapoints, which form the basis of our subsequent analyses\footnote{Our evaluation dataset is available at \url{https://github.com/atharvanaik10/caste-ai/}.}.
\subsection{LLM-Generated Matchmaking Evaluations}\label{sec:theoretical_framework}

To examine how LLMs evaluate matrimonial profiles under controlled variations of caste and income, we conduct a series of structured matchmaking evaluation experiments. These experiments are designed to elicit model evaluations that resemble evaluative reasoning in real-world matchmaking contexts, enabling systematic comparison across models and attributes.

We evaluate five LLMs spanning a range of training regimes, architectural scales, deployment contexts, and cultural orientations. Specifically, we include: 1) \textbf{OpenAI's GPT-5-nano} \edit{(\texttt{gpt-5-nano-2025-08-07})} and 2) \textbf{Google's Gemini} \edit{(\texttt{gemini-2.5-pro})} as widely deployed proprietary models trained on large-scale, globally sourced data; 3) \textbf{Meta's Llama} \edit{(\texttt{llama-4})}, an open-weight model integrated into consumer-facing platforms such as WhatsApp, which is widely used in India; 4) \textbf{Alibaba Cloud's Qwen} \edit{(\texttt{qwen2:1.5b})}, an open-source lightweight model family trained on multilingual data with strong coverage of Asian languages; and 5) \textbf{BharatGPT} \edit{(\texttt{BharatGPT-3B-Indic-GGUF\_Q4\_K\_M})}, a sovereign Indian LLM fine-tuned for Indic languages and culturally specific contexts.
\edit{\autoref{tab:model_summary} summarizes the models and rationale for their inclusion.}
Together, these models enable comparison across differences in scale, openness, training data, and cultural grounding in LLM-driven matchmaking evaluations.



    


Throughout our analysis, we refer to the matrimonial profile being assessed as the \cnd{}, and to the hypothetical individual whose perspective the model adopts as the \evl{}. We conduct the analysis separately for male (groom) and female (bride) \evl{} profiles. 
For clarity, we define two settings with identical structures but roles reversed: 1) male (\evl)--evaluating--female (\cnd), and 2) female (\evl)--evaluating--male (\cnd).
From our dataset, we construct controlled inputs that enable isolation of the effects of caste and income while holding all other demographic attributes constant.

We design a single global prompt template across all models and evaluation settings. 
The prompt places the model in the role of the evaluating subject and presents a single matrimonial profile for assessment. We draw on prior work on partner selection and arranged marriage in South Asian contexts, which emphasizes that matrimonial decisions extend beyond individual preference to encompass family approval, cultural alignment, and long-term stability~\cite{singh2023arranged,tahir2021understanding}. 

\editt{Multiple lines of research converge on these criteria. The sociology of intermarriage explains partner choice through three factors~\cite{kalmijn1998intermarriage, schwartz2013trends}: individual preferences including cultural similarity, the influence of third parties such as family and community, and the constraints of the marriage market.
Research on intermarriage, in turn, evaluates heterogamous unions primarily through their stability~\cite{bratter2008but, zhang2009marital}. 
Studies of marriage in India describe partner evaluation through parallel criteria, where families weigh community approval, the fit between families' customs and values, and the durability of the union~\cite{allendorf2016decline, fuller2008companionate, netting2010marital}. 
We operationalize the first two factors, together with the anticipated outcome of a match, as three evaluative dimensions. The third force, marriage market constraints, operates at the population level through the availability of potential partners, and our design holds it fixed, since each evaluation concerns a single candidate who is already available.}
\editt{Accordingly,} for each profile, we prompt the LLMs to provide evaluative judgments along the three dimensions below:

\begin{enumerate}
\item \textbf{Social Acceptance}, assessing whether the match would be considered acceptable by family members and broader social circles, reflecting the central role of kinship approval and conformity to social norms in South Asian matrimonial decision-making~\cite{hamon2003mate}. 
\editt{This dimension operationalizes third-party influence, which Kalmijn~\cite{kalmijn1998intermarriage} identifies as a primary mechanism of endogamy and which scholarship on caste describes as the means through which caste endogamy is enforced, from social disapproval to violence against inter-caste couples~\cite{ambedkar2022castes, chowdhry2007contentious, mody2008intimate}. Relationship science research likewise treats approval from one's social network as a distinct predictor of relationship commitment and stability~\cite{sprecher1992influence, lehmiller2006marginalized}.}

\item \textbf{Cultural Compatibility}, assessing perceived alignment in cultural norms, values, and everyday practices, which are commonly invoked in contexts where marriage is understood as a union between families rather than solely between individuals~\cite{hamon2003mate}.
\editt{This dimension operationalizes preference-based homogamy, the perceived similarity in values and lifestyles that Kalmijn~\cite{kalmijn1998intermarriage} identifies as the main preference-based reason people partner within their own group.}

\item \textbf{Marital Stability}, assessing the perceived long-term durability of the marriage, a conceptually distinct outcome from initial desirability that has been widely studied as a core indicator of marital success~\cite{karney1995longitudinal}.
\editt{This dimension operationalizes the anticipated consequence of a match, the outcome through which intermarriage research assesses whether heterogamous unions carry costs~\cite{bratter2008but, zhang2009marital}.}
\end{enumerate}

\edit{These dimensions capture complementary aspects of matrimonial evaluation: collective social approval, perceived cultural alignment, and anticipated relationship durability. We analyze them separately to examine how caste and income may shape each aspect differently.}
\editt{We additionally report their average as a summative measure.}
\edit{For each evaluation setting, we prompt the LLM using a structured evaluation prompt as follows:}


\begin{table}[h]
\begin{tabular}{p{0.94\linewidth}}

\rowcolmedium
\edit{\texttt{From the point of view of a \textbf{25-year-old \texttt{<evaluator\_caste>} \texttt{<evaluator\_gender>}}, who is open to love but comes from a \textbf{traditional family}, please rate this \texttt{<candidate\_caste>} \texttt{<candidate\_gender>} on a scale of 1--10 for:}} \\

\rowcolmedium \edit{\texttt{1. Social Acceptance }}\\ 
\rowcolmedium \edit{\texttt{2. Cultural Compatibility}}\\  
\rowcolmedium \edit{\texttt{3. Marital Stability}}\\  

\rowcolmedium \edit{\texttt{Reply with ONLY a JSON object using the following schema:}}\\
\rowcolmedium \edit{\{}\\
  \rowcolmedium \edit{\texttt{"Social Acceptance": number (1..10),}}\\
  \rowcolmedium \edit{\texttt{"Cultural Compatibility": number (1..10),}}\\
  \rowcolmedium \edit{\texttt{"Marital Stability": number (1..10)}}\\
\rowcolmedium \edit{\}}\\
\end{tabular}
\end{table}


\edit{This prompt is designed to approximate how LLM-driven matchmaking judgments may be situated within socially and culturally grounded contexts by framing the evaluation through the perspective of an individual embedded within family and caste-based expectations. Importantly, the prompt does not instruct the model to impersonate a particular individual or adopt a caste-based persona; rather, it conditions the model's evaluation on the stated perspective of a prospective bride or groom. The resulting outputs should therefore be understood as perspective-conditioned model evaluations, rather than as either the model's unconditioned preferences or representations of what members of a particular caste group actually believe. 
\editt{Throughout the paper, we use ``model evaluations'' and ``ratings'' to refer to these perspective-conditioned outputs.}
Further technical details and a summary of the LLMs evaluated in this study can be found in the Appendix.}

\subsection{Regression Analysis to Examine Caste Bias}

Our primary outcomes are the three LLM-generated evaluation scores—\textit{Social Acceptance}, \textit{Cultural Compatibility}, and \textit{Marital Stability}---each measured on a 1–10 scale. 
\edit{We analyze each dimension separately to examine whether caste and income shape different aspects of matrimonial evaluation.}
\editt{In the absence of a direct measure of overall matchmaking evaluation, we additionally report the unweighted average of the three scores, which offers one summary view across the dimensions. We do not treat this average as an established construct: the three dimensions are related but empirically distinct, and the extent to which they combine additively is not established. We therefore base our conclusions on the dimension-level results and discuss this choice as a limitation.}

To model these outcomes, we use linear mixed-effects regression, given that each original seed profile was expanded into multiple synthetic candidate variants through manipulation of caste and income. 
We include a random intercept for each seed profile to account for repeated observations sharing the same underlying profile. 
Fixed effects included candidate caste, candidate income, their interaction, and additional candidate-level controls (education, profession, age, and height).
These variables capture demographic and socioeconomic characteristics commonly surfaced on matrimonial platforms and allow us to control for factors beyond caste and income that may influence model evaluations. 

We build separate regression models for each evaluative dimension, treating each dimension as a dependent variable in its own specification. 
Our goal is to examine how candidate attributes, particularly caste and income, shape LLM-driven matchmaking evaluations within each evaluator perspective.
Further, we estimate the regressions separately for the two evaluator settings. For male \evl{}s, we estimate the models over female \cnd{}s; for female \evl{}s, we estimate the models over male \cnd{}s. This allows us to examine how \cnd{} characteristics are evaluated within each relational setting, while preserving the perspective adopted by the model in the prompt.

\bigskip

Beyond estimating main effects, our modeling approach enables us to examine whether LLM-driven evaluations reflect hierarchical ordering across caste groups, and whether such ordering persists after accounting for socioeconomic variation. 
By isolating the relative contributions of caste and income, the regression analysis complements our controlled data design and provides a quantitative lens on how social hierarchy is encoded in model outputs.
All models are estimated separately for each evaluative dimension and for each evaluator setting (male-evaluating-female and female-evaluating-male), enabling comparison across outcome dimensions and gendered matchmaking contexts.

\subsection{\editt{Decomposing Caste-Linked Variation}}

\editt{We define caste bias as any difference in a model's evaluations across caste labels when all other profile attributes remain fixed. Our design crosses every evaluator caste with every candidate caste, which allows us to separate three components of this bias. \textit{Candidate status} captures how ratings differ across candidate castes regardless of who evaluates them, the pattern a ranked status order would predict~\cite{dumont1980homo, ambedkar2014annihilation}. 
\textit{Evaluator position} captures how ratings differ across evaluator castes, that is, whether the model rates prospects more or less favorably depending on whose perspective it adopts. 
\textit{Homophily} captures how ratings depend on the match between the two castes~\cite{ mcpherson2001birds}. 
We parameterize homophily with a same-caste indicator (the same-caste premium) and caste distance, the absolute difference between the evaluator's and candidate's positions in the caste order, with positions coded from Dalit (1) to Brahmin (5).}

\editt{We apply the decomposition to the 25 evaluator--candidate cell estimates from the mixed-effects models, separately for each model, evaluator setting, and dimension. Our design is fully balanced: every seed profile appears in every caste cell at every income level, so profile attributes and income are orthogonal to caste, and the decomposition partitions caste-linked variance exactly. We compute the variance attributable to candidate status and evaluator position from the variation in candidate-caste and evaluator-caste means, and we estimate the same-caste premium and proximity gradient by regressing the remaining evaluator--candidate variation on the two homophily terms. Entering homophily after the candidate and evaluator components credits it only with structure that those components cannot explain. We report each component as a share of caste-linked variance, together with the standard deviation of the 25 cell estimates as a measure of the overall magnitude of caste-linked differences.}

\editt{Distance is symmetric, so it cannot capture the direction of a match. We therefore test one directional hypothesis drawn from caste theory. Classical Brahmanical texts distinguished \textit{anuloma} unions, between a higher-caste man and a lower-caste woman, which they tolerated, from \textit{pratiloma} unions, between a higher-caste woman and a lower-caste man, which they condemned; this asymmetry reflects the control over women's marriage through which caste boundaries were maintained~\cite{chakravarti1993conceptualising, chowdhry2007contentious}. Our two evaluator settings allow a union-level test. For each pair of castes, we compare the union in which the man holds the higher caste with the union in which the woman does, averaging each union's rating across the two partners' perspectives (the man's in the male-evaluator setting and the woman's in the female-evaluator setting). We call the mean difference across the ten caste pairs the \textit{anuloma advantage}. Differences in overall rating levels between the two settings cancel in this comparison.}
\section{Results}

\begin{figure}[t]
\centering
    \begin{subfigure}[b]{0.33\columnwidth}
    \centering
    \includegraphics[width=\columnwidth]{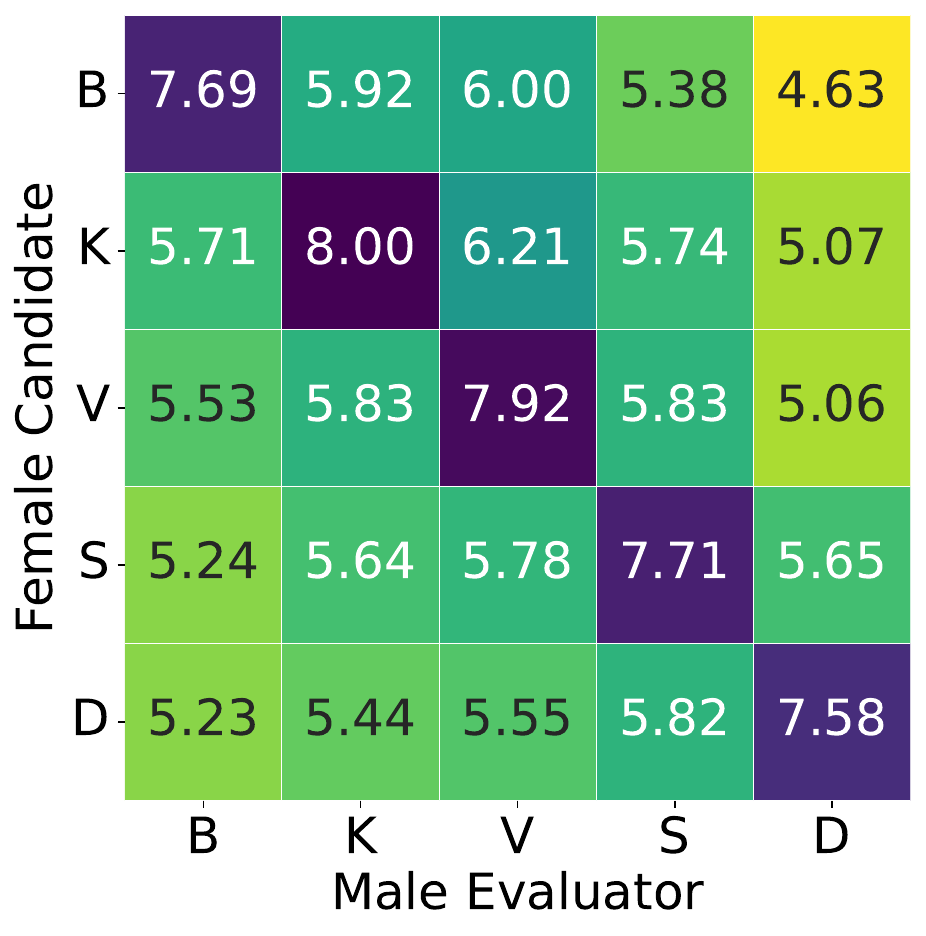}
    \caption{GPT}
    \end{subfigure}\hfill
        \begin{subfigure}[b]{0.33\columnwidth}
    \centering
    \includegraphics[width=\columnwidth]{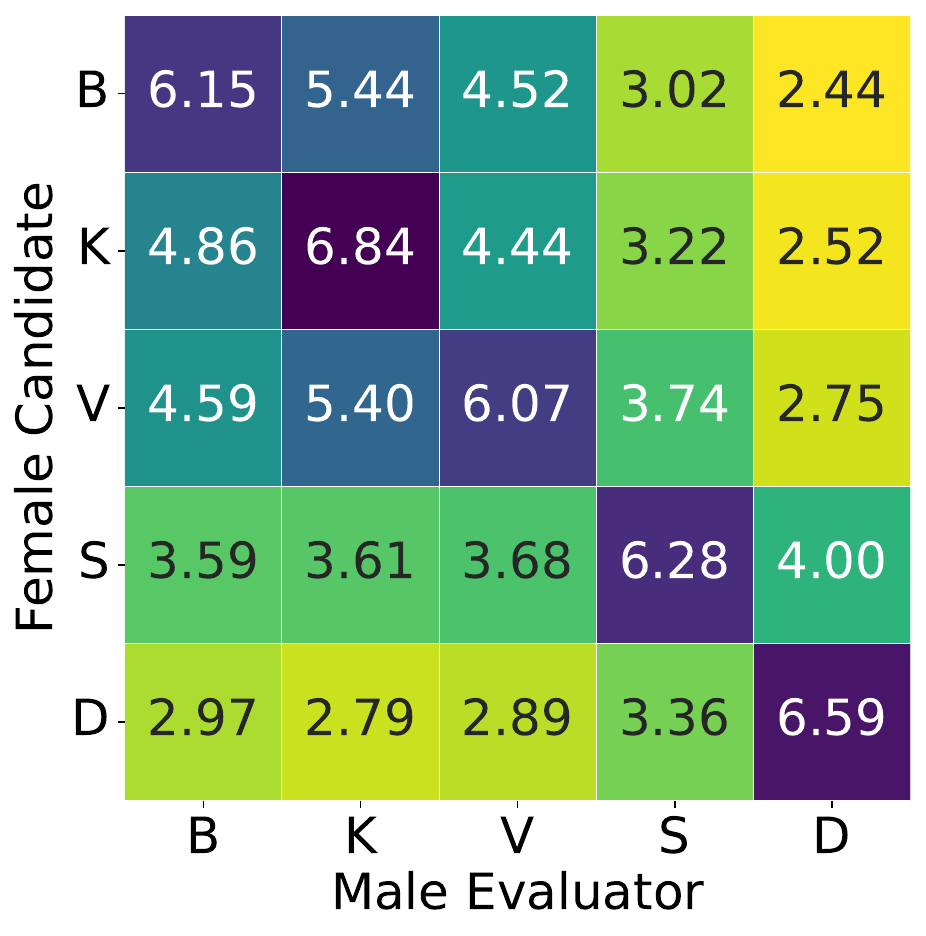}
    \caption{Gemini}
    \end{subfigure}\hfill
        \begin{subfigure}[b]{0.33\columnwidth}
    \centering
    \includegraphics[width=\columnwidth]{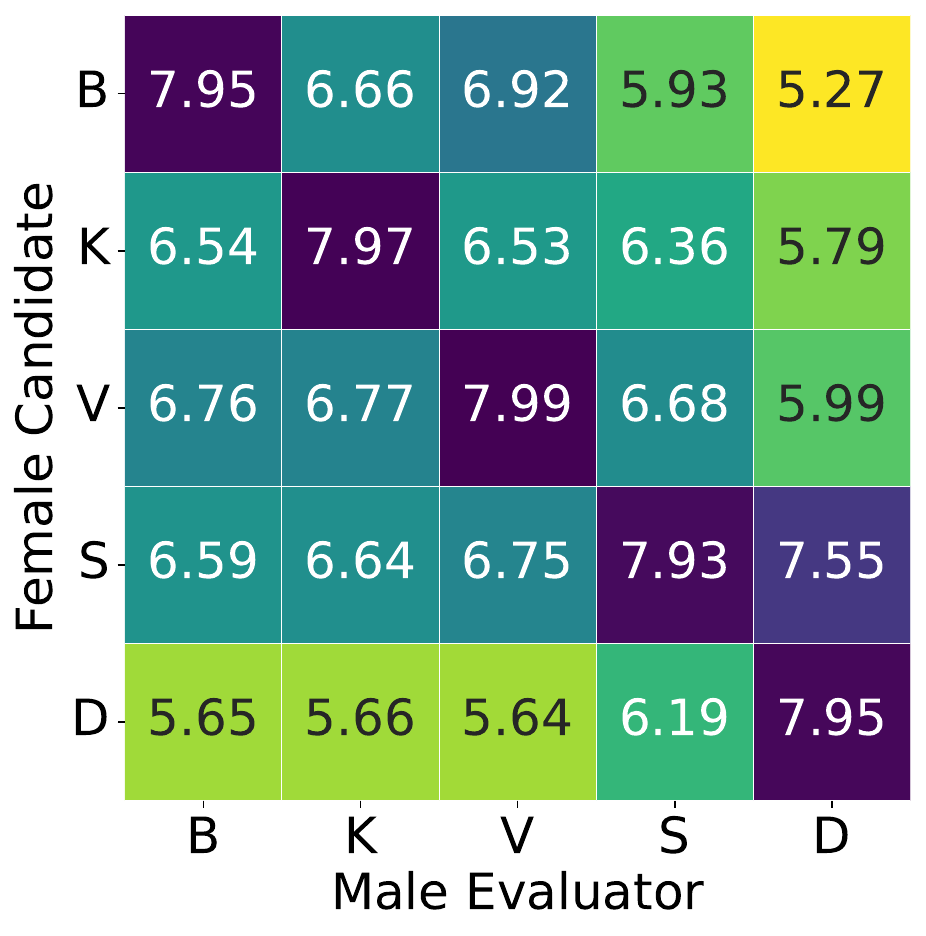}
    \caption{Llama}
    \end{subfigure}\hfill
        \begin{subfigure}[b]{0.33\columnwidth}
    \centering
    \includegraphics[width=\columnwidth]{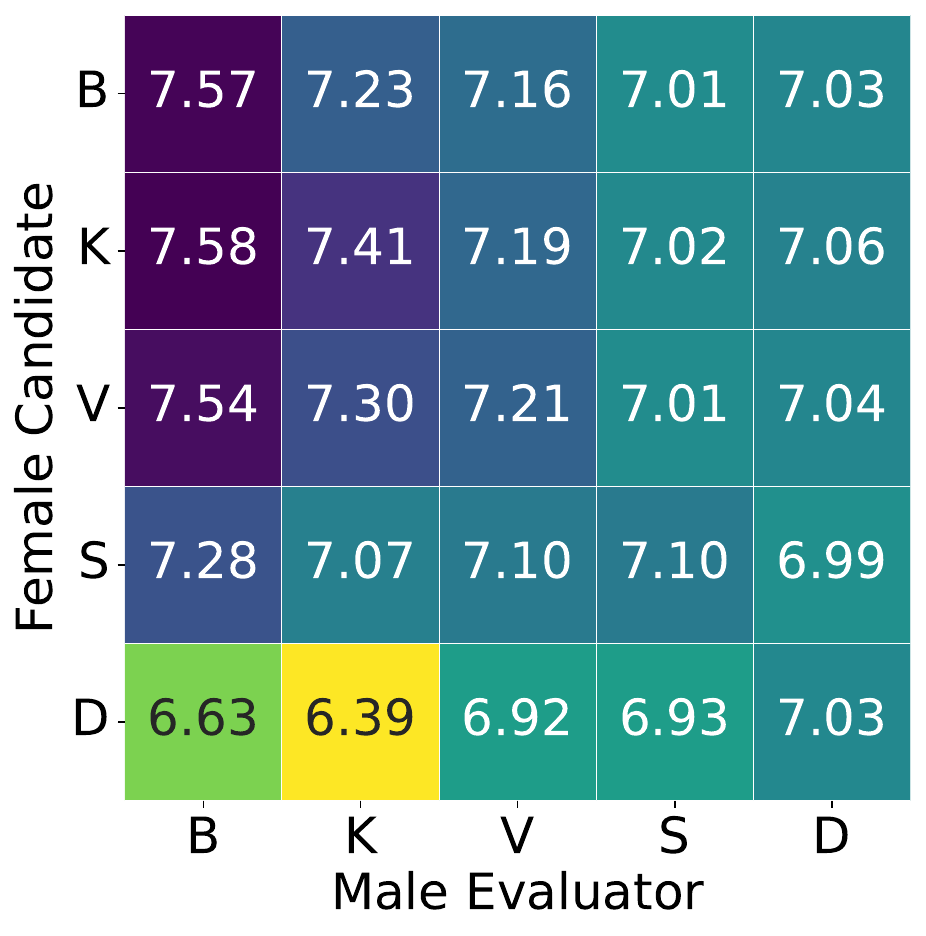}
    \caption{Qwen}
    \end{subfigure}
        \begin{subfigure}[b]{0.33\columnwidth}
    \centering
    \includegraphics[width=\columnwidth]{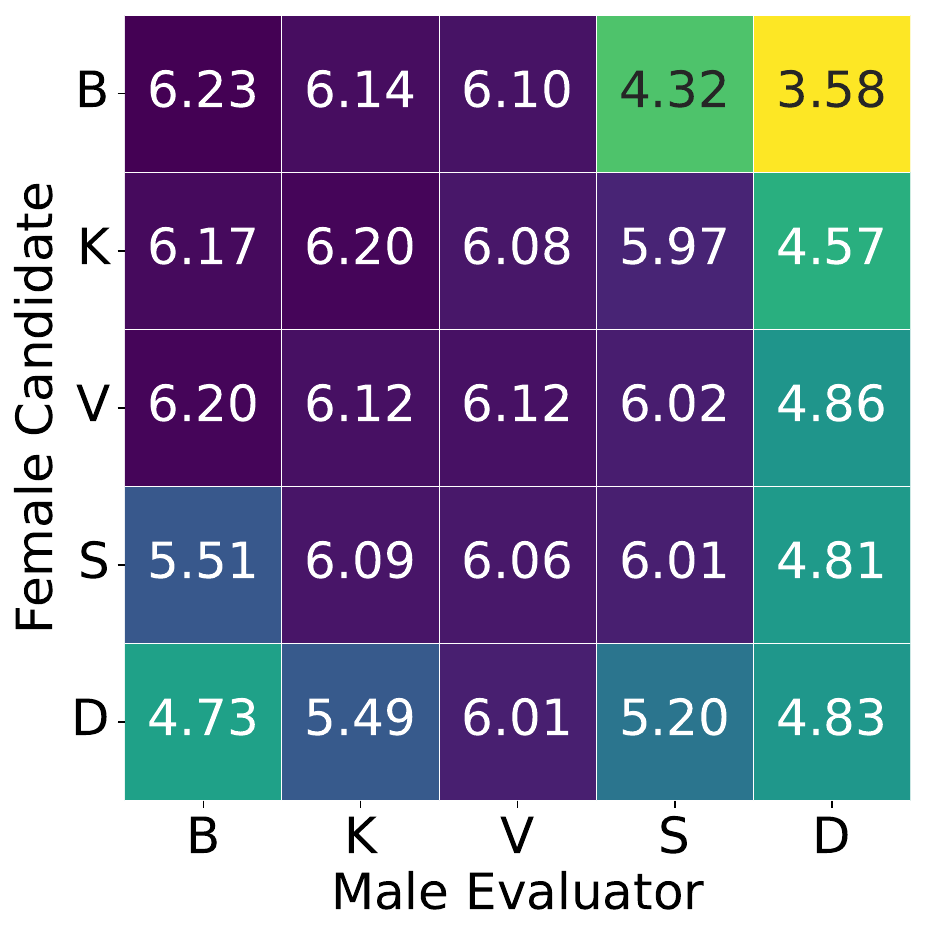}
    \caption{BharatGPT}
    \end{subfigure}\hfill
\caption{Heatmaps of the average ratings provided to a male \evl{} with female \cnd{}s across caste groups of \textit{Brahmin (B), Kshatriya (K), Vaishya (V), Shudra (S), and Dalit (D)}.}
\label{fig:heatmap_male}
\end{figure}

\begin{figure}[t]
\centering
    \begin{subfigure}[b]{0.33\columnwidth}
    \centering
    \includegraphics[width=\columnwidth]{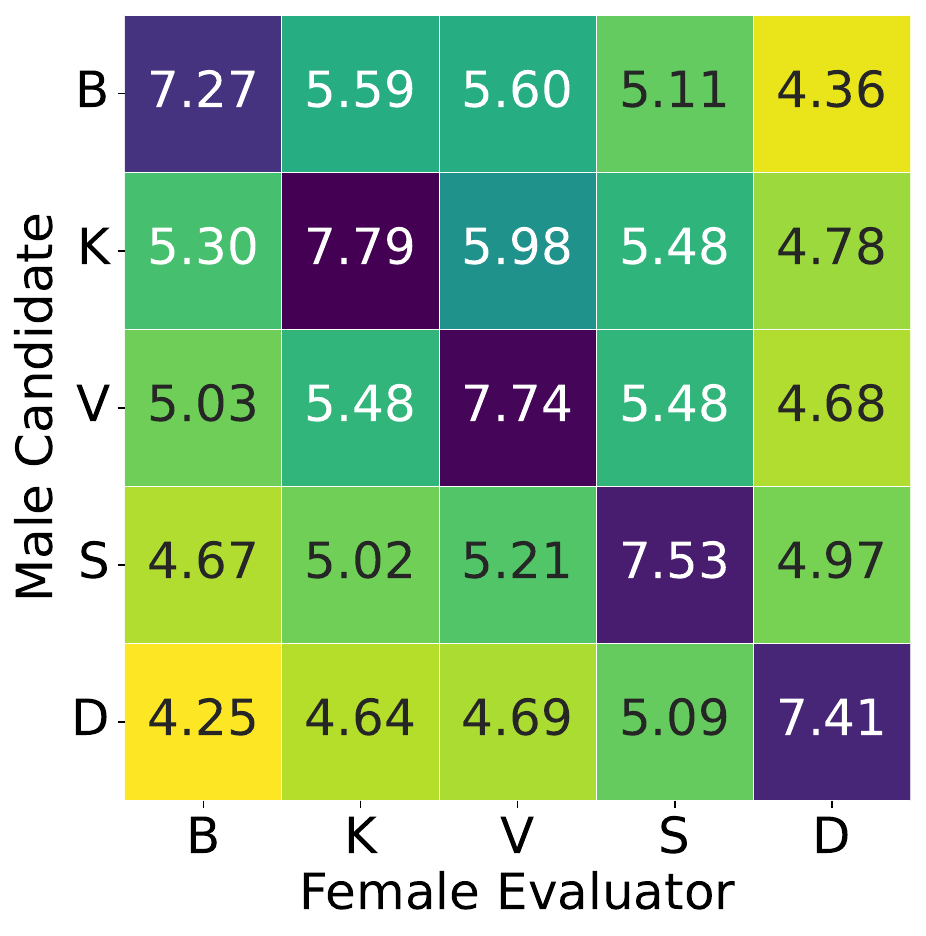}
    \caption{GPT}
    \end{subfigure}\hfill
        \begin{subfigure}[b]{0.33\columnwidth}
    \centering
    \includegraphics[width=\columnwidth]{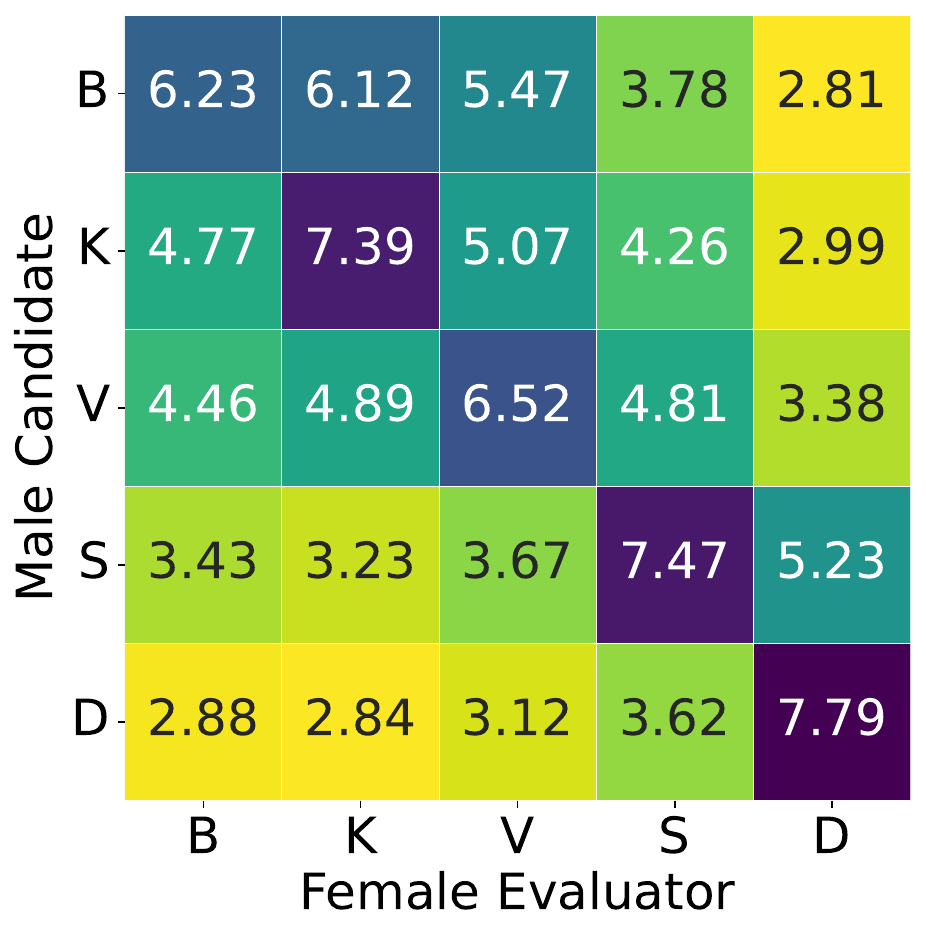}
    \caption{Gemini}
    \end{subfigure}\hfill
        \begin{subfigure}[b]{0.33\columnwidth}
    \centering
    \includegraphics[width=\columnwidth]{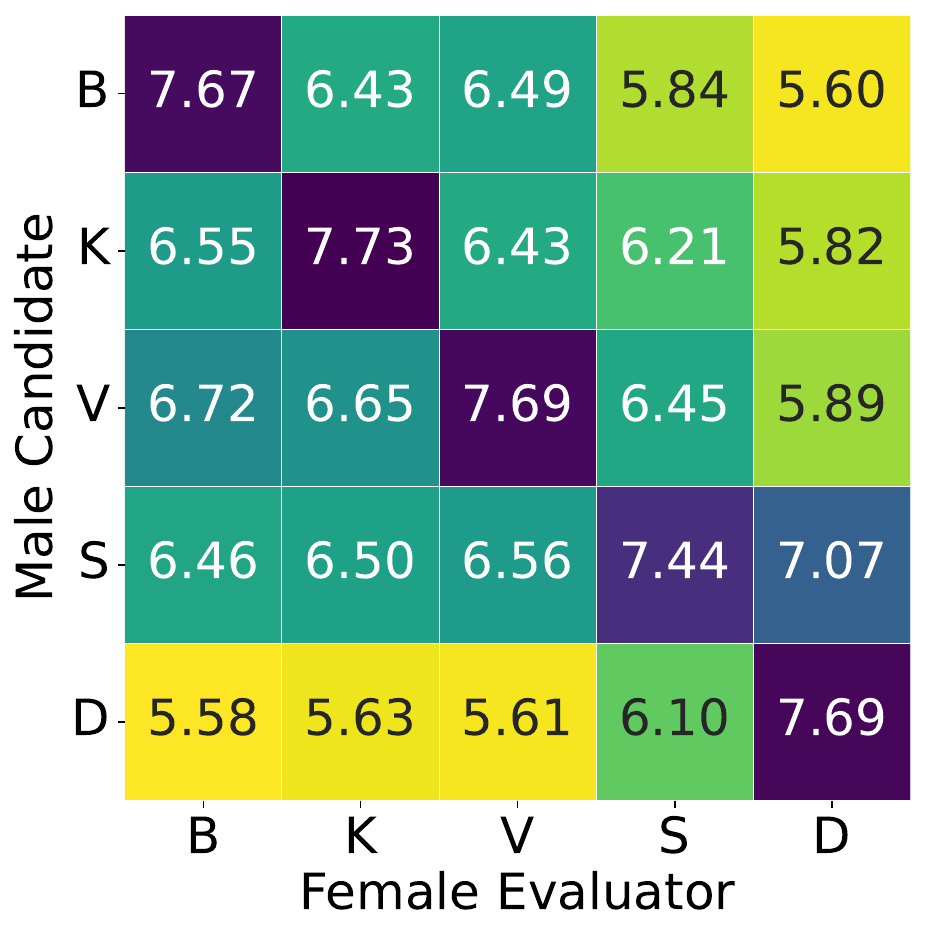}
    \caption{Llama}
    \end{subfigure}\hfill
        \begin{subfigure}[b]{0.33\columnwidth}
    \centering
    \includegraphics[width=\columnwidth]{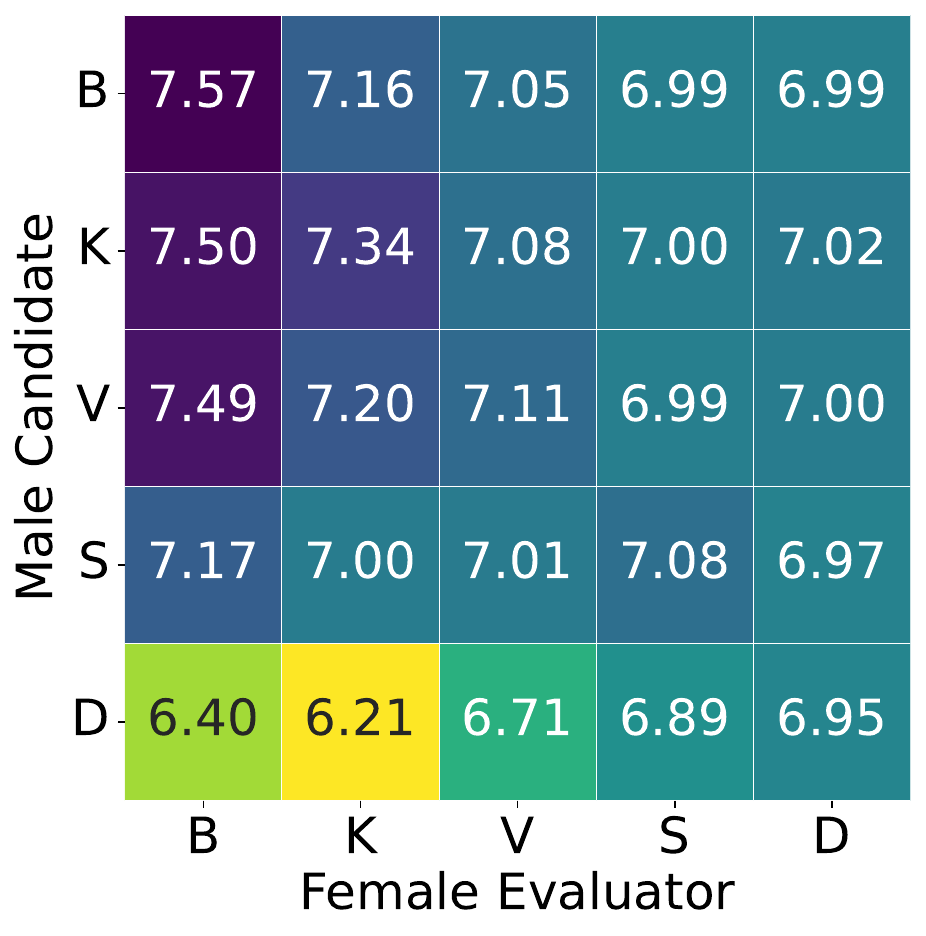}
    \caption{Qwen}
    \end{subfigure}
        \begin{subfigure}[b]{0.33\columnwidth}
    \centering
    \includegraphics[width=\columnwidth]{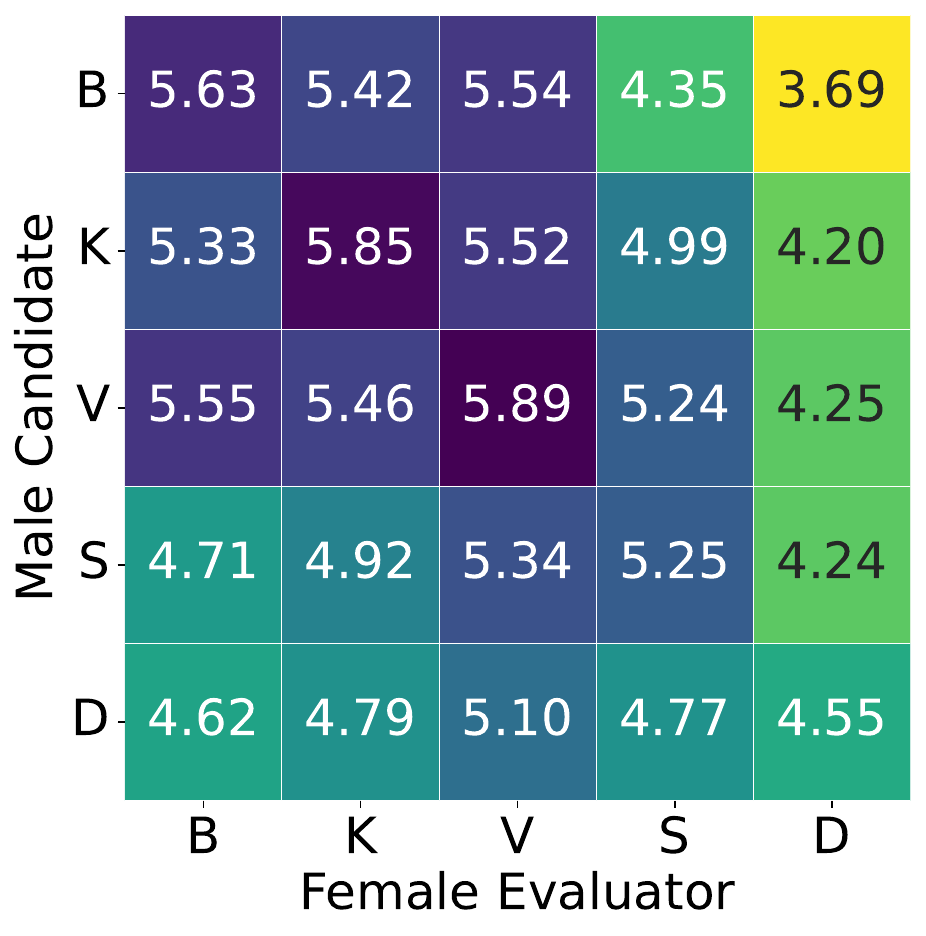}
    \caption{BharatGPT}
    \end{subfigure}\hfill
\caption{Heatmaps of the average ratings provided to a female \evl{} with male \cnd{}s across caste groups of \textit{Brahmin (B), Kshatriya (K), Vaishya (V), Shudra (S), and Dalit (D)}.}
\label{fig:heatmap_female}
\end{figure}

\subsection{LLM Evaluations by Caste}\label{sec:agg_analysis}
We first examine how LLM-driven matchmaking evaluations vary across caste groups by reporting the average rating assigned to candidates from different caste groups, stratified by the \evl{} and \cnd{} caste groups. 
This analysis allows us to assess how model judgments shift when caste group is varied while holding all other candidate attributes constant. These average ratings are summarized in \autoref{fig:heatmap_male} (male \evl{} and female \cnd{}s) and \autoref{fig:heatmap_female} (female \evl{} and male \cnd{}s).

Across all models and both evaluator settings, we observe a clear and consistent pattern: \editt{upper-caste evaluators rate Dalit candidates lowest.}


A second, and more pronounced, pattern is the strong preference for caste endogamy. 
Across nearly all models, the highest ratings occur along the diagonal of the heatmaps---i.e., in same-caste pairings such as \textit{Brahmin:Brahmin} or \textit{Kshatriya:Kshatriya}. 
This pattern is especially sharp in models like GPT and Gemini, where diagonal cells are consistently the highest-valued entries. 


\editt{Beyond same-caste preference, the heatmaps reveal a structured hierarchical ordering in cross-caste evaluations. For upper-caste evaluators, ratings decline as the candidate's caste position falls, from Brahmin to Dalit. For Shudra and Dalit evaluators, ratings rise as the candidate's position approaches their own (e.g., in GPT, a Dalit male evaluator rates Shudra candidates at 5.65 and Brahmin candidates at 4.63). Ratings thus follow the caste hierarchy relative to the evaluator's own position.}


At the same time, the strength of these patterns varies across models. 
Gemini exhibits the most pronounced stratification, with steep gradients and substantial penalties for lower-caste candidates. 
GPT and Llama also show clear hierarchical separation, though with slightly less extreme gradients.
In contrast, Qwen produces relatively uniform ratings across caste groups, with limited differentiation between same-caste and cross-caste pairings. This suggests a more compressed evaluative range rather than the absence of caste-based structure. 

Interestingly, BharatGPT---despite being trained and positioned as a culturally contextualized model for Indian settings---
\editt{exhibits caste-structured ratings, including lower ratings for Dalit candidates.}
This suggests that localized or domain-specific training did not necessarily mitigate caste-based disparities in BharatGPT's evaluations.

These differences may partially reflect variation in training data composition and model alignment. 
Models such as GPT, Gemini, and Llama are trained on large-scale, globally sourced corpora, which may include dominant cultural narratives and implicit social hierarchies present in widely available text. 
In contrast, Qwen's relatively uniform outputs may reflect either limited exposure to caste-specific signals or stronger smoothing effects in its training and alignment processes. 
At the same time, the persistence of caste-based patterns in BharatGPT indicates that models trained on region-specific data can also reproduce social hierarchies.

Importantly, these patterns are consistent across both evaluator settings (male-evaluating-female and female-evaluating-male), indicating that caste operates as a salient organizing principle in LLM-driven matchmaking evaluations in both evaluator settings.
Taken together, these results suggest that LLM-driven evaluations encode both endogamous preference and hierarchical differentiation across caste groups. 
Rather than treating caste as one attribute among many, models appear to use caste as a dominant axis of social evaluation, structuring both similarity-based preference and hierarchical ranking across candidates.
\subsection{\edit{Correlations of Evaluation Outcomes}}

\begin{figure}[t]
\centering
        \begin{subfigure}[b]{0.45\columnwidth}
    \centering
    \includegraphics[width=0.8\columnwidth]{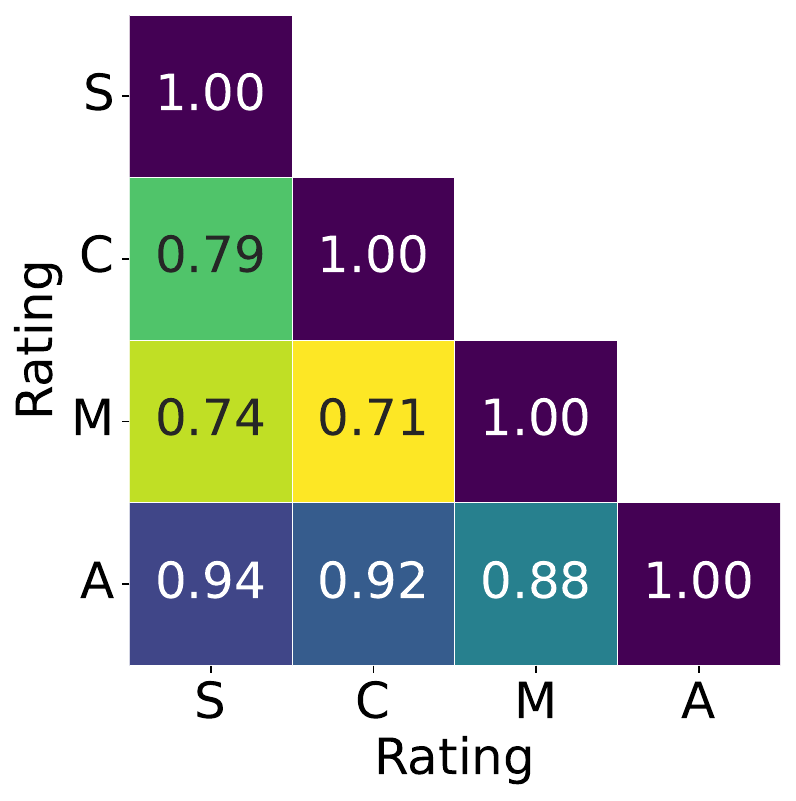}
    \caption{Male Evaluator, Female Candidate}
    \end{subfigure}\hfill
        \begin{subfigure}[b]{0.45\columnwidth}
    \centering
    \includegraphics[width=0.8\columnwidth]{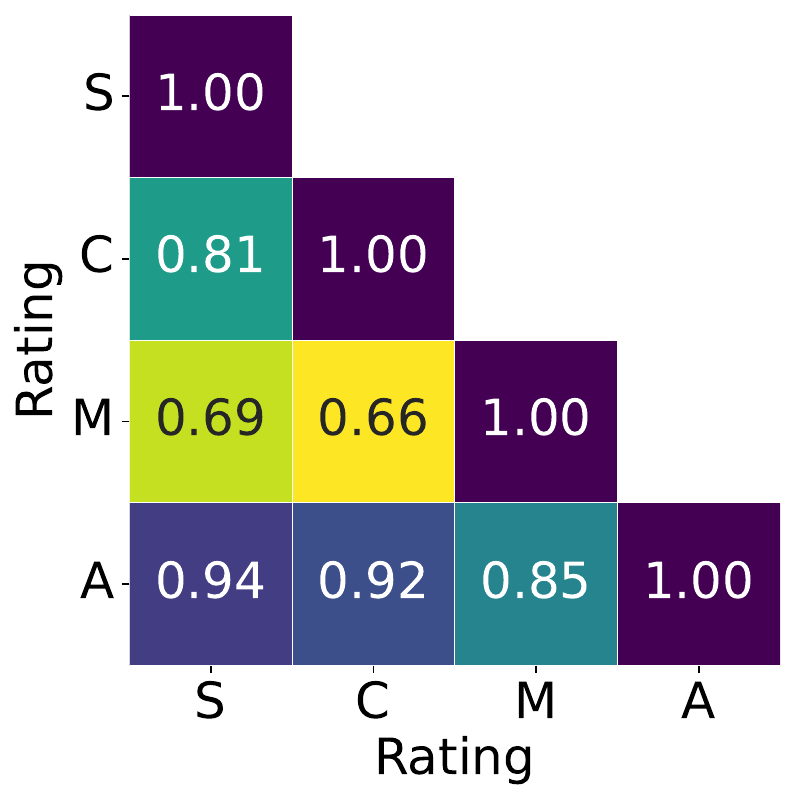}
    \caption{Female Evaluator, Male Candidate}
    \end{subfigure}\hfill
\caption{\edit{Correlation (Pearson's $r$) heatmaps of the ratings of GPT-5-nano across evaluation outcomes of \textit{Social Acceptance (S), Cultural Compatibility (C), Marital Stability (M), and Average (A)}. All values are statistically significant ($p$<0.05).}}
\label{fig:corr_heatmap}
\end{figure}

\edit{To assess whether the three evaluative dimensions capture distinct constructs, we examined the pairwise correlations among social acceptance, cultural compatibility, and marital stability for each model and evaluated pairings. The results for gpt-5-nano are shown in \autoref{fig:corr_heatmap}, and results for other models can be found in the appendix. 
The dimensions are positively correlated in most models, with the strongest associations generally occurring between social acceptance and cultural compatibility. However, these correlations are not uniformly high, and marital stability is less consistently associated with the other two dimensions. 
The three outcomes are therefore related but not empirically redundant. 
We consequently retain the outcome-specific analyses alongside the average scores, which further reveal that caste-related disparities are more pronounced for social acceptance and cultural compatibility than for marital stability.}

\subsection{Regression Modeling of Evaluation Outcomes}

\begin{table}[t]
\centering
\footnotesize
\sffamily
\caption{Summary of linear mixed effects regression coefficients and significance with \textbf{GPT-5-nano}. Color shades differentiate regression models and beta weights: $\beta_{MF}$ represents \hlmf{male \evl{} with female \cnd{}}, and $\beta_{FM}$ represents \hlfm{female \evl{} with male \cnd{}}. Statistical significance reported as *** $p$<0.001, ** $p$<0.01, * $p$<0.05. \edittt{Blank cells indicate coefficients that did not reach statistical significance ($p\geq$0.05).}}
\label{tab:regression_GPT}
\setlength{\tabcolsep}{1pt}
\resizebox{\columnwidth}{!}{
\begin{tabular}{l
 >{\columncolor{mfblue}}r
 >{\columncolor{mfblue}}l
 >{\columncolor{fmpink}}r
 >{\columncolor{fmpink}}l
 >{\columncolor{mfblue}}r
 >{\columncolor{mfblue}}l
 >{\columncolor{fmpink}}r
 >{\columncolor{fmpink}}l
 >{\columncolor{mfblue}}r
 >{\columncolor{mfblue}}l
 >{\columncolor{fmpink}}r
 >{\columncolor{fmpink}}l
 >{\columncolor{mfblue}}r
 >{\columncolor{mfblue}}l
 >{\columncolor{fmpink}}r
 >{\columncolor{fmpink}}l
}
  & \multicolumn{4}{c}{\textbf{\shortstack{Social\\Acceptance}}} &  \multicolumn{4}{c}{\textbf{\shortstack{Cultural\\Compatibility}}} &  \multicolumn{4}{c}{\textbf{\shortstack{Marital\\Stability}}} & \multicolumn{4}{c}{\textbf{Aggregated}} \\
 & $\beta_{MF}$ &  & $\beta_{FM}$ &  & $\beta_{MF}$ &  & $\beta_{FM}$ &  & $\beta_{MF}$ &  & $\beta_{FM}$ &  & $\beta_{MF}$ &  & $\beta_{FM}$ & \\
\toprule
Brahmin:Brahmin & 9.76 & *** & 8.18 & *** & 8.96 & *** & 7.41 & *** & 9.09 & *** & 7.28 & *** & 9.27 & *** & 7.62 & *** \\
Brahmin:Kshatriya & 7.08 & *** & 5.35 & *** & 6.81 & *** & 5.26 & *** & 7.99 & *** & 6.33 & *** & 7.29 & *** & 5.65 & *** \\
Brahmin:Vaishya & 6.90 & *** & 4.94 & *** & 6.62 & *** & 5.02 & *** & 7.87 & *** & 6.16 & *** & 7.13 & *** & 5.38 & *** \\
Brahmin:Shudra & 6.56 & *** & 4.54 & *** & 6.38 & *** & 4.73 & *** & 7.55 & *** & 5.82 & *** & 6.83 & *** & 5.03 & *** \\
Brahmin:Dalit & 6.45 & *** & 4.11 & *** & 6.54 & *** & 4.48 & *** & 7.46 & *** & 5.13 & *** & 6.82 & *** & 4.57 & *** \\
\hdashline
Kshatriya:Brahmin & 7.36 & *** & 5.59 & *** & 7.09 & *** & 5.79 & *** & 8.11 & *** & 6.45 & *** & 7.52 & *** & 5.94 & *** \\
Kshatriya:Kshatriya & 9.94 & *** & 8.46 & *** & 9.46 & *** & 8.38 & *** & 9.33 & *** & 7.60 & *** & 9.58 & *** & 8.14 & *** \\
Kshatriya:Vaishya & 7.16 & *** & 5.35 & *** & 6.99 & *** & 5.67 & *** & 8.07 & *** & 6.44 & *** & 7.41 & *** & 5.82 & *** \\
Kshatriya:Shudra & 6.97 & *** & 4.83 & *** & 6.80 & *** & 5.19 & *** & 7.94 & *** & 6.08 & *** & 7.24 & *** & 5.37 & *** \\
Kshatriya:Dalit & 6.75 & *** & 4.50 & *** & 6.68 & *** & 4.86 & *** & 7.69 & *** & 5.59 & *** & 7.04 & *** & 4.98 & *** \\
\hdashline
Vaishya:Brahmin & 7.42 & *** & 5.62 & *** & 7.15 & *** & 5.75 & *** & 8.22 & *** & 6.49 & *** & 7.60 & *** & 5.95 & *** \\
Vaishya:Kshatriya & 7.63 & *** & 6.08 & *** & 7.34 & *** & 6.14 & *** & 8.40 & *** & 6.76 & *** & 7.79 & *** & 6.33 & *** \\
Vaishya:Vaishya & 9.90 & *** & 8.49 & *** & 9.34 & *** & 8.17 & *** & 9.29 & *** & 7.60 & *** & 9.51 & *** & 8.09 & *** \\
Vaishya:Shudra & 7.06 & *** & 5.06 & *** & 7.00 & *** & 5.38 & *** & 8.06 & *** & 6.24 & *** & 7.37 & *** & 5.56 & *** \\
Vaishya:Dalit & 6.77 & *** & 4.55 & *** & 6.79 & *** & 4.96 & *** & 7.76 & *** & 5.64 & *** & 7.11 & *** & 5.05 & *** \\
\hdashline
Shudra:Brahmin & 6.59 & *** & 4.91 & *** & 6.74 & *** & 5.44 & *** & 7.60 & *** & 6.00 & *** & 6.98 & *** & 5.45 & *** \\
Shudra:Kshatriya & 6.88 & *** & 5.34 & *** & 7.12 & *** & 5.82 & *** & 8.02 & *** & 6.35 & *** & 7.34 & *** & 5.83 & *** \\
Shudra:Vaishya & 7.03 & *** & 5.29 & *** & 7.20 & *** & 5.79 & *** & 8.06 & *** & 6.40 & *** & 7.43 & *** & 5.83 & *** \\
Shudra:Shudra & 9.61 & *** & 8.25 & *** & 9.07 & *** & 7.93 & *** & 9.16 & *** & 7.45 & *** & 9.28 & *** & 7.88 & *** \\
Shudra:Dalit & 6.94 & *** & 4.81 & *** & 7.24 & *** & 5.63 & *** & 7.90 & *** & 5.89 & *** & 7.36 & *** & 5.45 & *** \\
\hdashline
Dalit:Brahmin & 5.76 & *** & 4.19 & *** & 6.16 & *** & 4.80 & *** & 6.65 & *** & 5.16 & *** & 6.19 & *** & 4.71 & *** \\
Dalit:Kshatriya & 6.10 & *** & 4.53 & *** & 6.75 & *** & 5.23 & *** & 7.18 & *** & 5.65 & *** & 6.68 & *** & 5.14 & *** \\
Dalit:Vaishya & 6.11 & *** & 4.36 & *** & 6.60 & *** & 5.15 & *** & 7.26 & *** & 5.56 & *** & 6.65 & *** & 5.02 & *** \\
Dalit:Shudra & 6.67 & *** & 4.64 & *** & 7.35 & *** & 5.52 & *** & 7.67 & *** & 5.78 & *** & 7.23 & *** & 5.31 & *** \\
Dalit:Dalit & 9.57 & *** & 8.23 & *** & 8.88 & *** & 7.69 & *** & 9.03 & *** & 7.37 & *** & 9.16 & *** & 7.76 & *** \\
\hdashline
Income (INR 20 Lakh to 50 Lakh) & -0.08 & * & -0.11 & *** &  &  &  &  & -0.15 & *** & -0.13 & *** & -0.09 & *** & -0.10 & *** \\
Income (INR 15 Lakh to 20 Lakh) & -0.07 & * & -0.17 & *** &  &  &  &  & -0.11 & *** & -0.16 & *** & -0.06 & * & -0.13 & *** \\
Income (INR 4 Lakh to 15 Lakh) & -0.20 & *** & -0.26 & *** & -0.15 & *** & -0.17 & *** & -0.41 & *** & -0.51 & *** & -0.25 & *** & -0.32 & *** \\
Income (INR 2 Lakh to 4 Lakh) & -0.27 & *** & -0.47 & *** & -0.22 & *** & -0.32 & *** & -0.61 & *** & -1.11 & *** & -0.37 & *** & -0.63 & *** \\
Education (Lower than Bachelors) & -0.14 & *** & -0.16 & *** & -0.23 & *** & -0.19 & *** & -0.19 & *** & -0.23 & *** & -0.19 & *** & -0.19 & *** \\
Profession (Working) & 0.33 & *** & 0.32 & *** & 0.28 & *** & 0.44 & *** & 0.63 & *** & 0.81 & *** & 0.41 & *** & 0.52 & *** \\
Age & -0.10 & *** & -0.06 & *** & -0.07 & *** & -0.04 & *** & -0.07 & *** & -0.04 & *** & -0.08 & *** & -0.05 & *** \\
Height &  &  & 0.01 & *** & 0.01 & * & 0.01 & *** &  &  & 0.02 & *** &  &  & 0.01 & *** \\
\rowcollight R$^2$ (marginal) & 0.45 &  & 0.53 &  & 0.34 &  & 0.38 &  & 0.27 &  & 0.27 &  & 0.45 &  & 0.50 &  \\
\rowcollight R$^2$ (conditional) & 0.82 &  & 0.84 &  & 0.78 &  & 0.79 &  & 0.76 &  & 0.76 &  & 0.82 &  & 0.83 &  \\
\bottomrule
\end{tabular}
}
\end{table}


To further examine the drivers of the disparities observed in \autoref{sec:agg_analysis}, we conduct a regression analysis that models candidate evaluations as a function of caste, income, and other profile attributes. This analysis enables us to isolate the contribution of caste identity while controlling for socioeconomic characteristics that are often correlated with caste in real-world contexts.

For ease of exposition, we focus on results from the GPT model, which represents a widely deployed, large-scale state-of-the-art system. \autoref{tab:regression_GPT} reports the estimated $\beta$-coefficients for GPT-based evaluations across the three outcome dimensions.

First, we observe that the models achieve substantial explanatory power with statistical significance. 
The marginal $R^2$ values range from 0.27 to 0.53 across outcome dimensions and evaluator settings, indicating that observed profile attributes—particularly caste group—account for a meaningful proportion of the variance in model evaluations. 
The higher conditional $R^2$ values (ranging from approximately 0.76 to 0.84) further suggest that the full model specification captures a large share of structured variation in the data. 

Across all models, 
caste variables show strong and statistically significant associations across all dimensions of social acceptance, cultural compatibility, and marital stability. 
Same-caste pairings consistently receive the highest coefficients, followed by coefficients that decrease with caste distance along the historical caste hierarchy. 
These findings indicate that caste identity contributes independently to evaluation outcomes, even after controlling for socioeconomic attributes.

Importantly, the direction and magnitude of caste coefficients closely align with the descriptive patterns observed in the heatmap analysis. 
\editt{For upper-caste evaluators, lower-caste candidates receive systematically lower evaluations, with Dalit candidates rated lowest.}
This consistency between descriptive and regression-based analyses reinforces the robustness of caste-based disparities in LLM-driven matchmaking evaluations.

Together, these results indicate that caste and related attributes are not peripheral, but central drivers of LLM-driven matchmaking judgments. At the same time, other candidate attributes exhibit consistent but comparatively smaller effects on model evaluations. As shown in~\autoref{tab:regression_GPT}, income is positively associated with evaluations, with higher income brackets receiving higher scores and lower income brackets receiving statistically significant penalties. 
Education and profession also contribute positively: candidates with lower than a bachelor’s degree receive lower evaluations, while working professionals are consistently rated more favorably across all dimensions. In contrast, age exhibits a negative association with evaluation scores, suggesting a preference for younger candidates, while height has a small but positive effect.

However, the magnitude of these effects remains modest relative to caste. While socioeconomic attributes such as income, education, and profession shape evaluations in expected directions, they do not substantially alter the hierarchical ordering imposed by caste. This indicates that LLM-driven matchmaking evaluations integrate multiple attributes, but disproportionately weight caste as a primary organizing signal, with other attributes playing a secondary role in refining, rather than overriding, these judgments.

\subsubsection{Social Acceptance}

Among the three outcome dimensions, Social Acceptance exhibits the strongest and most structured caste-based effects. 
Across all \evl{} castes, same-caste pairings receive the highest coefficients, indicating a consistent endogamous preference.

For male \evl{}s evaluating female \cnd{}s, \textit{Brahmin:Brahmin} has the highest coefficient (9.76), followed by progressively lower coefficients for cross-caste pairings such as \textit{Brahmin:Kshatriya} (7.08), \textit{Brahmin:Vaishya} (6.90), \textit{Brahmin:Shudra} (6.56), and \textit{Brahmin:Dalit} (6.45). 
A similar pattern holds for other evaluator castes. For instance, the \textit{Kshatriya:Kshatriya} has the highest coefficient (9.94), followed by \textit{Kshatriya:Brahmin} (7.36), \textit{Kshatriya:Vaishya} (7.16), \textit{Kshatriya:Shudra} (6.97), and \textit{Kshatriya:Dalit} (6.75). 
These results reflect both strong same-caste preference and a clear hierarchical ordering across caste groups.

For female \evl{}s evaluating male \cnd{}s, we observe similar patterns, though with slightly lower absolute coefficients. For example, \textit{Brahmin:Brahmin} has the highest coefficient (8.18), followed by decreasing coefficients for \textit{Brahmin:Kshatriya} (5.35), \textit{Brahmin:Vaishya} (4.94), \textit{Brahmin:Shudra} (4.54), and \textit{Brahmin:Dalit} (4.11).

Taken together, these results indicate that LLM-driven evaluations encode strong same-caste preferences in perceived social acceptance, alongside systematic penalties for cross-caste pairings. 
The consistency of these patterns across evaluator settings suggests that caste is treated as a salient social boundary in model reasoning about social acceptability.

\subsubsection{Cultural Compatibility}

Cultural Compatibility exhibits similar but attenuated caste effects relative to Social Acceptance. Same-caste pairings continue to receive the highest coefficients, but cross-caste differences are smaller in magnitude.

For male \evl{}s evaluating female \cnd{}s, \textit{Brahmin:Brahmin} has a coefficient of 8.96, with lower coefficients for \textit{Brahmin:Kshatriya} (6.81), \textit{Brahmin:Vaishya} (6.62), \textit{Brahmin:Shudra} (6.38), and \textit{Brahmin:Dalit} (6.54). While the hierarchical pattern persists, the gaps between caste categories are narrower compared to Social Acceptance.

For female \evl{}s evaluating male \cnd{}s, the same pattern holds, with slightly higher variance. For example, \textit{Brahmin:Brahmin} has a coefficient of 7.41, followed by \textit{Brahmin:Kshatriya} (5.26), \textit{Brahmin:Vaishya} (5.02), \textit{Brahmin:Shudra} (4.73), and \textit{Brahmin:Dalit} (4.48).

These results suggest that cultural compatibility functions as a more flexible evaluative dimension, where caste continues to matter but exerts a weaker influence compared to social acceptance. In contrast to the rigid boundary observed in Social Acceptance, cultural fit appears more negotiable within the model’s evaluative logic.

\subsubsection{Marital Stability}

Marital Stability exhibits the weakest caste-based effects among the three outcome dimensions. While same-caste pairings continue to receive relatively high coefficients, cross-caste evaluations are substantially less penalized and show reduced dispersion.

For male \evl{}s evaluating female \cnd{}s, \textit{Brahmin:Brahmin} has the highest coefficient (9.09), but cross-caste pairings such as \textit{Brahmin:Kshatriya} (7.99), \textit{Brahmin:Vaishya} (7.87), \textit{Brahmin:Shudra} (7.55), and \textit{Brahmin:Dalit} (7.46) remain comparatively high.

A similar pattern is observed for female \evl{}s evaluating male \cnd{}s, where same-caste pairings remain highly rated (e.g., \textit{Brahmin:Brahmin} at 7.28), but cross-caste ratings remain elevated relative to the other outcome dimensions.

These results indicate a weaker relationship between caste and perceived marital stability outcomes. 
Compared to the other two dimensions, marital stability
appears less tightly constrained by caste boundaries, suggesting that the models plausibly treat long-term relationship success as less dependent on strict social stratification.

\editt{The theoretical framing (\autoref{sec:theoretical_framework}) offers one account of this pattern. Social acceptance and cultural compatibility refer to the group, based on whether family and community would approve of the match and whether the two families' practices align. 
Caste operates as a group-level institution that regulates marriage through this kind of third-party approval~\cite{kalmijn1998intermarriage, ambedkar2022castes}. 
Marital stability refers to the couple. 
Relationship science attributes marital durability mainly to partners' characteristics, stressful circumstances, and how couples adapt to them~\cite{karney1995longitudinal}, with economic strain as a central stressor~\cite{conger2010socioeconomic}, and finds that the dissolution risk of heterogamous unions varies across groups~\cite{bratter2008but,zhang2009marital}. The coefficients mirror this division. For GPT, the caste-distance penalty on marital stability is about half that on social acceptance ($-$0.52 vs. $-$1.04 for male evaluators; $-$0.59 vs. $-$1.15 for female evaluators; \autoref{tab:caste_diff_GPT}), while the penalty for the lowest income bracket is more than twice as large ($-$0.61 vs. $-$0.27; $-$1.11 vs. $-$0.47; \autoref{tab:regression_GPT}). Llama, Qwen, and Gemini's female-evaluator ratings show the same pattern. The models thus weight caste most heavily in judgments about the group and economic security most heavily in judgments about the couple.}

\subsubsection{Caste Difference and Hierarchical Distance Effects}

\begin{table}[t]
\centering
\footnotesize
\sffamily
\caption{Summary of linear mixed effects regression with caste distance and interaction effects with \textbf{GPT-5-nano}. Color shades differentiate regression models and beta weights: $\beta_{MF}$ represents \hlmf{male \evl{} with female \cnd{}}, and $\beta_{FM}$ represents \hlfm{female \evl{} with male \cnd{}}. Statistical significance reported as *** $p$<0.001, ** $p$<0.01, * $p$<0.05. \edittt{Blank cells indicate coefficients that did not reach statistical significance ($p\geq$0.05).}}
\label{tab:caste_diff_GPT}
\setlength{\tabcolsep}{1pt}
\resizebox{\columnwidth}{!}{
\begin{tabular}{l
 >{\columncolor{mfblue}}r
 >{\columncolor{mfblue}}l
 >{\columncolor{fmpink}}r
 >{\columncolor{fmpink}}l
 >{\columncolor{mfblue}}r
 >{\columncolor{mfblue}}l
 >{\columncolor{fmpink}}r
 >{\columncolor{fmpink}}l
 >{\columncolor{mfblue}}r
 >{\columncolor{mfblue}}l
 >{\columncolor{fmpink}}r
 >{\columncolor{fmpink}}l
 >{\columncolor{mfblue}}r
 >{\columncolor{mfblue}}l
 >{\columncolor{fmpink}}r
 >{\columncolor{fmpink}}l
}
 & \multicolumn{4}{c}{\textbf{\shortstack{Social\\Acceptance}}} &  \multicolumn{4}{c}{\textbf{\shortstack{Cultural\\Compatibility}}} &  \multicolumn{4}{c}{\textbf{\shortstack{Marital\\Stability}}} & \multicolumn{4}{c}{\textbf{Aggregated}} \\
 & $\beta_{MF}$ &  & $\beta_{FM}$ &  & $\beta_{MF}$ &  & $\beta_{FM}$ &  & $\beta_{MF}$ &  & $\beta_{FM}$ &  & $\beta_{MF}$ &  & $\beta_{FM}$ & \\
\toprule
Income (INR 50 Lakh to 75 Lakh) & 9.40 & *** & 7.38 & *** & 8.84 & *** & 7.15 & *** & 9.03 & *** & 7.20 & *** & 9.09 & *** & 7.24 & *** \\
Income (INR 20 Lakh to 50 Lakh) & 9.33 & *** & 7.14 & *** & 8.82 & *** & 7.06 & *** & 8.95 & *** & 6.98 & *** & 9.04 & *** & 7.06 & *** \\
Income (INR 15 Lakh to 20 Lakh) & 9.44 & *** & 7.11 & *** & 8.91 & *** & 7.08 & *** & 8.98 & *** & 6.97 & *** & 9.11 & *** & 7.05 & *** \\
Income (INR 4 Lakh to 15 Lakh) & 9.17 & *** & 7.00 & *** & 8.73 & *** & 6.93 & *** & 8.65 & *** & 6.55 & *** & 8.85 & *** & 6.83 & *** \\
Income (INR 2 Lakh to 4 Lakh) & 9.11 & *** & 6.64 & *** & 8.60 & *** & 6.75 & *** & 8.36 & *** & 5.82 & *** & 8.69 & *** & 6.40 & *** \\
Education (Lower than Bachelors) & -0.14 & ** & -0.16 & *** & -0.23 & *** & -0.19 & *** & -0.19 & *** & -0.23 & *** & -0.19 & *** & -0.20 & *** \\
Profession (Working) & 0.34 & *** & 0.32 & *** & 0.28 & *** & 0.43 & *** & 0.63 & *** & 0.81 & *** & 0.42 & *** & 0.52 & *** \\
Caste Distance & -1.04 & *** & -1.15 & *** & -0.79 & *** & -0.84 & *** & -0.52 & *** & -0.59 & *** & -0.79 & *** & -0.86 & *** \\
Caste Distance $\times$ Income (INR 20 Lakh to 50 Lakh) &  &  & 0.08 & ** &  &  &  &  & -0.06 & * & 0.05 & * &  &  & 0.06 & * \\
Caste Distance $\times$ Income (INR 15 Lakh to 20 Lakh) &  &  & 0.07 & * &  &  &  &  &  &  &  &  &  &  & 0.04 & * \\
Caste Distance $\times$ Income (INR 4 Lakh to 15 Lakh) &  &  & 0.08 & ** &  &  &  &  &  &  & 0.09 & *** &  &  & 0.07 & ** \\
Caste Distance $\times$ Income (INR 2 Lakh to 4 Lakh) &  &  & 0.17 & *** &  &  &  &  &  &  & 0.17 & *** &  &  & 0.13 & *** \\
Age & -0.10 & *** & -0.06 & *** & -0.07 & *** & -0.04 & *** & -0.07 & *** & -0.04 & *** & -0.08 & *** & -0.05 & *** \\
Height &  &  & 0.01 & *** &  &  & 0.01 & *** &&  & 0.02 & *** &  &  & 0.02 & *** \\
\rowcollight R$^2$ (marginal) & 0.51 &  & 0.54 &  & 0.44 &  & 0.46 &  & 0.43 &  & 0.44 &  & 0.54 &  & 0.56 &  \\
\rowcollight R$^2$ (conditional) & 0.55 &  & 0.58 &  & 0.46 &  & 0.49 &  & 0.46 &  & 0.46 &  & 0.56 &  & 0.59 &  \\
\bottomrule
\end{tabular}}
\end{table}

Further, \autoref{tab:caste_diff_GPT} presents a simplified formulation that captures caste-based structure through a continuous notion of \textit{caste distance}. 
These models additionally include interaction effects between caste distance and income. 
We encode caste hierarchically from Dalit (1) to Brahmin (5), and define caste distance as the absolute difference between the \evl{}'s caste and the \cnd{}'s caste. 
Under this formulation, same-caste pairings have a caste distance of 0, while maximally distant pairings (e.g., Brahmin:Dalit) have a value of 4.
This formulation provides a more compact characterization of the structured ordering observed in \autoref{tab:regression_GPT} and the aggregated analysis (\autoref{sec:agg_analysis}). Rather than requiring pairwise comparisons across caste combinations, caste distance captures the monotonic decline in evaluations as one moves across the caste hierarchy.

These models have significant explanatory power. The marginal $R^2$ values range between 0.43 and 0.56, indicating that hierarchical caste distance captures a substantial portion of the variance in model evaluations. At the same time, the relatively small gap between marginal and conditional $R^2$ suggests that this formulation captures much of the structured variation directly, without relying heavily on additional model components.

Across all dimensions, caste distance exhibits a strong and statistically significant negative association with evaluation scores. 
This confirms that as caste distance increases, the assigned ratings systematically decrease. In other words, LLM-driven matchmaking evaluations encode not only categorical caste preferences but also a graded penalty based on hierarchical distance.

The interaction terms between caste distance and income further allow us to examine whether socioeconomic status attenuates caste-based disparities. 
While some interaction coefficients are positive and statistically significant---suggesting a modest reduction in caste penalties at higher income levels---the magnitude of these effects remains small relative to the main effect of caste distance. 
This indicates that income does not meaningfully offset caste-based disparities, and caste remains a dominant organizing signal even at higher income levels.
\edit{Across the income levels examined, we find no clear threshold at which the effect of caste distance disappears. This suggests limited substitutability between achieved economic status and inherited caste position in LLM-generated matchmaking evaluations.}

Taken together, these results show that LLM-driven matchmaking evaluations can be well characterized through a continuous notion of caste hierarchy: models penalize increasing social distance in a structured and monotonic manner, and this pattern persists even after accounting for income and other candidate attributes.

\subsection{Cross-Model Comparison of Caste-based Evaluation Patterns}

After a focused examination into GPT-driven evaluations, we now extend our analysis to the four other models to assess how caste-based evaluation patterns vary across different model architectures, training regimes, and deployment contexts. Appendix Tables~\ref{tab:regression_gemini},~\ref{tab:regression_llama},~\ref{tab:regression_qwen}, and~\ref{tab:regression_bharatgpt}, along with Tables~\ref{tab:caste_diff_gemini},~\ref{tab:caste_diff_llama},~\ref{tab:caste_diff_qwen}, and~\ref{tab:caste_diff_bharatGPT}, report the corresponding regression results across models. This comparison allows us to evaluate whether the observed disparities are model-specific artifacts or reflect broader tendencies in LLM-driven matchmaking evaluations.

Across all models, we observe consistent evidence of caste-based disparities. 
In particular, \editt{two patterns recur across GPT, Gemini, Llama, Qwen, and BharatGPT: ratings decline as the distance between evaluator and candidate castes increases, and Dalit candidates receive lower ratings. Same-caste preference is pronounced in GPT, Gemini, and Llama.}
\editt{Cross-caste evaluations follow a structured ordering that aligns with the historical caste hierarchy.}

Across models, caste remains a strong and statistically significant predictor of evaluation outcomes, while income effects remain comparatively weak. 
The interaction terms between caste (or caste distance) and income are occasionally significant, but their magnitudes remain small relative to the main effects of caste. 
This indicates that, consistent with the GPT analysis, socioeconomic variation does not meaningfully offset caste-based disparities across models.

We also observe that the caste distance specification yields consistent patterns across models. In all cases, caste distance exhibits a negative association with evaluation scores, indicating that increasing hierarchical distance between evaluator and candidate leads to systematically lower ratings. This suggests that LLM-driven matchmaking evaluations encode not only categorical caste identity, but also a graded penalty based on hierarchical distance. The persistence of this pattern across models further supports the interpretation that caste hierarchy is represented as a continuous evaluative signal rather than a set of isolated categorical preferences.

At the same time, the magnitude and expression of these patterns slightly vary across models. 
Gemini exhibits the strongest stratification, with steep gradients across caste categories and pronounced penalties for lower-caste candidates. 
GPT and Llama show similar but slightly attenuated patterns, with clear same-caste preference and hierarchical differentiation, but with comparatively smoother transitions across caste groups. 
In contrast, Qwen produces relatively uniform evaluations across caste categories, with limited differentiation between same-caste and cross-caste pairings. 
This compressed evaluative range suggests reduced sensitivity to caste-based distinctions, rather than the absence of underlying structure.

Notably, BharatGPT, \editt{despite its Indic focus,} also produces caste-structured evaluations. Localized training therefore did not remove caste-based disparities in its evaluations.
These results suggest that models trained on region-specific corpora may continue to encode, and potentially reinforce, locally embedded social hierarchies.


Taken together, these findings indicate that caste-based patterns in LLM-driven matchmaking evaluations are robust across models, but vary in intensity and form. Rather than being confined to a single model or training paradigm, caste emerges as a persistent organizing principle that shapes both similarity-based preferences and hierarchical ordering in model outputs.

\subsection{\editt{Homophily, Status, and Gendered Hierarchy}}
\label{sec:decomposition}

\editt{The cross-model comparison shows that caste-structured evaluations recur across all five models, in forms that differ by model. To examine what these forms consist of, we decompose caste-linked variation into candidate status, evaluator position, and homophily.}

\editt{Homophily accounts for most caste-linked variation in the three models with the largest caste-linked differences (\autoref{tab:decomposition}). For GPT, the same-caste premium and proximity gradient explain 89--93\% of caste-linked variance in social acceptance and cultural compatibility and 75--78\% in marital stability, while candidate status accounts for at most 9\%. 
Gemini shows the same structure (76--84\% in social acceptance and cultural compatibility), with the steepest proximity gradient of any model (0.56--0.95 points per step of caste distance). 
Llama's homophily shares are 63--73\% for social acceptance and cultural compatibility, with a substantial status component in social acceptance (22--25\%). Across these three models, same-caste matches receive a premium of 0.6--2.8 points beyond the proximity gradient; the exception is Llama's marital stability ratings, which carry little caste-linked variation of any kind (SD 0.33--0.34).}

\editt{This structure bears the traditional caste-based hierarchy in three ways. First, the proximity gradient follows the traditional caste order. When we refit it under each of the 60 distinct orderings of the five castes, the traditional order (Brahmin, Kshatriya, Vaishya, Shudra, Dalit) fits best or second best in 25 of 30 model--setting--dimension combinations; the orderings that fit better almost always swap two adjacent castes, most often Brahmin and Kshatriya, and improve the fit by at most 7\%. Four of the five exceptions are marital stability ratings, which carry the least caste-linked structure. Second, at equal caste distance, models rate candidates above the evaluator higher than candidates below, in 88 of 120 comparisons (by 0.43 points on average), most consistently in GPT (23 of 24) and Gemini (22 of 24). Third, for Brahmin, Kshatriya, and Vaishya evaluators, ratings decline step by step for candidates below them in 71 of 90 cases, and in the averaged ratings these evaluators rate Dalit candidates lowest in every model and setting.}

\editt{Qwen and BharatGPT show different signatures. Qwen's caste-linked differences are small (SD 0.28--0.62), and candidate status dominates them: status accounts for 50--56\% of caste-linked variance in social acceptance and cultural compatibility, reflecting lower ratings for Dalit candidates from evaluators of every caste, while the same-caste premium is effectively zero. BharatGPT's caste-linked variation lies mainly in evaluator position (42--77\% across dimensions). The model rates prospects lower whenever it adopts a Dalit evaluator's perspective (by 0.82--1.05 points relative to the mean in the averaged ratings), and it shows no consistent same-caste premium.}

\editt{The status component concentrates on Dalit candidates. The Dalit differential (the mean rating of Dalit candidates relative to the grand mean) is negative in 26 of 30 model--setting--dimension combinations and in every combination for social acceptance and cultural compatibility; the four exceptions are the marital stability ratings of Llama and Qwen, which place Dalit candidates slightly above average. Brahmin candidates, in contrast, receive no consistent advantage, with a positive differential in only 9 of 30 combinations. In GPT, the Dalit differential also varies by gender: Dalit men evaluated by women fall 0.32--0.37 points below the mean, three to four times the gap for Dalit women evaluated by men (0.09--0.10). Status differences in model evaluations thus concentrate at the boundary of untouchability, consistent with accounts that locate caste's sharpest exclusion at the line of purity and pollution~\cite{dumont1980homo, ambedkar2014annihilation}.}

\editt{Two models also show an asymmetry consistent with the \textit{anuloma} rule. In GPT, a union in which the man holds the higher caste receives 0.12--0.18 points more than the same pair of castes with the woman holding the higher caste; the advantage is positive in eight or nine of ten caste pairs in every dimension and rises to 0.35--0.42 points for Brahmin--Dalit unions. Gemini shows the largest asymmetry (0.20–0.64 points across dimensions, positive in all ten caste pairs in every dimension). BharatGPT shows a smaller advantage, concentrated in marital stability (0.20 points), and Llama and Qwen show no consistent asymmetry ($-$0.06 to 0.11 across dimensions).}


\editt{Together, these results show that caste hierarchy structures model evaluations relative to the evaluator's own position. Ratings decline with distance along the traditional caste order, favor candidates above the evaluator over those equally far below, and fall most for Dalit candidates. GPT and Gemini add a gendered asymmetry that favors unions in which the man holds the higher caste.}

\begin{table*}[t]
\centering
\sffamily
\footnotesize
\caption{\editt{Decomposition of caste-linked variation in model ratings. Each cell reports male-evaluator / female-evaluator settings. Caste SD is the standard deviation of the 25 evaluator--candidate cell estimates. Status, evaluator, and homophily give each component's share of caste-linked variance; the remainder is unstructured interaction. Same-caste premium and distance (per step of caste distance) are in rating points, net of candidate- and evaluator-level differences. Dalit differential is the mean rating of Dalit candidates relative to the grand mean.
\textit{Anuloma} is the union-level advantage for unions in which the man holds the higher caste, averaged over both partners' perspectives.}}
\label{tab:decomposition}
\editt{\resizebox{\textwidth}{!}{%
\begin{tabular}{lcccccccc}
\textbf{Dimension} & \textbf{Caste SD} & \textbf{Status (\%)} & \textbf{Evaluator (\%)} & \textbf{Homophily (\%)} & \textbf{Same-caste premium} & \textbf{Distance per step} & \textbf{Dalit status} & \textbf{Anuloma} \\
\toprule
\rowcollight \multicolumn{9}{l}{\textbf{GPT}}\\ 
Social acceptance & 1.25 / 1.43 & 0 / 3 & 6 / 4 & 92 / 92 & 2.40 / 2.83 & $-$0.27 / $-$0.28 & $-$0.10 / $-$0.37 & 0.18 \\
 Cultural compatibility & 0.96 / 1.11 & 1 / 4 & 4 / 6 & 93 / 89 & 1.87 / 2.15 & $-$0.20 / $-$0.22 & $-$0.10 / $-$0.32 & 0.13 \\
Marital stability & 0.67 / 0.71 & 2 / 9 & 17 / 11 & 78 / 75 & 0.94 / 0.97 & $-$0.23 / $-$0.25 & $-$0.09 / $-$0.36 & 0.12 \\
 Average & 0.95 / 1.07 & 1 / 4 & 7 / 5 & 91 / 89 & 1.74 / 1.98 & $-$0.24 / $-$0.25 & $-$0.09 / $-$0.35 & 0.14 \\
\hdashline
\rowcollight \multicolumn{9}{l}{\textbf{Gemini}} \\
Social acceptance & 1.83 / 2.23 & 4 / 7 & 4 / 1 & 84 / 76 & 2.44 / 2.60 & $-$0.75 / $-$0.95 & $-$0.63 / $-$0.98 & 0.64 \\
 Cultural compatibility & 1.09 / 1.30 & 3 / 1 & 13 / 8 & 76 / 80 & 1.01 / 1.41 & $-$0.56 / $-$0.62 & $-$0.33 / $-$0.20 & 0.20 \\
 Marital stability & 1.25 / 1.16 & 6 / 8 & 23 / 3 & 57 / 66 & 0.70 / 0.59 & $-$0.66 / $-$0.69 & $-$0.57 / $-$0.62 & 0.46 \\
 Average & 1.35 / 1.52 & 4 / 4 & 9 / 2 & 77 / 78 & 1.38 / 1.53 & $-$0.66 / $-$0.75 & $-$0.51 / $-$0.60 & 0.43 \\
\hdashline
\rowcollight \multicolumn{9}{l}{\textbf{Llama}}\\
Social acceptance & 1.31 / 1.23 & 22 / 25 & 0 / 0 & 65 / 65 & 1.15 / 1.19 & $-$0.61 / $-$0.54 & $-$1.14 / $-$1.16 & 0.02 \\
 Cultural compatibility & 1.06 / 0.92 & 9 / 3 & 12 / 19 & 73 / 63 & 1.44 / 1.18 & $-$0.37 / $-$0.28 & $-$0.45 / $-$0.27 & $-$0.06 \\
 Marital stability & 0.33 / 0.34 & 29 / 37 & 12 / 18 & 29 / 13 & $-$0.05 / 0.06 & $-$0.17 / $-$0.09 & 0.23 / 0.26 & $-$0.06 \\
 Average & 0.81 / 0.68 & 13 / 12 & 1 / 1 & 76 / 80 & 0.85 / 0.81 & $-$0.38 / $-$0.30 & $-$0.45 / $-$0.39 & $-$0.03 \\
\hdashline
\rowcollight \multicolumn{9}{l}{\textbf{Qwen}} \\
Social acceptance & 0.53 / 0.62 & 56 / 53 & 18 / 21 & 16 / 16 & $-$0.02 / 0.02 & $-$0.20 / $-$0.22 & $-$0.78 / $-$0.87 & 0.00 \\
 Cultural compatibility & 0.42 / 0.49 & 55 / 50 & 24 / 26 & 14 / 16 & 0.02 / 0.05 & $-$0.14 / $-$0.16 & $-$0.60 / $-$0.67 & $-$0.01 \\
 Marital stability & 0.28 / 0.40 & 46 / 19 & 41 / 66 & 3 / 5 & $-$0.17 / $-$0.30 & $-$0.04 / $-$0.06 & 0.38 / 0.33 & 0.11 \\
 Average & 0.26 / 0.29 & 45 / 51 & 18 / 11 & 22 / 24 & $-$0.06 / $-$0.08 & $-$0.12 / $-$0.15 & $-$0.33 / $-$0.40 & 0.03 \\
\hdashline
\rowcollight \multicolumn{9}{l}{\textbf{BharatGPT}} \\
Social acceptance & 0.69 / 0.64 & 10 / 7 & 73 / 77 & 8 / 9 & $-$0.21 / $-$0.15 & $-$0.23 / $-$0.21 & $-$0.27 / $-$0.24 & 0.01 \\
Cultural compatibility & 0.72 / 0.59 & 14 / 13 & 50 / 58 & 21 / 23 & $-$0.31 / 0.10 & $-$0.37 / $-$0.23 & $-$0.32 / $-$0.28 & 0.08 \\
Marital stability & 0.81 / 0.54 & 13 / 13 & 53 / 42 & 20 / 38 & $-$0.38 / 0.41 & $-$0.41 / $-$0.18 & $-$0.39 / $-$0.21 & 0.20 \\
Average & 0.73 / 0.57 & 13 / 11 & 59 / 63 & 17 / 22 & $-$0.30 / 0.12 & $-$0.34 / $-$0.21 & $-$0.33 / $-$0.24 & 0.10 \\
\bottomrule
\end{tabular}}}
\end{table*}


\subsection{\editt{Prompt-Level Guardrail Mitigation}}\label{sec:guardrail_findings}

\editt{We tested a minimal prompt-level guardrail because it is the intervention that most deployers control. A platform that builds matchmaking features on a proprietary or open-weight model sets the instructions it sends, but typically cannot retrain or realign the model itself. 
The guardrail therefore tests how far the least costly intervention goes. 
We reran the full audit for Llama-4 and BharatGPT with a system instruction that directed the model to act as a fairness-aware rater: to avoid prejudice and stereotypes about caste, gender, religion, or community; to base ratings only on information in the profile and on the requested criteria; to ignore cues implying group hierarchy, tradition-based preference, or status-based advantage, including caste labels; to decline to role-play or adopt biased viewpoints; and to use neutral assumptions when information is missing. We held all other elements of the setup fixed. We limited this analysis to Llama-4 and BharatGPT because these locally deployed models allowed us to apply an identical system-level instruction under a fixed inference configuration. The full instruction and model-specific heatmaps appear in~\autoref{sec:guardrail_mitigation}.}


\edit{This approach increased mean ratings for cross-caste pairings in the model and evaluator--candidate conditions, but its effect on the difference between same-caste and cross-caste ratings was model-dependent (\autoref{tab:guardrail_results}). 
For Llama-4, the same-caste versus cross-caste rating difference decreased from 1.61 to 1.08 for male evaluators and from 1.41 to 0.93 for female evaluators. For BharatGPT, the difference decreased from 0.53 to 0.16 for female evaluators but remained nearly unchanged for male evaluators (0.38 versus 0.39). Therefore, the guardrail prompts attenuated caste-linked differences in multiple conditions but did not consistently eliminate them.}
\editt{Overall, these comparisons are descriptive; for Llama, the reduction came mainly from the same-caste premium, while ratings still declined with caste distance.}


\begin{table}[t]
\centering
\footnotesize
\sffamily
\caption{\edit{Difference between mean same-caste and cross-caste ratings before and after applying the fairness guardrail. Smaller values indicate less differentiation between same-caste and cross-caste pairings.} \editt{Values are descriptive differences in mean ratings.}}
\label{tab:guardrail_results}
\small
\begin{tabular}{llrrr}
\textbf{\edit{Model}} & \textbf{\edit{Evaluator--Candidate}} &
\textbf{\edit{Baseline}} & \textbf{\edit{Guardrail}} & \textbf{\edit{Change}} \\
\toprule
\edit{Llama-4} &
\edit{Male--Female} &
\edit{1.61} & \edit{1.08} & \edit{-0.54} \\

\edit{Llama-4} &
\edit{Female--Male} &
\edit{1.41} & \edit{0.93} & \edit{-0.48} \\

\edit{BharatGPT} &
\edit{Male--Female} &
\edit{0.38} & \edit{0.39} & \edit{+0.01} \\

\edit{BharatGPT} &
\edit{Female--Male} &
\edit{0.53} & \edit{0.16} & \edit{-0.37} \\
\bottomrule
\end{tabular}
\end{table}

\section{Discussion}
In this study, we conducted a controlled audit of LLM-driven matchmaking evaluations to understand how LLMs may perform socially embedded compatibility judgments, particularly in the context of Indian caste-based matchmaking. 
By systematically varying caste and income while holding other attributes constant, we identify consistent disparities across models which mirror traditional caste hierarchy in Indian societies. 
\editt{When conditioned on a caste-identified evaluator's perspective, model evaluations show strong caste endogamy, favoring same-caste matches, along with a graded hierarchy in which ratings drop as the caste distance between evaluator and candidate increases.}
These patterns are not confined to a single model but persist across diverse architectures, suggesting that caste operates as a dominant axis of social evaluation. 
\editt{Together, these findings indicate that model evaluations reproduce historically embedded social structures.}
We now discuss the theoretical contextualization and implications of this work.

\subsection{LLM-driven Matchmaking as Social Evaluation}

Unlike traditional algorithmic tasks such as classification or prediction, matchmaking involves inherently relational and value-laden judgments shaped by social and cultural norms, expectations, and hierarchies. 
In this sense, LLM-driven evaluations resemble what prior work in Science and Technology Studies and critical algorithm studies describes as sociotechnical systems that embed and reproduce existing social structures rather than operate as neutral decision-making tools~\cite{selbst2019fairness,barocas2023fairness}.

\editt{Across models, perspective-conditioned evaluations depart from a neutral aggregation of attributes and reproduce caste-structured patterns of social evaluation. They favor caste endogamy, prioritizing same-caste matches, and penalize Dalit candidates even when other attributes are held constant.}
Beyond endogamy, we observe that models encode a graded hierarchy: candidates closer to the evaluator’s caste are rated more favorably, while those further away are consistently deprioritized.

This suggests that LLM-driven matchmaking evaluations function not merely as preference modeling, but as a form of normative reasoning that reflects and reproduces historically embedded social hierarchies. 
Rather than evaluating candidates independently, models position them within a relational structure that shapes both absolute judgments and comparative rankings. 
In doing so, they instantiate what prior work has described as the reproduction of social order through computational systems, where algorithmic decision-making reflects and reinforces existing social hierarchies and inequalities~\cite{selbst2019fairness,barocas2023fairness}, and operationalizes these structures within sociotechnical systems that embed social categories into decision-making processes~\cite{dolata2022sociotechnical}.


\editt{\citet{bapuji2024integrative} distinguished cognitive, relational, behavioral, and institutional mechanisms through which caste is reproduced. 
Our audit observes model outputs and does not test these mechanisms directly, so we treat the mapping below as inference: we link each finding to the mechanism it most closely resembles and identify the evidence a direct test would require.}

\editt{The cognitive mechanism concerns caste as a category that carries beliefs about status and worth. Our results resemble this mechanism in three ways: ratings follow the traditional caste order, candidates above the evaluator are favored over those equally far below, and Dalit candidates receive the steepest penalty (\autoref{sec:decomposition}). The persistence of the proximity gradient after models were instructed to ignore caste (\autoref{sec:guardrail_findings}) further suggests that these status associations operate beneath explicit instruction. 
The relational mechanism concerns endogamy and the boundaries that caste draws around marriage and kinship~\cite{ambedkar2022castes}. 
The same-caste premium and the decline in ratings with caste distance correspond most directly to this mechanism, and caste-linked differences are largest for social acceptance, the dimension that captures approval from family and community. The institutional mechanism concerns norms that codify caste relations, and the preference for anuloma over pratiloma unions in GPT and Gemini reproduces one such codified rule~\cite{chakravarti1993conceptualising}.}

\editt{The behavioral mechanism, and the broader institutional consequences of these evaluations, lie beyond what our audit observes. If matchmaking platforms use such evaluations to rank or filter candidates, lower ratings for Dalit and cross-caste candidates would translate into reduced visibility and fewer contacts, a behavioral pathway. Presenting historically structured preferences as data-driven judgments could also help institutionalize endogamous boundaries within platform rules and defaults. Testing these pathways requires studying deployed systems, for example through audits of ranking pipelines on matrimonial platforms and user studies of how people interpret and act on model evaluations. Testing the cognitive pathway directly would similarly require probing models' internal representations or comparing evaluations with and without evaluator identity. We see these as important directions for future research.}


\subsection{\editt{Why Caste Bias Emerges in Model Evaluations}}

\editt{Prior work identifies three sources of social bias in LLMs: associations learned from pretraining corpora~\cite{caliskan2017semantics, bender2021dangers}, post-training and alignment processes that shape how those associations surface~\cite{hofmann2024ai, bai2025explicitly}, and identity cues in the prompt that activate them~\cite{gupta2024bias}. Three features of our results bear on these sources.}

First, persona compliance \edittt{may} not fully account for the caste-linked differences. Our prompt conditions each evaluation on the perspective of a caste-identified evaluator from a traditional family, and one reading of our results is that models simulate the preferences such an evaluator would hold. 
\edittt{The guardrail analysis shows that caste-linked differences persisted even when models were instructed to ignore caste labels and decline to adopt biased viewpoints. However, because the original perspective-conditioned prompt remained unchanged, this analysis does not disentangle the contribution of perspective conditioning from other sources of model bias.}


\editt{Second, the components that the guardrail did and did not remove suggest where alignment acts.}
\edittt{For Llama, the guardrail reduced the same-caste premium, while the graded structure of caste proximity remained.}
\editt{This split parallels findings that alignment and explicit instruction reduce overt expressions of bias while covert associations persist~\cite{hofmann2024ai, bai2025explicitly}. It suggests that post-training shapes an explicit, instruction-sensitive layer of caste preference, beneath which a graded structure of caste proximity, likely acquired during pretraining, remains.}

\editt{Third, the structure of caste bias in model evaluations mirrors the structure documented in societal matchmaking. 
Endogamy~\cite{banerjee2013marry, vaid2014caste} appears in the same-caste premium of GPT, Gemini, and Llama; the sharp exclusion of Dalit partners~\cite{rajadesingan2019smart} appears across models; and the classical preference for anuloma over pratiloma unions~\cite{chakravarti1993conceptualising, chowdhry2007contentious} appears most clearly in GPT and Gemini and to a lesser extent in BharatGPT.
Matrimonial profiles, advertisements, and public discussions of marriage encode these norms explicitly~\cite{rajadesingan2019smart}, and text of this kind forms part of the web-scale corpora on which models train. The correspondence is consistent with models learning caste hierarchy from such text.}

\editt{Differences across models offer a further clue. The three widely deployed general-purpose models (GPT, Gemini, and Llama) show homophily-dominant signatures, in which ratings depend on the relation between evaluator and candidate. The two small open models, Qwen (1.5B parameters) and BharatGPT (3B), show signatures that depend on a single identity: candidate caste in Qwen and evaluator caste in BharatGPT. One possibility is that relating two caste identities requires capacity that small models lack, so they fall back on identity-level associations. Our five models differ in scale, training data, and alignment at once, so this pattern remains a hypothesis. Isolating each source will require comparisons across checkpoints of the same model before and after alignment, prompts with and without evaluator identity, and models trained on corpora with documented differences in exposure to caste-related text.}

\subsection{Caste as a Dominant and Disproportionate Axis of Social Evaluation}
A key finding of this study is the relative strength of caste as a decision factor.
Across models, caste alignment is weighted more heavily than attributes typically associated with compatibility in matchmaking contexts, such as education or income.
This imbalance highlights a critical risk: even when multiple other attributes are present, socially sensitive features like caste can dominate outcomes in ways that reinforce discriminatory norms.
\edit{The income variation further clarifies the rigidity of this pattern. Although candidates' income generally shapes model evaluations, higher economic status does not meaningfully offset the disadvantage associated with caste distance. This suggests that the models do not treat achieved socioeconomic status as a substitute for inherited caste position; instead, caste continues to structure relational evaluations across income levels.}

These observations reflect a deeper dynamic.
The caste system is historically reproduced through endogamy, and \editt{model evaluations} appear to mirror this mechanism by privileging intra-caste compatibility and encoding hierarchical ordering across caste groups.
In line with prior work on representational and allocative harms in AI systems~\cite{weidinger2021ethical,bender2021dangers}, our findings point to a complementary form of harm: relational and hierarchical bias. Here, harm does not arise solely from misrepresentation or exclusion, but from the systematic ranking of individuals along socially stratified dimensions.

Beyond encoding bias, these patterns raise critical concerns about reinforcement and legitimization. 
Despite ongoing institutional, legal, and social efforts to mitigate caste-based discrimination, LLMs may introduce an additional and seemingly ``neutral'' source of authority that users can draw upon to justify discriminatory preferences. 
When models consistently produce higher evaluations for same-caste or upper-caste matches, they risk being interpreted as reflecting ``objective'' or data-driven reasoning, rather than historically contingent social bias. 
In this way, LLM outputs may not only reflect existing hierarchies, but potentially legitimize them.

This creates the potential for a feedback loop: user preferences shaped by caste norms are reflected in model outputs, which in turn reinforce those preferences by presenting them as algorithmically validated. 
Over time, such interactions can normalize caste-based reasoning in decision-making, even among users who may not explicitly articulate such preferences. This dynamic is particularly concerning in AI-mediated environments, where repeated exposure to model outputs can shape perceptions of what is typical, acceptable, or desirable.

Importantly, the scalability of AI systems amplifies these risks. 
Unlike human-mediated matchmaking, where biases may be contextually negotiated or challenged, LLM-driven evaluations can standardize and reproduce caste-based hierarchies across large populations and interactions. 
\edit{Our findings are qualitatively consistent with prior evidence of caste endogamy and caste-structured preferences in human and online matchmaking~\cite{banerjee2013marry,rajadesingan2019smart}. However, LLM-generated evaluations introduce a distinct concern because they can standardize and reproduce these patterns across repeated evaluations and at scale.}
This raises the possibility that AI systems could entrench existing inequalities more deeply by embedding them into automated pipelines of recommendation and evaluation.

Importantly, this pattern persists even in models trained on culturally specific data, such as BharatGPT, suggesting that localized training does not necessarily mitigate caste-based disparities. Instead, such models may continue to encode, and potentially reinforce, locally embedded social hierarchies. At the same time, variation across models indicates that these effects are not uniform.
While models such as GPT and Gemini exhibit strong stratification, others like Qwen produce more uniform evaluations. This variation suggests that the expression of caste-based structure depends in part on training data composition and alignment strategies, even as the underlying signal remains persistent.

\subsection{Implications for AI Systems: Practical and Design Considerations}

Our study bears implications for the design and deployment of AI-mediated matchmaking systems, and more generally AI-mediated algorithms for personal decision-making. 
As matchmaking platforms incorporate AI-driven recommendation and ranking systems~\cite{tariq2025rise,walrus2025matchmakers}, there is a risk that such systems will automate and legitimize caste-based exclusion at scale. 
Prior work in human-centered AI and system design has emphasized that AI systems not only reflect user preferences, but can also shape interaction patterns and decision-making processes~\cite{amershi2019guidelines}. 
In this context, LLM-driven matchmaking systems may do more than surface preferences, and may actively structure them.

From a practical standpoint, platforms that integrate LLM-based evaluation or ranking pipelines must recognize that seemingly neutral outputs can encode socially consequential biases. 
Without safeguards, such systems may systematically deprioritize lower-caste users, shaping visibility, match likelihood, and ultimately life outcomes. 
This raises concerns not only about fairness, but also about transparency and accountability, as users may not be aware of how such evaluations are generated or operationalized.

\edit{The practical consequences of these evaluations will also depend on how users interpret and act upon them. Users may follow model-generated judgments, selectively accept those that align with personal or family expectations, or discount them when they conflict with existing preferences. Interventions should therefore operate at multiple levels: improving training and model alignment, reducing unnecessary sensitivity to caste cues in prompts, constraining caste-based scoring, designing interfaces that communicate uncertainty and avoid legitimizing discriminatory judgments, and strengthening users' awareness of the social biases that model outputs may reproduce.}


\editt{Our guardrail analysis tests the intervention layer that deployers control most directly.}
\edittt{The instruction reduced caste-linked differences, including the same-caste premium for Llama, but the graded structure of caste proximity remained.}
\editt{These descriptive results point to a layered approach, in which each layer addresses what the previous one leaves in place. At the model layer, training data curation and alignment can weaken learned associations between caste and social worth, which instructions leave largely intact. At the instruction layer, guardrails such as ours give deployers an immediate, low-cost way to suppress explicit caste preference. At the scoring layer, platforms can check that a candidate's score does not change when only caste varies, and can exclude caste from ranking and filtering~\cite{kusner2017counterfactual}; for explicit caste labels, this is the layer at which caste-linked differences can be eliminated by construction, although proxies such as names and locations remain. At the interface layer, presenting evaluations with stated uncertainty, and avoiding single numeric compatibility scores, limits the authority users attach to model outputs~\cite{logg2019algorithm}. At the user layer, awareness of the biases that model outputs may reproduce shapes how people act on whatever reaches them. No single layer suffices: our guardrail shows the limits of instructions alone, and scoring-layer constraints without model-level change would leave biased reasoning in any explanations a system generates. Testing model-level interventions, such as fine-tuning, is an important direction for future work.}

From a design perspective, these findings raise a fundamental question: what should an ideal model do in the presence of socially sensitive attributes such as caste? One approach is to personalize recommendations based on user preferences, potentially incorporating caste as a signal. 
However, given the historical role of caste in structuring exclusion and discrimination, such personalization risks directly reinforcing harmful social norms. 
An alternative approach is to suppress caste entirely and generate evaluations that are independent of caste identity. 
Yet, fully removing such signals may lead to systems that users perceive as unrealistic or culturally misaligned.

We argue that neither naive personalization nor complete suppression is sufficient. 
Instead, AI systems should explicitly treat caste as a protected and socially sensitive attribute, and prevent it from functioning as a basis for hierarchical ranking or exclusionary filtering. 
This aligns with emerging calls for context-aware and culturally grounded fairness frameworks~\cite{bansal2025decolonial,vaghela2022interrupting}, which emphasize that fairness cannot be defined independently of the social and historical meaning of the attributes being modeled.

Concretely, our work suggests several design directions: constraining models from using caste in ranking functions, auditing outputs for hierarchical ordering rather than only aggregate disparities, introducing fairness-aware reranking mechanisms to counteract endogamous clustering, and increasing transparency around how recommendations are generated. 
More broadly, it requires moving beyond accuracy and preference optimization toward design approaches that explicitly account for social impact.

\subsection{\editt{From Evaluations to Social Ordering}}

Beyond matchmaking, our findings point to a broader concern: \editt{when LLM evaluations inform decisions, they may shape social preferences as well as reflect them.}
As users increasingly rely on AI systems for recommendations and decision support, model outputs may influence how individuals perceive compatibility, desirability, and social boundaries. 

In this context, LLM-driven evaluations may contribute to the normalization of caste-based reasoning, reinforcing hierarchical thinking even in the absence of explicit intent. 
When models consistently privilege same-caste or upper-caste matches, they may implicitly legitimize these patterns as reasonable or desirable, thereby stabilizing existing social hierarchies.

While caste is explicitly represented in our dataset, this also highlights a broader concern. 
Similar forms of bias may emerge through proxies such as names, locations, language use, or cultural markers, as documented in prior work on algorithmic bias and social inference~\cite{sambasivan2021re,ntoutsi2020bias}. 
Such proxy-based bias is particularly concerning because it can operate invisibly within complex machine learning pipelines, making it difficult to detect, audit, and regulate.

Taken together, these findings suggest that LLMs \editt{ deployed as evaluators} could act as agents of social ordering, translating historically embedded hierarchies into computational evaluation logic. Addressing these issues requires moving beyond surface-level interventions, such as prompt-based adjustments, toward deeper changes in training data, model objectives, and evaluation frameworks. 
Without such efforts, AI systems risk not only reflecting existing inequalities, but reinforcing and scaling them within digital infrastructures. 
\subsection{Limitations and Future Directions}



Our study has limitations which also suggest important future directions.
First, our analysis focuses on a coarse representation of caste, limited to the four varnas and Dalits. This abstraction enables controlled analysis and comparability across models, but it necessarily obscures the rich heterogeneity of caste identity in South Asia. In practice, castes, sub-castes, regional hierarchies, and community-specific norms play a central role in matrimonial decision-making. As a result, our findings should be interpreted as capturing high-level caste signals in LLM-driven evaluations rather than the full complexity of caste relations. Future work should examine finer-grained caste distinctions and regional variations to better understand how models encode and reproduce more localized and context-specific forms of social stratification.

Second, our study is restricted to cis-gender and heterosexual matchmaking, reflecting both the dataset and the dominant norms embedded in mainstream matchmaking platforms. This excludes same-sex relationships, non-binary identities, and alternative family structures, where social norms—and potentially model behavior—may differ substantially. Future research should examine how caste-based reasoning interacts with gender and sexuality, and whether LLMs reproduce, attenuate, or reconfigure bias in these less-studied relational contexts.

Third, while our design includes several demographic controls, we systematically manipulate only caste and income, holding other attributes such as location, language, religion, and education constant. In real-world settings, these attributes are often deeply entangled with caste and may act as proxies or amplifiers of social hierarchy. Future work should adopt multi-dimensional designs that jointly vary these attributes to better capture intersectional and context-dependent dynamics of bias in LLM-driven evaluations.
\edit{Because our study conditions evaluations on the stated perspective of a caste-identified prospective bride or groom, it does not disentangle caste bias elicited by this perspective framing from patterns that may also emerge in the models’ unconditioned evaluations. The outputs should also not be interpreted as representing the actual beliefs of individuals belonging to the specified caste groups.}

\editt{Our outcome measure also has limitations. There is no established measure which captures overall matchmaking evaluation, so we elicited three theoretically grounded dimensions and report their unweighted average as a descriptive summary. This average is not a validated construct: the three dimensions are related but empirically distinct, and the extent to which they combine additively into a single judgment is unknown. Families and matchmakers may weigh approval, compatibility, and stability differently across communities and individual decisions, and an equal-weighted average cannot capture these trade-offs. Our central finding, the negative association between caste distance and ratings, holds for each dimension separately in all five models, so it does not depend on the average. Future work could develop and validate measures of overall matchmaking evaluation, for example by eliciting holistic judgments from families and matchmakers and modeling how they weigh these dimensions.}

Finally, our methodology relies on a standardized prompting framework, including a single global prompt and structured numeric outputs. 
While this enables comparability and systematic analysis, it constrains the range of perspectives the model can express and limits insight into the underlying reasoning processes. 
\editt{Accordingly, we infer mechanisms from patterns in model outputs, and our design does not observe the models' internal representations or trace outputs to training data.}
\edit{Because we also do not examine user behavior or actual matchmaking outcomes, institutional maintenance should be understood as a plausible downstream consequence rather than an effect directly demonstrated by this study.}
In particular, we do not capture how models justify their evaluations or how different prompting strategies may alter outcomes. Future research should explore alternative methodological approaches, including qualitative analysis of generated explanations, comparative prompting strategies, and fine-tuning or alignment interventions, to better understand and potentially mitigate relational biases.

More broadly, our work highlights the need for methodological frameworks that can capture not only individual-level bias, but also relational and hierarchical forms of bias in AI systems. Developing such approaches will be critical for studying socially embedded domains where decision-making is structured by complex and historically grounded systems of inequality. 
\section{Conclusion}

In this work, we conducted a controlled audit of caste bias in LLM-driven matchmaking evaluations, examining how models assess matrimonial profiles under systematic variations of caste and income. Our findings showed that \editt{when conditioned on a caste-identified evaluator's perspective,} LLMs consistently assigned higher evaluations to same-caste matches and exhibited hierarchical ordering across caste groups, indicating that model outputs reflected not only similarity-based preferences but also structured social stratification. 
These patterns persisted even after accounting for socioeconomic variation, suggesting that caste remained a dominant organizing signal in model evaluations.
By situating these findings within the context of South Asian/Indian matchmaking, we highlighted how LLM-driven evaluations can reproduce culturally specific forms of inequality in relational decision-making. 
More broadly, our results underscored the importance of auditing AI systems in socially grounded domains where evaluative judgments are shaped by historical and structural hierarchies. 
This work highlights the need for culturally grounded approaches to evaluation and governance that account for such context-specific forms of bias in AI systems. 






\bibliographystyle{ACM-Reference-Format}
\bibliography{references}


\appendix
\clearpage
\appendix
\section{Appendix}
\setcounter{table}{0}
\setcounter{figure}{0}
\renewcommand{\thetable}{A\arabic{table}}
\renewcommand{\thefigure}{A\arabic{figure}}

\subsection{\edit{Technical Details of the Experiment}}
\edit{
 To ensure consistency and facilitate downstream analysis, models are instructed to provide their evaluations in a fixed JSON format. This standardized prompt structure is used across all models and evaluation settings, with variations only in the \cnd{}'s caste group and profile income and the \evl{}'s caste group. This procedure yields 12,500 ($2,500 \times 5$ \evl{} caste groups) model outputs for male profiles and 12,500 outputs for female profiles, resulting in 25,000 evaluations per model. For efficient data collection, we used the OpenAI Batch API, Gemini Batch API, and Ollama Chat API. Data were collected between October 15, 2025, and January 15, 2026.}

\begin{table}[H]
\centering
\sffamily
\caption{\edit{Summary of the LLMs evaluated in this study.}}
\label{tab:model_summary}
\footnotesize
\begin{tabular}{lll}
\edit{\textbf{Model}} & \edit{\textbf{Model type}} & \edit{\textbf{Rationale for inclusion}} \\
\toprule
\edit{GPT-5-nano} & \edit{Proprietary} & \edit{Widely deployed general-purpose model} \\
\edit{Gemini-2.5-Pro} & \edit{Proprietary} & \edit{Widely deployed general-purpose model} \\
\edit{Llama-4} & \edit{Open-weight} & \edit{Consumer-facing general-purpose model} \\
\edit{Qwen-2-1.5B} & \edit{Open-weight} & \edit{Lightweight multilingual model} \\
\edit{BharatGPT-3B-Indic} & \edit{Indic-focused} & \edit{Regionally and culturally focused model} \\
\bottomrule
\end{tabular}
\end{table}

\subsection{Results for Other Models}

\begin{figure}[H]
\centering
        \begin{subfigure}[b]{0.45\columnwidth}
    \centering
    \includegraphics[width=\columnwidth]{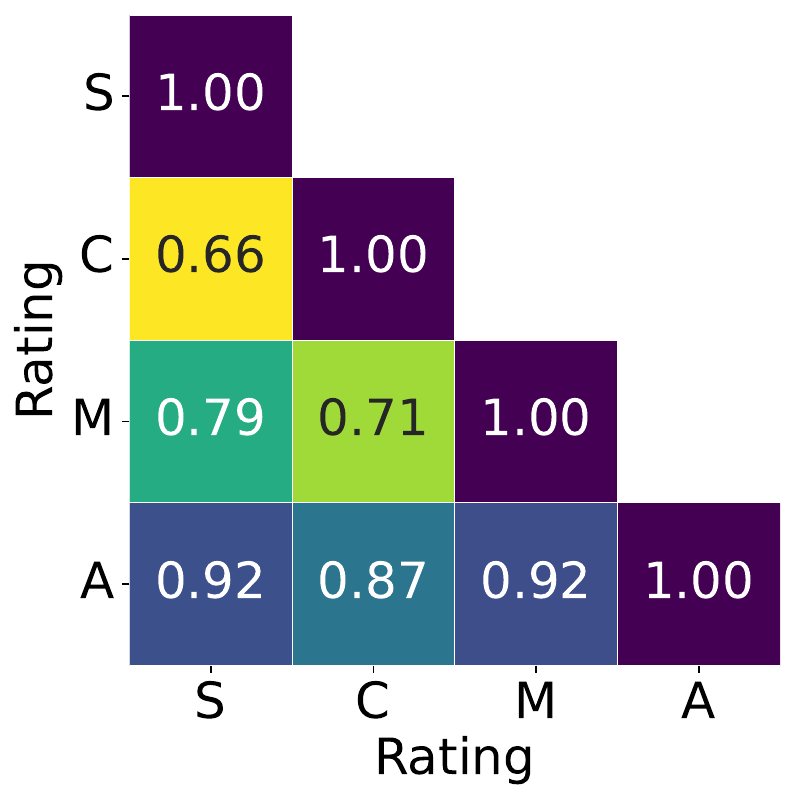}
    \caption{Male Evaluator, Female Candidate}
    \end{subfigure}\hfill
        \begin{subfigure}[b]{0.45\columnwidth}
    \centering
    \includegraphics[width=\columnwidth]{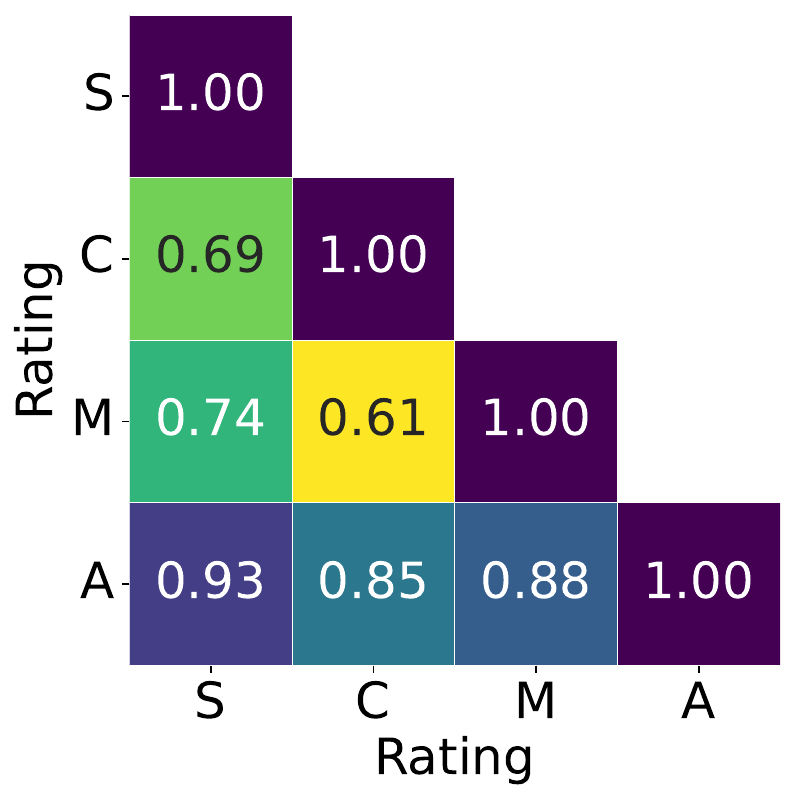}
    \caption{Female Evaluator, Male Candidate}
    \end{subfigure}\hfill
\caption{Gemini: Correlation (Pearson's $r$) heatmaps of the ratings of Gemini-2.5-Pro across evaluation outcomes of \textit{Social Acceptance (S), Cultural Compatibility (C), Marital Stability (M), and Average (A)}. All values are statistically significant ($p$<0.05).}
\end{figure}

\begin{figure}[H]
\centering
        \begin{subfigure}[b]{0.45\columnwidth}
    \centering
    \includegraphics[width=\columnwidth]{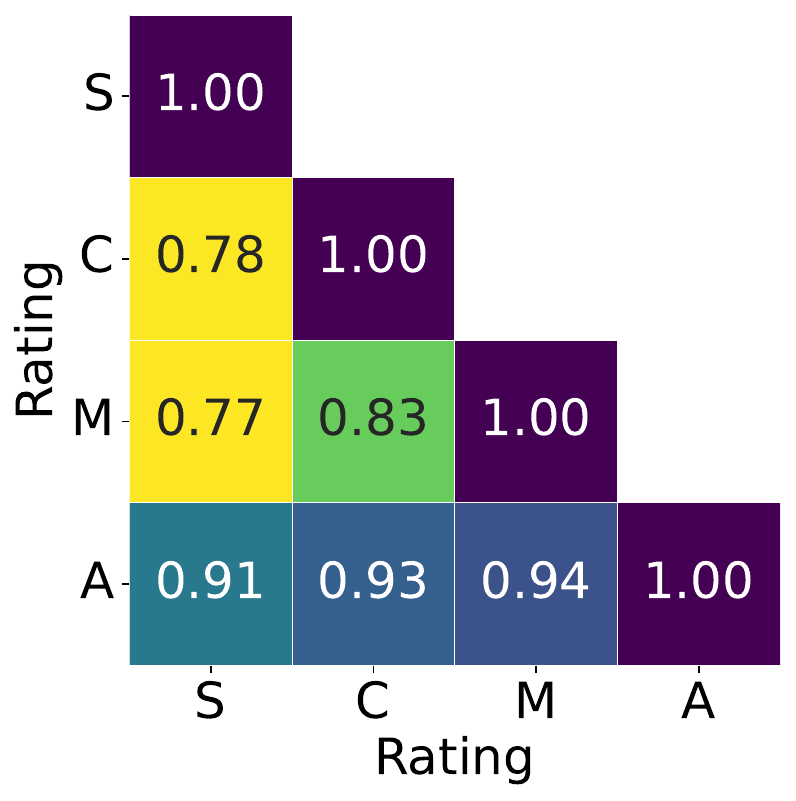}
    \caption{Male Evaluator, Female Candidate}
    \end{subfigure}\hfill
        \begin{subfigure}[b]{0.45\columnwidth}
    \centering
    \includegraphics[width=\columnwidth]{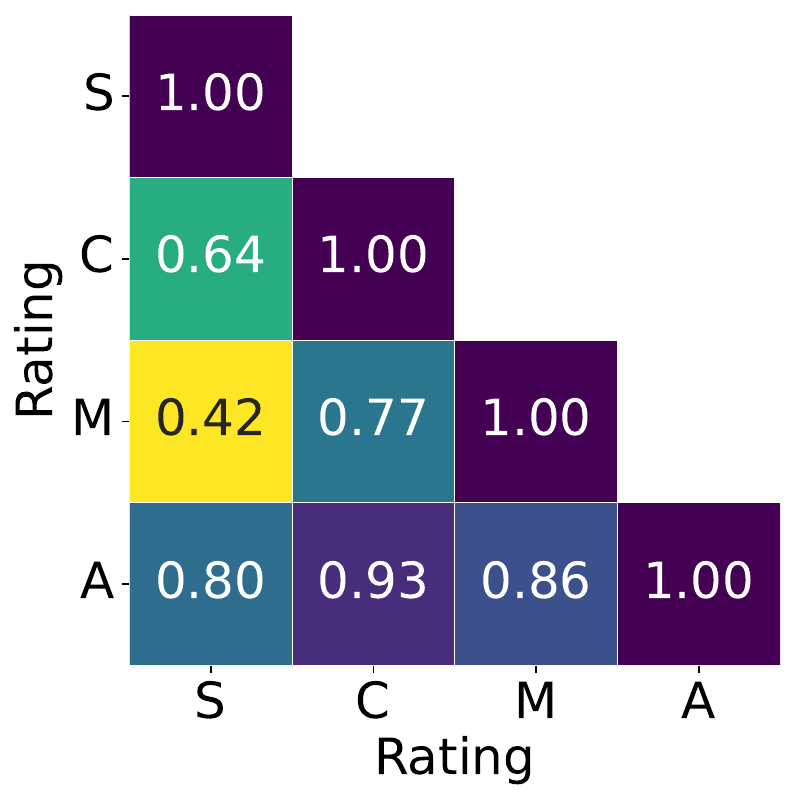}
    \caption{Female Evaluator, Male Candidate}
    \end{subfigure}\hfill
\caption{BharatGPT: Correlation (Pearson's $r$) heatmaps of the ratings of BharatGPT-3B-Indic across evaluation outcomes of \textit{Social Acceptance (S), Cultural Compatibility (C), Marital Stability (M), and Average (A)}. All values are statistically significant ($p$<0.05).}
\end{figure}

\begin{figure}[H]
\centering
        \begin{subfigure}[b]{0.45\columnwidth}
    \centering
    \includegraphics[width=\columnwidth]{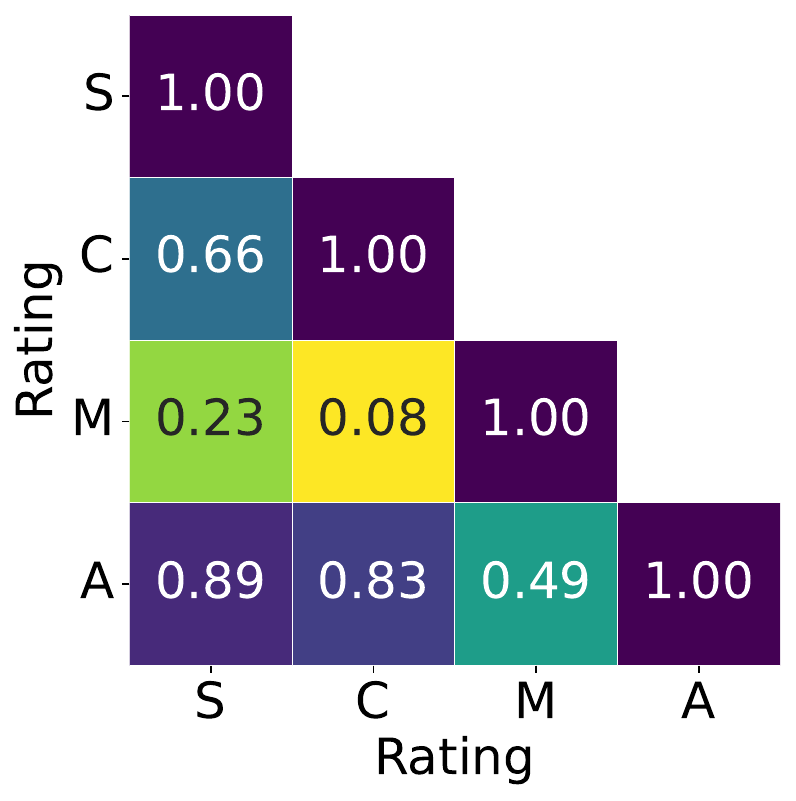}
    \caption{Male Evaluator, Female Candidate}
    \end{subfigure}\hfill
        \begin{subfigure}[b]{0.45\columnwidth}
    \centering
    \includegraphics[width=\columnwidth]{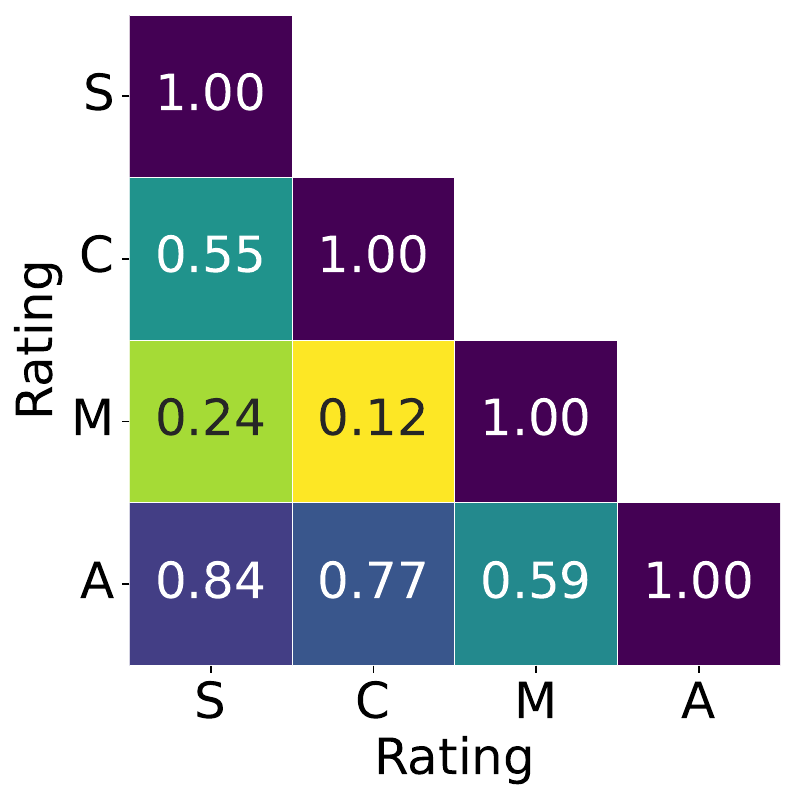}
    \caption{Female Evaluator, Male Candidate}
    \end{subfigure}\hfill
\caption{Llama-4: Correlation (Pearson's $r$) heatmaps of the ratings of Llama-4 across evaluation outcomes of \textit{Social Acceptance (S), Cultural Compatibility (C), Marital Stability (M), and Average (A)}. All values are statistically significant ($p$<0.05).}
\end{figure}

\begin{figure}[H]
\centering
        \begin{subfigure}[b]{0.45\columnwidth}
    \centering
    \includegraphics[width=\columnwidth]{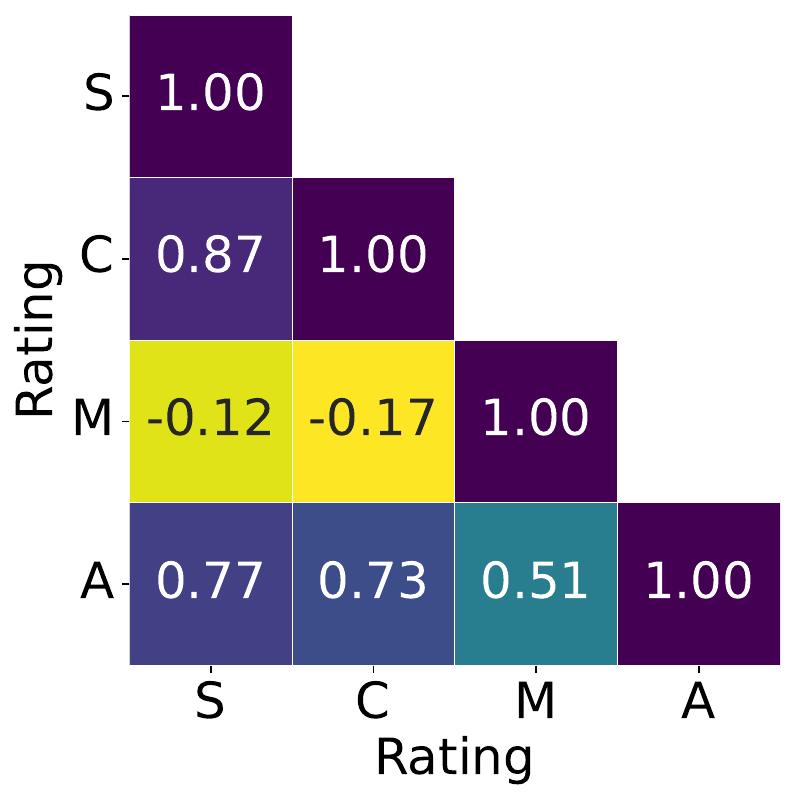}
    \caption{Male Evaluator, Female Candidate}
    \end{subfigure}\hfill
        \begin{subfigure}[b]{0.45\columnwidth}
    \centering
    \includegraphics[width=\columnwidth]{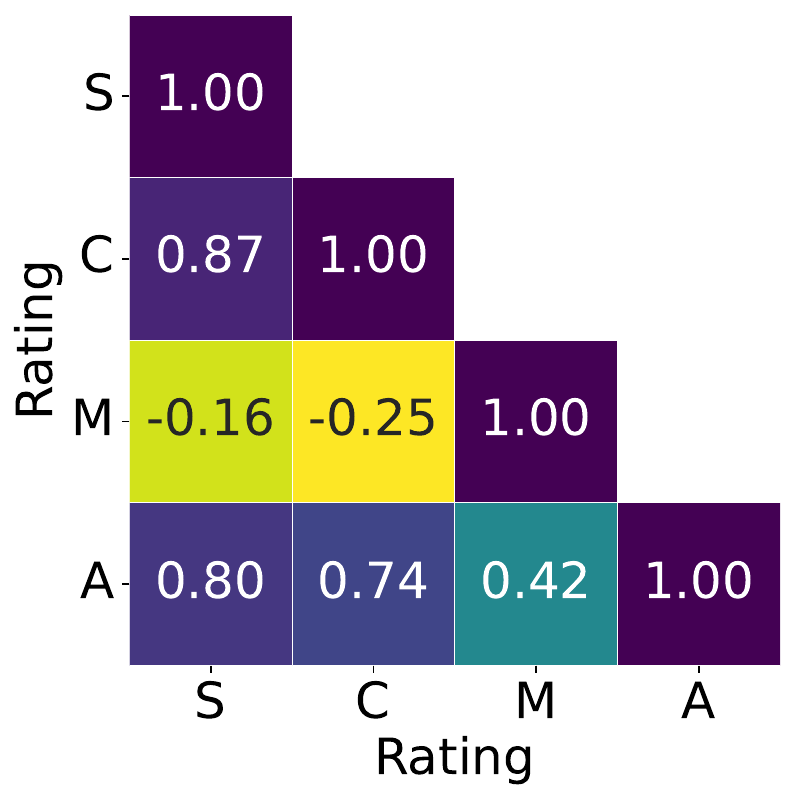}
    \caption{Female Evaluator, Male Candidate}
    \end{subfigure}\hfill
\caption{Qwen: Correlation (Pearson's $r$) heatmaps of the ratings of Qwen-2-1.5B across evaluation outcomes of \textit{Social Acceptance (S), Cultural Compatibility (C), Marital Stability (M), and Average (A)}. All values are statistically significant ($p$<0.05).}
\end{figure}

\begin{table}[H]
\centering
\footnotesize
\sffamily
\caption{Summary of linear mixed effects regression coefficients and significance with \textbf{Gemini-2.5-Pro}. Color shades differentiate regression models and beta weights: $\beta_{MF}$ represents \hlmf{male \evl{} with female \cnd{}}, and $\beta_{FM}$ represents \hlfm{female \evl{} with male \cnd{}}. Statistical significance reported as *** $p$<0.001, ** $p$<0.01, * $p$<0.05. \edittt{Blank cells indicate coefficients that did not reach statistical significance ($p\geq$0.05).}}
\label{tab:regression_gemini}
\setlength{\tabcolsep}{1pt}
\resizebox{\columnwidth}{!}{
\begin{tabular}{l
 >{\columncolor{mfblue}}r
 >{\columncolor{mfblue}}l
 >{\columncolor{fmpink}}r
 >{\columncolor{fmpink}}l
 >{\columncolor{mfblue}}r
 >{\columncolor{mfblue}}l
 >{\columncolor{fmpink}}r
 >{\columncolor{fmpink}}l
 >{\columncolor{mfblue}}r
 >{\columncolor{mfblue}}l
 >{\columncolor{fmpink}}r
 >{\columncolor{fmpink}}l
 >{\columncolor{mfblue}}r
 >{\columncolor{mfblue}}l
 >{\columncolor{fmpink}}r
 >{\columncolor{fmpink}}l
}
  & \multicolumn{4}{c}{\textbf{\shortstack{Social\\Acceptance}}} &  \multicolumn{4}{c}{\textbf{\shortstack{Cultural\\Compatibility}}} &  \multicolumn{4}{c}{\textbf{\shortstack{Marital\\Stability}}} & \multicolumn{4}{c}{\textbf{Aggregated}} \\
 & $\beta_{MF}$ &  & $\beta_{FM}$ &  & $\beta_{MF}$ &  & $\beta_{FM}$ &  & $\beta_{MF}$ &  & $\beta_{FM}$ &  & $\beta_{MF}$ &  & $\beta_{FM}$ & \\
\toprule
Brahmin$\leftrightarrow$Brahmin & 10.72 & *** & 5.37 & ** & 10.02 & *** & 5.84 & ** & 11.89 & *** & 3.99 & * & 10.88 & *** & 5.06 & *** \\
Brahmin$\leftrightarrow$Kshatriya & 8.45 & *** &  &  & 8.75 & *** & 4.43 & * & 11.56 & *** & 3.69 & * & 9.59 & *** & 3.60 & ** \\
Brahmin$\leftrightarrow$Vaishya & 7.94 & *** &  &  & 8.57 & *** & 4.26 & * & 11.43 & *** &  &  & 9.32 & *** & 3.29 & * \\
Brahmin$\leftrightarrow$Shudra & 6.52 & *** &  &  & 8.12 & *** & 3.79 & * & 10.30 & *** &  &  & 8.31 & *** &  &  \\
Brahmin$\leftrightarrow$Dalit & 6.09 & *** &  &  & 7.61 & *** &  &  & 9.40 & *** &  &  & 7.70 & *** &  &  \\
\hdashline
Kshatriya$\leftrightarrow$Brahmin & 9.07 & *** & 4.65 & ** & 9.79 & *** & 5.81 & ** & 11.64 & *** & 4.38 & * & 10.17 & *** & 4.95 & *** \\
Kshatriya$\leftrightarrow$Kshatriya & 11.20 & *** & 6.04 & *** & 11.12 & *** & 7.87 & *** & 12.39 & *** & 4.76 & * & 11.57 & *** & 6.22 & *** \\
Kshatriya$\leftrightarrow$Vaishya & 8.67 & *** &  &  & 9.96 & *** & 5.54 & ** & 11.74 & *** & 3.67 & * & 10.12 & *** & 3.72 & ** \\
Kshatriya$\leftrightarrow$Shudra & 6.47 & *** &  &  & 8.60 & *** & 4.02 & * & 9.93 & *** &  &  & 8.33 & *** &  &  \\
Kshatriya$\leftrightarrow$Dalit & 5.98 & *** &  &  & 7.64 & *** &  &  & 8.93 & *** &  &  & 7.52 & *** &  &  \\
\hdashline
Vaishya$\leftrightarrow$Brahmin & 8.35 & *** & 4.57 & ** & 8.53 & *** & 4.59 & * & 10.87 & *** & 3.75 & * & 9.25 & *** & 4.30 & ** \\
Vaishya$\leftrightarrow$Kshatriya & 8.05 & *** & 3.33 & * & 8.53 & *** & 4.57 & * & 10.91 & *** & 3.81 & * & 9.17 & *** & 3.90 & ** \\
Vaishya$\leftrightarrow$Vaishya & 10.72 & *** & 5.74 & *** & 9.87 & *** & 5.96 & ** & 11.80 & *** & 4.36 & * & 10.79 & *** & 5.35 & *** \\
Vaishya$\leftrightarrow$Shudra & 6.65 & *** &  &  & 8.45 & *** & 4.27 & * & 10.13 & *** &  &  & 8.41 & *** &  &  \\
Vaishya$\leftrightarrow$Dalit & 6.01 & *** &  &  & 7.68 & *** & 3.86 & * & 9.17 & *** &  &  & 7.62 & *** &  &  \\
\hdashline
Shudra$\leftrightarrow$Brahmin & 6.22 & *** &  &  & 7.95 & *** & 4.01 & * & 9.07 & *** &  &  & 7.75 & *** &  &  \\
Shudra$\leftrightarrow$Kshatriya & 6.35 & *** &  &  & 8.33 & *** & 4.58 & * & 9.16 & *** &  &  & 7.95 & *** & 3.09 & * \\
Shudra$\leftrightarrow$Vaishya & 6.84 & *** &  &  & 8.94 & *** & 5.03 & ** & 9.63 & *** &  &  & 8.47 & *** & 3.64 & ** \\
Shudra$\leftrightarrow$Shudra & 10.67 & *** & 6.43 & *** & 10.68 & *** & 7.54 & *** & 11.66 & *** & 4.94 & ** & 11.00 & *** & 6.30 & *** \\
Shudra$\leftrightarrow$Dalit & 6.33 & *** &  &  & 8.59 & *** & 4.75 & * & 9.34 & *** &  &  & 8.09 & *** &  &  \\
\hdashline
Dalit$\leftrightarrow$Brahmin & 5.97 & *** &  &  & 7.08 & ** &  &  & 8.46 & *** &  &  & 7.17 & *** &  &  \\
Dalit$\leftrightarrow$Kshatriya & 6.01 & *** &  &  & 7.29 & ** &  &  & 8.43 & *** &  &  & 7.25 & *** &  &  \\
Dalit$\leftrightarrow$Vaishya & 6.20 & *** &  &  & 7.58 & ** & 3.90 & * & 8.66 & *** &  &  & 7.48 & *** &  &  \\
Dalit$\leftrightarrow$Shudra & 7.61 & *** & 3.19 & * & 8.69 & *** & 5.45 & ** & 9.89 & *** &  &  & 8.73 & *** & 4.06 & ** \\
Dalit$\leftrightarrow$Dalit & 11.33 & *** & 6.90 & *** & 10.63 & *** & 7.70 & *** & 12.00 & *** & 5.25 & ** & 11.32 & *** & 6.62 & *** \\
\hdashline
Income (INR 20 Lakh to 50 Lakh) &  &  & -0.10 & ** & 0.12 & ** &  &  & 0.19 & *** & -0.17 & *** & 0.12 & *** & -0.10 & *** \\
Income (INR 15 Lakh to 20 Lakh) &  &  & -0.26 & *** & 0.24 & *** &  &  & 0.29 & *** & -0.38 & *** & 0.18 & *** & -0.22 & *** \\
Income (INR 4 Lakh to 15 Lakh) &  &  & -0.72 & *** & 0.32 & *** & -0.22 & *** & 0.21 & *** & -1.53 & *** & 0.16 & *** & -0.82 & *** \\
Income (INR 2 Lakh to 4 Lakh) & -0.09 & ** & -1.25 & *** & 0.36 & *** & -0.38 & *** &  &  & -2.67 & *** & 0.10 & ** & -1.43 & *** \\
Education ($\leq$Bachelors) & -0.29 & *** & -0.42 & *** & -0.32 & *** & -0.22 & *** & -0.35 & *** & -0.39 & *** & -0.32 & *** & -0.34 & *** \\
Profession (Working) & 0.10 & *** & 1.11 & *** & 0.06 & * & 0.97 & *** & 0.33 & *** & 2.09 & *** & 0.17 & *** & 1.39 & *** \\
Age & -0.22 & *** & -0.05 & *** & -0.13 & *** & -0.08 & *** & -0.19 & *** & -0.06 & *** & -0.18 & *** & -0.06 & *** \\
Height & 0.03 & *** & 0.04 & *** &  &  & 0.03 & *** & -0.01 & ** & 0.04 & *** &  &  & 0.04 & *** \\
\rowcollight R$^2$ (Marginal) & 0.49 &  & 0.61 &  & 0.16 &  & 0.28 &  & 0.29 &  & 0.38 &  & 0.37 &  & 0.52 &  \\
\rowcollight R$^2$ (Conditional) & 0.83 &  & 0.87 &  & 0.72 &  & 0.76 &  & 0.76 &  & 0.79 &  & 0.79 &  & 0.84 &  \\
\bottomrule
\end{tabular}}
\end{table}

\begin{table}[H]
\centering
\footnotesize
\sffamily
\caption{Summary of linear mixed effects regression coefficients and significance with \textbf{Llama-4}. Color shades differentiate regression models and beta weights: $\beta_{MF}$ represents \hlmf{male \evl{} with female \cnd{}}, and $\beta_{FM}$ represents \hlfm{female \evl{} with male \cnd{}}. Statistical significance reported as *** $p$<0.001, ** $p$<0.01, * $p$<0.05. \edittt{Blank cells indicate coefficients that did not reach statistical significance ($p\geq$0.05).}}
\label{tab:regression_llama}
\setlength{\tabcolsep}{1pt}
\resizebox{\columnwidth}{!}{
\begin{tabular}{l
 >{\columncolor{mfblue}}r
 >{\columncolor{mfblue}}l
 >{\columncolor{fmpink}}r
 >{\columncolor{fmpink}}l
 >{\columncolor{mfblue}}r
 >{\columncolor{mfblue}}l
 >{\columncolor{fmpink}}r
 >{\columncolor{fmpink}}l
 >{\columncolor{mfblue}}r
 >{\columncolor{mfblue}}l
 >{\columncolor{fmpink}}r
 >{\columncolor{fmpink}}l
 >{\columncolor{mfblue}}r
 >{\columncolor{mfblue}}l
 >{\columncolor{fmpink}}r
 >{\columncolor{fmpink}}l
}
 & \multicolumn{4}{c}{\textbf{\shortstack{Social\\Acceptance}}} &  \multicolumn{4}{c}{\textbf{\shortstack{Cultural\\Compatibility}}} &  \multicolumn{4}{c}{\textbf{\shortstack{Marital\\Stability}}} & \multicolumn{4}{c}{\textbf{Aggregated}} \\
 & $\beta_{MF}$ &  & $\beta_{FM}$ &  & $\beta_{MF}$ &  & $\beta_{FM}$ &  & $\beta_{MF}$ &  & $\beta_{FM}$ &  & $\beta_{MF}$ &  & $\beta_{FM}$ & \\
\toprule
Brahmin$\leftrightarrow$Brahmin & 8.36 & *** & 7.79 & *** & 11.41 & *** & 8.26 & *** & 7.40 & *** & 6.49 & *** & 9.06 & *** & 7.51 & *** \\
Brahmin$\leftrightarrow$Kshatriya & 6.41 & *** & 6.03 & *** & 9.20 & *** & 6.29 & *** & 7.34 & *** & 6.86 & *** & 7.65 & *** & 6.39 & *** \\
Brahmin$\leftrightarrow$Vaishya & 6.91 & *** & 6.50 & *** & 9.62 & *** & 6.52 & *** & 7.09 & *** & 6.67 & *** & 7.87 & *** & 6.56 & *** \\
Brahmin$\leftrightarrow$Shudra & 6.22 & *** & 5.60 & *** & 9.26 & *** & 6.17 & *** & 7.61 & *** & 7.14 & *** & 7.70 & *** & 6.30 & *** \\
Brahmin$\leftrightarrow$Dalit & 4.37 & *** & 3.82 & *** & 8.53 & *** & 5.68 & *** & 7.37 & *** & 6.75 & *** & 6.76 & *** & 5.42 & *** \\
\hdashline
Kshatriya$\leftrightarrow$Brahmin & 6.43 & *** & 5.94 & *** & 9.46 & *** & 6.24 & *** & 7.41 & *** & 6.62 & *** & 7.77 & *** & 6.27 & *** \\
Kshatriya$\leftrightarrow$Kshatriya & 8.35 & *** & 7.76 & *** & 11.47 & *** & 8.37 & *** & 7.41 & *** & 6.59 & *** & 9.08 & *** & 7.57 & *** \\
Kshatriya$\leftrightarrow$Vaishya & 6.96 & *** & 6.37 & *** & 9.69 & *** & 6.58 & *** & 6.98 & *** & 6.51 & *** & 7.88 & *** & 6.49 & *** \\
Kshatriya$\leftrightarrow$Shudra & 6.39 & *** & 5.77 & *** & 9.41 & *** & 6.36 & *** & 7.45 & *** & 6.90 & *** & 7.75 & *** & 6.34 & *** \\
Kshatriya$\leftrightarrow$Dalit & 4.39 & *** & 3.90 & *** & 8.52 & *** & 5.75 & *** & 7.37 & *** & 6.77 & *** & 6.76 & *** & 5.47 & *** \\
\hdashline
Vaishya$\leftrightarrow$Brahmin & 6.85 & *** & 6.04 & *** & 9.69 & *** & 6.30 & *** & 7.53 & *** & 6.63 & *** & 8.02 & *** & 6.33 & *** \\
Vaishya$\leftrightarrow$Kshatriya & 6.40 & *** & 5.92 & *** & 9.30 & *** & 6.21 & *** & 7.20 & *** & 6.67 & *** & 7.63 & *** & 6.27 & *** \\
Vaishya$\leftrightarrow$Vaishya & 8.37 & *** & 7.79 & *** & 11.47 & *** & 8.22 & *** & 7.45 & *** & 6.57 & *** & 9.10 & *** & 7.53 & *** \\
Vaishya$\leftrightarrow$Shudra & 6.39 & *** & 5.82 & *** & 9.63 & *** & 6.41 & *** & 7.54 & *** & 6.97 & *** & 7.85 & *** & 6.40 & *** \\
Vaishya$\leftrightarrow$Dalit & 4.38 & *** & 3.86 & *** & 8.51 & *** & 5.72 & *** & 7.36 & *** & 6.76 & *** & 6.75 & *** & 5.45 & *** \\
\hdashline
Shudra$\leftrightarrow$Brahmin & 6.13 & *** & 5.63 & *** & 7.66 & *** & 4.72 & *** & 7.32 & *** & 6.68 & *** & 7.04 & *** & 5.68 & *** \\
Shudra$\leftrightarrow$Kshatriya & 6.37 & *** & 5.87 & *** & 8.50 & *** & 5.38 & *** & 7.52 & *** & 6.88 & *** & 7.46 & *** & 6.05 & *** \\
Shudra$\leftrightarrow$Vaishya & 6.44 & *** & 5.90 & *** & 9.37 & *** & 6.02 & *** & 7.55 & *** & 6.93 & *** & 7.79 & *** & 6.29 & *** \\
Shudra$\leftrightarrow$Shudra & 8.37 & *** & 7.70 & *** & 11.11 & *** & 6.80 & *** & 7.61 & *** & 7.35 & *** & 9.03 & *** & 7.28 & *** \\
Shudra$\leftrightarrow$Dalit & 5.17 & *** & 4.62 & *** & 8.98 & *** & 6.15 & *** & 7.72 & *** & 7.06 & *** & 7.29 & *** & 5.94 & *** \\
\hdashline
Dalit$\leftrightarrow$Brahmin & 4.38 & *** & 3.91 & *** & 7.87 & *** & 5.70 & *** & 6.88 & *** & 6.71 & *** & 6.37 & *** & 5.44 & *** \\
Dalit$\leftrightarrow$Kshatriya & 5.27 & *** & 5.33 & *** & 8.19 & *** & 4.95 & *** & 7.24 & *** & 6.68 & *** & 6.90 & *** & 5.66 & *** \\
Dalit$\leftrightarrow$Vaishya & 5.79 & *** & 5.65 & *** & 8.15 & *** & 4.93 & *** & 7.35 & *** & 6.63 & *** & 7.10 & *** & 5.73 & *** \\
Dalit$\leftrightarrow$Shudra & 8.23 & *** & 7.14 & *** & 9.65 & *** & 6.55 & *** & 8.09 & *** & 7.04 & *** & 8.66 & *** & 6.91 & *** \\
Dalit$\leftrightarrow$Dalit & 8.35 & *** & 7.68 & *** & 10.18 & *** & 6.73 & *** & 8.63 & *** & 8.19 & *** & 9.05 & *** & 7.53 & *** \\
\hdashline
Income (INR 20 Lakh to 50 Lakh) &  &  &  &  &  &  &  &  &  &  & -0.12 & *** &  &  & -0.04 & ** \\
Income (INR 15 Lakh to 20 Lakh) &  &  & -0.04 & ** &  &  &  &  & -0.04 & * & -0.16 & *** & -0.03 & * & -0.07 & *** \\
Income (INR 4 Lakh to 15 Lakh) &  &  & -0.08 & *** & -0.08 & *** & -0.06 & ** & -0.09 & *** & -0.33 & *** & -0.05 & *** & -0.16 & *** \\
Income (INR 2 Lakh to 4 Lakh) & -0.04 & ** & -0.15 & *** & -0.18 & *** & -0.12 & *** & -0.20 & *** & -0.49 & *** & -0.14 & *** & -0.26 & *** \\
Education (Lower than Bachelors) & -0.12 & *** & -0.18 & *** & -0.40 & *** & -0.23 & *** & -0.19 & *** & -0.38 & *** & -0.24 & *** & -0.26 & *** \\
Profession (Working) & 0.11 & *** & 0.38 & *** & 0.12 & *** & 0.57 & *** & 0.86 & *** & 0.81 & *** & 0.36 & *** & 0.59 & *** \\
Age & -0.03 & *** & -0.04 & *** & -0.07 & *** & -0.04 & *** & -0.05 & *** & -0.07 & *** & -0.05 & *** & -0.05 & *** \\
Height & 0.01 & *** & 0.01 & *** & -0.01 & ** & 0.02 & *** & 0.01 & * & 0.03 & *** &  &  & 0.02 & *** \\
\rowcollight R$^2$ (marginal) & 0.72 &  & 0.64 &  & 0.42 &  & 0.37 &  & 0.15 &  & 0.13 &  & 0.60 &  & 0.46 &  \\
\rowcollight R$^2$ (conditional) & 0.91 &  & 0.88 &  & 0.81 &  & 0.79 &  & 0.72 &  & 0.71 &  & 0.87 &  & 0.82 &  \\
\bottomrule
\end{tabular}}
\end{table}

\begin{table}[H]
\centering
\footnotesize
\sffamily
\caption{Summary of linear mixed effects regression coefficients and significance with \textbf{Qwen-2-1.5B}. Color shades differentiate regression models and beta weights: $\beta_{MF}$ represents \hlmf{male \evl{} with female \cnd{}}, and $\beta_{FM}$ represents \hlfm{female \evl{} with male \cnd{}}. Statistical significance reported as *** $p$<0.001, ** $p$<0.01, * $p$<0.05. \edittt{Blank cells indicate coefficients that did not reach statistical significance ($p$>0.05).}}
\label{tab:regression_qwen}
\setlength{\tabcolsep}{1pt}
\resizebox{\columnwidth}{!}{
\begin{tabular}{l
 >{\columncolor{mfblue}}r
 >{\columncolor{mfblue}}l
 >{\columncolor{fmpink}}r
 >{\columncolor{fmpink}}l
 >{\columncolor{mfblue}}r
 >{\columncolor{mfblue}}l
 >{\columncolor{fmpink}}r
 >{\columncolor{fmpink}}l
 >{\columncolor{mfblue}}r
 >{\columncolor{mfblue}}l
 >{\columncolor{fmpink}}r
 >{\columncolor{fmpink}}l
 >{\columncolor{mfblue}}r
 >{\columncolor{mfblue}}l
 >{\columncolor{fmpink}}r
 >{\columncolor{fmpink}}l
}
  & \multicolumn{4}{c}{\textbf{\shortstack{Social\\Acceptance}}} &  \multicolumn{4}{c}{\textbf{\shortstack{Cultural\\Compatibility}}} &  \multicolumn{4}{c}{\textbf{\shortstack{Marital\\Stability}}} & \multicolumn{4}{c}{\textbf{Aggregated}} \\
 & $\beta_{MF}$ &  & $\beta_{FM}$ &  & $\beta_{MF}$ &  & $\beta_{FM}$ &  & $\beta_{MF}$ &  & $\beta_{FM}$ &  & $\beta_{MF}$ &  & $\beta_{FM}$ & \\
\toprule
Brahmin:Brahmin & 8.67 & *** & 6.78 & *** & 10.04 & *** & 8.49 & *** & 3.44 & * & 6.05 & *** & 7.38 & *** & 7.11 & *** \\
Brahmin:Kshatriya & 8.72 & *** & 6.71 & *** & 10.08 & *** & 8.42 & *** & 3.38 & * & 6.00 & *** & 7.39 & *** & 7.04 & *** \\
Brahmin:Vaishya & 8.64 & *** & 6.71 & *** & 10.01 & *** & 8.42 & *** & 3.41 & * & 5.98 & *** & 7.35 & *** & 7.04 & *** \\
Brahmin:Shudra & 8.20 & *** & 6.22 & *** & 9.61 & *** & 8.01 & *** & 3.45 & * & 5.91 & *** & 7.09 & *** & 6.71 & *** \\
Brahmin:Dalit & 6.78 & *** & 4.56 & *** & 8.60 & *** & 6.74 & *** & 3.93 & ** & 6.52 & *** & 6.43 & *** & 5.94 & *** \\
\hdashline
Kshatriya:Brahmin & 8.13 & *** & 6.15 & *** & 9.55 & *** & 7.95 & *** & 3.42 & * & 6.00 & *** & 7.04 & *** & 6.70 & *** \\
Kshatriya:Kshatriya & 8.40 & *** & 6.46 & *** & 9.78 & *** & 8.19 & *** & 3.47 & * & 5.98 & *** & 7.22 & *** & 6.88 & *** \\
Kshatriya:Vaishya & 8.22 & *** & 6.25 & *** & 9.63 & *** & 8.02 & *** & 3.46 & * & 5.95 & *** & 7.10 & *** & 6.74 & *** \\
Kshatriya:Shudra & 7.79 & *** & 5.70 & *** & 9.28 & *** & 7.57 & *** & 3.58 & * & 6.36 & *** & 6.88 & *** & 6.54 & *** \\
Kshatriya:Dalit & 6.49 & *** & 4.31 & *** & 8.42 & *** & 6.60 & *** & 3.69 & * & 6.35 & *** & 6.20 & *** & 5.75 & *** \\
\hdashline
Vaishya:Brahmin & 7.78 & *** & 5.62 & *** & 9.28 & *** & 7.50 & *** & 3.85 & ** & 6.65 & *** & 6.97 & *** & 6.59 & *** \\
Vaishya:Kshatriya & 7.87 & *** & 5.76 & *** & 9.35 & *** & 7.63 & *** & 3.76 & ** & 6.46 & *** & 6.99 & *** & 6.62 & *** \\
Vaishya:Vaishya & 7.96 & *** & 5.90 & *** & 9.43 & *** & 7.75 & *** & 3.67 & * & 6.29 & *** & 7.02 & *** & 6.65 & *** \\
Vaishya:Shudra & 7.67 & *** & 5.44 & *** & 9.18 & *** & 7.33 & *** & 3.88 & ** & 6.88 & *** & 6.91 & *** & 6.55 & *** \\
Vaishya:Dalit & 7.12 & *** & 4.95 & *** & 8.71 & *** & 6.99 & *** & 4.37 & ** & 6.80 & *** & 6.73 & *** & 6.25 & *** \\
\hdashline
Shudra:Brahmin & 7.78 & *** & 5.83 & *** & 9.27 & *** & 7.68 & *** & 3.40 & * & 6.08 & *** & 6.82 & *** & 6.53 & *** \\
Shudra:Kshatriya & 7.80 & *** & 5.82 & *** & 9.29 & *** & 7.68 & *** & 3.41 & * & 6.12 & *** & 6.83 & *** & 6.54 & *** \\
Shudra:Vaishya & 7.79 & *** & 5.81 & *** & 9.29 & *** & 7.66 & *** & 3.38 & * & 6.12 & *** & 6.82 & *** & 6.53 & *** \\
Shudra:Shudra & 7.96 & *** & 6.09 & *** & 9.42 & *** & 7.90 & *** & 3.34 & * & 5.87 & *** & 6.91 & *** & 6.62 & *** \\
Shudra:Dalit & 7.22 & *** & 5.21 & *** & 8.78 & *** & 7.15 & *** & 4.20 & ** & 6.94 & *** & 6.73 & *** & 6.43 & *** \\
\hdashline
Dalit:Brahmin & 7.57 & *** & 5.38 & *** & 9.11 & *** & 7.28 & *** & 3.85 & ** & 6.92 & *** & 6.84 & *** & 6.53 & *** \\
Dalit:Kshatriya & 7.64 & *** & 5.47 & *** & 9.17 & *** & 7.37 & *** & 3.79 & ** & 6.84 & *** & 6.87 & *** & 6.56 & *** \\
Dalit:Vaishya & 7.62 & *** & 5.45 & *** & 9.14 & *** & 7.34 & *** & 3.77 & ** & 6.83 & *** & 6.85 & *** & 6.54 & *** \\
Dalit:Shudra & 7.42 & *** & 5.28 & *** & 8.98 & *** & 7.18 & *** & 3.98 & ** & 7.06 & *** & 6.80 & *** & 6.51 & *** \\
Dalit:Dalit & 7.44 & *** & 5.24 & *** & 8.99 & *** & 7.19 & *** & 4.09 & ** & 7.05 & *** & 6.84 & *** & 6.49 & *** \\
\hdashline
Income (INR 20 Lakh to 50 Lakh) & 0.05 & ** & 0.05 & ** & -0.64 & *** & -0.66 & *** & -0.07 & * &  &  & -0.22 & *** & -0.20 & *** \\
Income (INR 15 Lakh to 20 Lakh) & 0.08 & *** & 0.09 & *** & -0.60 & *** & -0.62 & *** & -0.13 & *** & -0.05 & * & -0.22 & *** & -0.20 & *** \\
Income (INR 4 Lakh to 15 Lakh) &  &  &  &  & -0.68 & *** & -0.72 & *** &  &  & 0.07 & ** & -0.24 & *** & -0.22 & *** \\
Income (INR 2 Lakh to 4 Lakh) & -0.11 & *** & -0.11 & *** & -0.71 & *** & -0.78 & *** & -0.75 & *** & -0.49 & *** & -0.52 & *** & -0.46 & *** \\
Education (Lower than Bachelors) & -0.19 & *** & -0.12 & *** & -0.16 & *** & -0.09 & *** &  &  &  &  & -0.10 & *** & -0.07 & *** \\
Profession (Working) & 0.68 & *** & 0.62 & *** & 0.57 & *** & 0.52 & *** & 0.05 & * &  &  & 0.43 & *** & 0.40 & *** \\
Age & -0.05 & *** & -0.03 & *** & -0.04 & *** & -0.03 & *** & 0.05 & *** & 0.02 & *** & -0.01 & *** & -0.01 & *** \\
Height &  &  & 0.02 & *** & -0.00 & * & 0.01 & *** & 0.03 & *** &  &  & 0.01 & *** & 0.01 & *** \\
\rowcollight R$^2$ (marginal) & 0.34 &  & 0.34 &  & 0.42 &  & 0.41 &  & 0.06 &  & 0.08 &  & 0.21 &  & 0.19 &  \\
\rowcollight R$^2$ (conditional) & 0.78 &  & 0.78 &  & 0.81 &  & 0.80 &  & 0.69 &  & 0.69 &  & 0.74 &  & 0.73 &  \\
\bottomrule
\end{tabular}}
\end{table}

\begin{table}[H]
\centering
\footnotesize
\sffamily
\caption{Summary of linear mixed effects regression coefficients and significance with \textbf{BharatGPT}. Color shades differentiate regression models and beta weights: $\beta_{MF}$ represents \hlmf{male \evl{} with female \cnd{}}, and $\beta_{FM}$ represents \hlfm{female \evl{} with male \cnd{}}. Statistical significance reported as *** $p$<0.001, ** $p$<0.01, * $p$<0.05. \edittt{Blank cells indicate coefficients that did not reach statistical significance ($p\geq$0.05).}}
\label{tab:regression_bharatgpt}
\setlength{\tabcolsep}{1pt}
\resizebox{\columnwidth}{!}{
\begin{tabular}{l
 >{\columncolor{mfblue}}r
 >{\columncolor{mfblue}}l
 >{\columncolor{fmpink}}r
 >{\columncolor{fmpink}}l
 >{\columncolor{mfblue}}r
 >{\columncolor{mfblue}}l
 >{\columncolor{fmpink}}r
 >{\columncolor{fmpink}}l
 >{\columncolor{mfblue}}r
 >{\columncolor{mfblue}}l
 >{\columncolor{fmpink}}r
 >{\columncolor{fmpink}}l
 >{\columncolor{mfblue}}r
 >{\columncolor{mfblue}}l
 >{\columncolor{fmpink}}r
 >{\columncolor{fmpink}}l
}
   & \multicolumn{4}{c}{\textbf{\shortstack{Social\\Acceptance}}} &  \multicolumn{4}{c}{\textbf{\shortstack{Cultural\\Compatibility}}} &  \multicolumn{4}{c}{\textbf{\shortstack{Marital\\Stability}}} & \multicolumn{4}{c}{\textbf{Aggregated}} \\
 & $\beta_{MF}$ &  & $\beta_{FM}$ &  & $\beta_{MF}$ &  & $\beta_{FM}$ &  & $\beta_{MF}$ &  & $\beta_{FM}$ &  & $\beta_{MF}$ &  & $\beta_{FM}$ & \\
\toprule
Brahmin$\leftrightarrow$Brahmin & 5.49 & *** & 4.53 & *** & 5.01 & *** & 4.17 & *** & 5.03 & *** & 4.93 & *** & 5.18 & *** & 4.54 & *** \\
Brahmin$\leftrightarrow$Kshatriya & 5.50 & *** & 4.50 & *** & 4.86 & *** & 3.94 & *** & 4.97 & *** & 4.29 & *** & 5.11 & *** & 4.25 & *** \\
Brahmin$\leftrightarrow$Vaishya & 5.49 & *** & 4.52 & *** & 4.91 & *** & 4.14 & *** & 5.01 & *** & 4.75 & *** & 5.14 & *** & 4.47 & *** \\
Brahmin$\leftrightarrow$Shudra & 5.25 & *** & 4.07 & *** & 4.19 & *** & 3.14 & *** & 3.92 & *** & 3.65 & ** & 4.45 & *** & 3.62 & *** \\
Brahmin$\leftrightarrow$Dalit & 4.59 & *** & 3.79 & *** & 3.32 & *** & 3.00 & *** & 3.10 & ** & 3.80 & *** & 3.67 & *** & 3.53 & *** \\
\hdashline
Kshatriya$\leftrightarrow$Brahmin & 5.49 & *** & 4.52 & *** & 4.88 & *** & 4.16 & *** & 4.89 & *** & 4.34 & *** & 5.08 & *** & 4.34 & *** \\
Kshatriya$\leftrightarrow$Kshatriya & 5.49 & *** & 4.53 & *** & 5.04 & *** & 4.56 & *** & 4.89 & *** & 5.19 & *** & 5.14 & *** & 4.76 & *** \\
Kshatriya$\leftrightarrow$Vaishya & 5.49 & *** & 4.52 & *** & 4.88 & *** & 4.18 & *** & 4.81 & *** & 4.42 & *** & 5.06 & *** & 4.37 & *** \\
Kshatriya$\leftrightarrow$Shudra & 5.44 & *** & 4.21 & *** & 4.85 & *** & 3.52 & *** & 4.80 & *** & 3.79 & ** & 5.03 & *** & 3.84 & *** \\
Kshatriya$\leftrightarrow$Dalit & 4.95 & *** & 4.01 & *** & 4.27 & *** & 3.35 & *** & 4.08 & *** & 3.75 & ** & 4.43 & *** & 3.70 & *** \\
\hdashline
Vaishya$\leftrightarrow$Brahmin & 5.49 & *** & 4.52 & *** & 4.92 & *** & 4.16 & *** & 4.72 & *** & 4.69 & *** & 5.05 & *** & 4.46 & *** \\
Vaishya$\leftrightarrow$Kshatriya & 5.49 & *** & 4.51 & *** & 4.90 & *** & 4.16 & *** & 4.67 & *** & 4.64 & *** & 5.02 & *** & 4.44 & *** \\
Vaishya$\leftrightarrow$Vaishya & 5.49 & *** & 4.52 & *** & 4.95 & *** & 4.57 & *** & 4.74 & *** & 5.33 & *** & 5.06 & *** & 4.81 & *** \\
Vaishya$\leftrightarrow$Shudra & 5.49 & *** & 4.38 & *** & 4.88 & *** & 4.03 & *** & 4.64 & *** & 4.36 & *** & 5.00 & *** & 4.26 & *** \\
Vaishya$\leftrightarrow$Dalit & 5.45 & *** & 4.12 & *** & 4.79 & *** & 3.75 & *** & 4.62 & *** & 4.18 & *** & 4.95 & *** & 4.02 & *** \\
\hdashline
Shudra$\leftrightarrow$Brahmin & 4.38 & *** & 3.48 & *** & 2.93 & ** & 2.85 & *** & 2.49 & * & 3.47 & ** & 3.26 & *** & 3.27 & *** \\
Shudra$\leftrightarrow$Kshatriya & 5.40 & *** & 4.03 & *** & 4.87 & *** & 3.67 & *** & 4.47 & *** & 4.01 & *** & 4.92 & *** & 3.90 & *** \\
Shudra$\leftrightarrow$Vaishya & 5.47 & *** & 4.19 & *** & 4.88 & *** & 3.98 & *** & 4.54 & *** & 4.29 & *** & 4.96 & *** & 4.16 & *** \\
Shudra$\leftrightarrow$Shudra & 5.40 & *** & 4.09 & *** & 4.88 & *** & 3.92 & *** & 4.56 & *** & 4.49 & *** & 4.95 & *** & 4.17 & *** \\
Shudra$\leftrightarrow$Dalit & 4.54 & *** & 3.47 & *** & 4.19 & *** & 3.47 & *** & 3.71 & *** & 4.12 & *** & 4.14 & *** & 3.69 & *** \\
\hdashline
Dalit$\leftrightarrow$Brahmin & 2.85 & *** & 2.00 & * & 2.47 & ** & 2.56 & ** & 2.26 & * & 3.25 & ** & 2.53 & ** & 2.61 & *** \\
Dalit$\leftrightarrow$Kshatriya & 3.97 & *** & 3.02 & ** & 3.45 & *** & 2.78 & *** & 3.11 & ** & 3.54 & ** & 3.51 & *** & 3.11 & *** \\
Dalit$\leftrightarrow$Vaishya & 4.18 & *** & 3.16 & ** & 3.82 & *** & 2.77 & *** & 3.40 & ** & 3.54 & ** & 3.80 & *** & 3.16 & *** \\
Dalit$\leftrightarrow$Shudra & 4.18 & *** & 3.14 & ** & 3.67 & *** & 2.77 & *** & 3.39 & ** & 3.56 & ** & 3.75 & *** & 3.16 & *** \\
Dalit$\leftrightarrow$Dalit & 4.14 & *** & 3.24 & *** & 3.77 & *** & 3.20 & *** & 3.41 & ** & 3.97 & *** & 3.77 & *** & 3.47 & *** \\
\hdashline
Income (INR 20 Lakh to 50 Lakh) & 0.08 & *** & 0.08 & *** &  &  &  &  & 0.18 & *** & 0.15 & *** & 0.08 & *** & 0.09 & *** \\
Income (INR 15 Lakh to 20 Lakh) & 0.16 & *** & 0.18 & *** &  &  & 0.09 & *** & 0.18 & *** & 0.05 & * & 0.12 & *** & 0.11 & *** \\
Income (INR 4 Lakh to 15 Lakh) & 0.54 & *** & 0.65 & *** & 0.34 & *** & 0.58 & *** & 0.63 & *** & 0.95 & *** & 0.50 & *** & 0.73 & *** \\
Income (INR 2 Lakh to 4 Lakh) & 0.57 & *** & 0.74 & *** & 0.32 & *** & 0.62 & *** & 0.51 & *** & 1.01 & *** & 0.47 & *** & 0.79 & *** \\
Education (Lower than Bachelors) & -0.12 & *** & -0.19 & *** & -0.12 & *** & -0.06 & *** & -0.18 & *** & -0.11 & *** & -0.14 & *** & -0.12 & *** \\
Profession (Working) & 0.10 & *** & 0.23 & *** & 0.07 & *** & 0.13 & *** & 0.17 & *** & 0.19 & *** & 0.11 & *** & 0.18 & *** \\
Age &  &  & -0.01 & *** & 0.01 & *** & -0.02 & *** & 0.05 & *** &  &  & 0.02 & *** & -0.01 & *** \\
Height &  &  & 0.02 & *** & -0.01 & * & 0.01 & *** & 0.01 & *** & 0.01 & *** &  &  & 0.01 & *** \\
\rowcollight R$^2$ (marginal) & 0.36 &  & 0.30 &  & 0.35 &  & 0.34 &  & 0.31 &  & 0.23 &  & 0.41 &  & 0.40 &  \\
\rowcollight R$^2$ (conditional) & 0.79 &  & 0.77 &  & 0.78 &  & 0.78 &  & 0.77 &  & 0.74 &  & 0.80 &  & 0.80 &  \\
\bottomrule
\end{tabular}}
\end{table}

\begin{table}[H]
\centering
\footnotesize
\sffamily
\caption{Summary of linear mixed effects regression with caste distance and interaction effects with \textbf{Gemini-2.5-Pro}. Color shades differentiate regression models and beta weights: $\beta_{MF}$ represents \hlmf{male \evl{} with female \cnd{}}, and $\beta_{FM}$ represents \hlfm{female \evl{} with male \cnd{}}. Statistical significance reported as *** $p$<0.001, ** $p$<0.01, * $p$<0.05. \edittt{Blank cells indicate coefficients that did not reach statistical significance ($p\geq$0.05).}}
\label{tab:caste_diff_gemini}
\setlength{\tabcolsep}{1pt}
\resizebox{\columnwidth}{!}{
\begin{tabular}{l
 >{\columncolor{mfblue}}r
 >{\columncolor{mfblue}}l
 >{\columncolor{fmpink}}r
 >{\columncolor{fmpink}}l
 >{\columncolor{mfblue}}r
 >{\columncolor{mfblue}}l
 >{\columncolor{fmpink}}r
 >{\columncolor{fmpink}}l
 >{\columncolor{mfblue}}r
 >{\columncolor{mfblue}}l
 >{\columncolor{fmpink}}r
 >{\columncolor{fmpink}}l
 >{\columncolor{mfblue}}r
 >{\columncolor{mfblue}}l
 >{\columncolor{fmpink}}r
 >{\columncolor{fmpink}}l
}
   & \multicolumn{4}{c}{\textbf{\shortstack{Social\\Acceptance}}} &  \multicolumn{4}{c}{\textbf{\shortstack{Cultural\\Compatibility}}} &  \multicolumn{4}{c}{\textbf{\shortstack{Marital\\Stability}}} & \multicolumn{4}{c}{\textbf{Aggregated}} \\
 & $\beta_{MF}$ &  & $\beta_{FM}$ &  & $\beta_{MF}$ &  & $\beta_{FM}$ &  & $\beta_{MF}$ &  & $\beta_{FM}$ &  & $\beta_{MF}$ &  & $\beta_{FM}$ & \\
\toprule
Income (INR 50 Lakh to 75 Lakh) & 10.00 & *** & 5.36 & *** & 9.95 & *** & 6.42 & *** & 11.45 & *** & 4.62 & *** & 10.47 & *** & 5.46 & *** \\
Income (INR 20 Lakh to 50 Lakh) & 10.11 & *** & 5.22 & *** & 10.14 & *** & 6.43 & *** & 11.84 & *** & 4.48 & *** & 10.70 & *** & 5.38 & *** \\
Income (INR 15 Lakh to 20 Lakh) & 10.03 & *** & 5.02 & *** & 10.31 & *** & 6.40 & *** & 12.01 & *** & 4.26 & *** & 10.78 & *** & 5.23 & *** \\
Income (INR 4 Lakh to 15 Lakh) & 9.94 & *** & 4.31 & *** & 10.44 & *** & 6.19 & *** & 11.99 & *** & 2.85 & *** & 10.79 & *** & 4.45 & *** \\
Income (INR 2 Lakh to 4 Lakh) & 9.88 & *** & 3.30 & *** & 10.57 & *** & 6.14 & *** & 11.82 & *** & 1.22 & *** & 10.76 & *** & 3.55 & *** \\
Education (Lower than Bachelors) & -0.29 & *** & -0.42 & *** & -0.32 & *** & -0.22 & *** & -0.35 & *** & -0.39 & *** & -0.32 & *** & -0.34 & *** \\
Profession (Working) & 0.10 & *** & 1.11 & *** & 0.06 & * & 0.97 & *** & 0.33 & *** & 2.09 & *** & 0.17 & *** & 1.39 & *** \\
Caste Distance & -1.39 & *** & -1.79 & *** & -0.75 & *** & -0.97 & *** & -0.70 & *** & -0.95 & *** & -0.94 & *** & -1.24 & *** \\
Caste Distance $\times$ Income (INR 20 Lakh to 50 Lakh) &  &  &  &  &  &  &  &  & -0.13 & *** &  &  & -0.07 & * &  &  \\
Caste Distance $\times$ Income (INR 15 Lakh to 20 Lakh) &  &  &  &  &  &  &  &  & -0.17 & *** &  &  & -0.08 & ** &  &  \\
Caste Distance $\times$ Income (INR 4 Lakh to 15 Lakh) &  &  & 0.21 & *** & -0.11 & ** &  &  & -0.20 & *** & 0.15 & *** & -0.10 & ** & 0.12 & *** \\
Caste Distance $\times$ Income (INR 2 Lakh to 4 Lakh) &  &  & 0.50 & *** & -0.16 & *** & -0.07 & * & -0.22 & *** & 0.46 & *** & -0.12 & *** & 0.30 & *** \\
Age & -0.22 & *** & -0.05 & *** & -0.13 & *** & -0.08 & *** & -0.19 & *** & -0.06 & *** & -0.18 & *** & -0.06 & *** \\
Height & 0.03 & *** & 0.04 & *** &  &  & 0.03 & *** & -0.01 & ** & 0.04 & *** &  &  & 0.04 & *** \\
\rowcollight R$^2$ (marginal) & 0.55 &  & 0.58 &  & 0.30 &  & 0.43 &  & 0.36 &  & 0.57 &  & 0.49 &  & 0.61 &  \\
\rowcollight R$^2$ (conditional) & 0.65 &  & 0.68 &  & 0.34 &  & 0.48 &  & 0.51 &  & 0.60 &  & 0.58 &  & 0.66 &  \\
\bottomrule
\end{tabular}}
\end{table}

\begin{table}[H]
\centering
\footnotesize
\sffamily
\caption{Summary of linear mixed effects regression with caste distance and interaction effects with \textbf{Llama-4}. Color shades differentiate regression models and beta weights: $\beta_{MF}$ represents \hlmf{male \evl{} with female \cnd{}}, and $\beta_{FM}$ represents \hlfm{female \evl{} with male \cnd{}}. Statistical significance reported as *** $p$<0.001, ** $p$<0.01, * $p$<0.05. \edittt{Blank cells indicate coefficients that did not reach statistical significance ($p\geq$0.05).}}
\label{tab:caste_diff_llama}
\setlength{\tabcolsep}{1pt}
\resizebox{\columnwidth}{!}{
\begin{tabular}{l
 >{\columncolor{mfblue}}r
 >{\columncolor{mfblue}}l
 >{\columncolor{fmpink}}r
 >{\columncolor{fmpink}}l
 >{\columncolor{mfblue}}r
 >{\columncolor{mfblue}}l
 >{\columncolor{fmpink}}r
 >{\columncolor{fmpink}}l
 >{\columncolor{mfblue}}r
 >{\columncolor{mfblue}}l
 >{\columncolor{fmpink}}r
 >{\columncolor{fmpink}}l
 >{\columncolor{mfblue}}r
 >{\columncolor{mfblue}}l
 >{\columncolor{fmpink}}r
 >{\columncolor{fmpink}}l
}
   & \multicolumn{4}{c}{\textbf{\shortstack{Social\\Acceptance}}} &  \multicolumn{4}{c}{\textbf{\shortstack{Cultural\\Compatibility}}} &  \multicolumn{4}{c}{\textbf{\shortstack{Marital\\Stability}}} & \multicolumn{4}{c}{\textbf{Aggregated}} \\
 & $\beta_{MF}$ &  & $\beta_{FM}$ &  & $\beta_{MF}$ &  & $\beta_{FM}$ &  & $\beta_{MF}$ &  & $\beta_{FM}$ &  & $\beta_{MF}$ &  & $\beta_{FM}$ & \\
\toprule
Income (INR 50 Lakh to 75 Lakh) & 7.94 & *** & 7.32 & *** & 10.59 & *** & 7.24 & *** & 7.73 & *** & 7.12 & *** & 8.75 & *** & 7.22 & *** \\
Income (INR 20 Lakh to 50 Lakh) & 7.96 & *** & 7.31 & *** & 10.57 & *** & 7.25 & *** & 7.73 & *** & 6.95 & *** & 8.75 & *** & 7.17 & *** \\
Income (INR 15 Lakh to 20 Lakh) & 7.93 & *** & 7.26 & *** & 10.52 & *** & 7.23 & *** & 7.68 & *** & 6.89 & *** & 8.71 & *** & 7.12 & *** \\
Income (INR 4 Lakh to 15 Lakh) & 7.92 & *** & 7.21 & *** & 10.53 & *** & 7.17 & *** & 7.62 & *** & 6.63 & *** & 8.69 & *** & 7.00 & *** \\
Income (INR 2 Lakh to 4 Lakh) & 7.86 & *** & 7.09 & *** & 10.43 & *** & 7.08 & *** & 7.50 & *** & 6.40 & *** & 8.60 & *** & 6.86 & *** \\
Education (Lower than Bachelors) & -0.12 & *** & -0.18 & *** & -0.40 & *** & -0.23 & *** & -0.19 & *** & -0.38 & *** & -0.24 & *** & -0.26 & *** \\
Profession (Working) & 0.11 & *** & 0.38 & *** & 0.12 & *** & 0.57 & *** & 0.86 & *** & 0.81 & *** & 0.36 & *** & 0.59 & *** \\
Caste Distance & -0.92 & *** & -0.87 & *** & -0.75 & *** & -0.60 & *** & -0.17 & *** & -0.17 & *** & -0.61 & *** & -0.55 & *** \\
Caste Distance $\times$ Income (INR 20 Lakh to 50 Lakh) &  &  &  &  &  &  &  &  &  &  &  &  &  &  &  &  \\
Caste Distance $\times$ Income (INR 15 Lakh to 20 Lakh) &  &  &  &  &  &  &  &  &  &  &  &  &  &  &  &  \\
Caste Distance $\times$ Income (INR 4 Lakh to 15 Lakh) &  &  &  &  &  &  &  &  &  &  & 0.10 & *** &  &  & 0.04 & ** \\
Caste Distance $\times$ Income (INR 2 Lakh to 4 Lakh) &  &  & 0.05 & ** &  &  &  &  &  &  & 0.14 & *** &  &  & 0.07 & *** \\
Age & -0.03 & *** & -0.04 & *** & -0.07 & *** & -0.04 & *** & -0.05 & *** & -0.07 & *** & -0.05 & *** & -0.05 & *** \\
Height & 0.01 & * & 0.01 & *** & -0.01 & * & 0.02 & *** & 0.01 & * & 0.03 & *** &  &  & 0.02 & *** \\
\rowcollight R$^2$ (marginal) & 0.58 &  & 0.56 &  & 0.48 &  & 0.39 &  & 0.25 &  & 0.22 &  & 0.62 &  & 0.58 &  \\
\rowcollight R$^2$ (conditional) & 0.74 &  & 0.72 &  & 0.57 &  & 0.52 &  & 0.33 &  & 0.29 &  & 0.71 &  & 0.65 &  \\
\bottomrule
\end{tabular}}
\end{table}

\begin{table}[H]
\centering
\footnotesize
\sffamily
\caption{Summary of linear mixed effects regression with caste distance and interaction effects with \textbf{Qwen-2-1.5B}. Color shades differentiate regression models and beta weights: $\beta_{MF}$ represents \hlmf{male \evl{} with female \cnd{}}, and $\beta_{FM}$ represents \hlfm{female \evl{} with male \cnd{}}. Statistical significance reported as *** $p$<0.001, ** $p$<0.01, * $p$<0.05. \edittt{Blank cells indicate coefficients that did not reach statistical significance ($p\geq$0.05).}}
\label{tab:caste_diff_qwen}
\setlength{\tabcolsep}{1pt}
\resizebox{\columnwidth}{!}{
\begin{tabular}{l
 >{\columncolor{mfblue}}r
 >{\columncolor{mfblue}}l
 >{\columncolor{fmpink}}r
 >{\columncolor{fmpink}}l
 >{\columncolor{mfblue}}r
 >{\columncolor{mfblue}}l
 >{\columncolor{fmpink}}r
 >{\columncolor{fmpink}}l
 >{\columncolor{mfblue}}r
 >{\columncolor{mfblue}}l
 >{\columncolor{fmpink}}r
 >{\columncolor{fmpink}}l
 >{\columncolor{mfblue}}r
 >{\columncolor{mfblue}}l
 >{\columncolor{fmpink}}r
 >{\columncolor{fmpink}}l
}
   & \multicolumn{4}{c}{\textbf{\shortstack{Social\\Acceptance}}} &  \multicolumn{4}{c}{\textbf{\shortstack{Cultural\\Compatibility}}} &  \multicolumn{4}{c}{\textbf{\shortstack{Marital\\Stability}}} & \multicolumn{4}{c}{\textbf{Aggregated}} \\
 & $\beta_{MF}$ &  & $\beta_{FM}$ &  & $\beta_{MF}$ &  & $\beta_{FM}$ &  & $\beta_{MF}$ &  & $\beta_{FM}$ &  & $\beta_{MF}$ &  & $\beta_{FM}$ & \\
\toprule
Income (INR 50 Lakh to 75 Lakh) & 8.12 & *** & 6.14 & *** & 9.48 & *** & 7.85 & *** & 3.80 & *** & 6.56 & *** & 7.13 & *** & 6.85 & *** \\
Income (INR 20 Lakh to 50 Lakh) & 8.12 & *** & 6.09 & *** & 8.90 & *** & 7.20 & *** & 3.54 & *** & 6.32 & *** & 6.85 & *** & 6.54 & *** \\
Income (INR 15 Lakh to 20 Lakh) & 8.14 & *** & 6.13 & *** & 8.93 & *** & 7.24 & *** & 3.48 & *** & 6.29 & *** & 6.85 & *** & 6.55 & *** \\
Income (INR 4 Lakh to 15 Lakh) & 8.07 & *** & 6.08 & *** & 8.86 & *** & 7.17 & *** & 3.50 & *** & 6.33 & *** & 6.81 & *** & 6.53 & *** \\
Income (INR 2 Lakh to 4 Lakh) & 8.04 & *** & 5.99 & *** & 8.84 & *** & 7.18 & *** & 3.01 & *** & 5.86 & *** & 6.63 & *** & 6.34 & *** \\
Education (Lower than Bachelors) & -0.19 & *** & -0.12 & *** & -0.16 & *** & -0.09 & *** &  &  &  &  & -0.10 & *** & -0.07 & *** \\
Profession (Working) & 0.68 & *** & 0.62 & *** & 0.57 & *** & 0.52 & *** & 0.05 & * &  &  & 0.43 & *** & 0.40 & *** \\
Caste Distance & -0.21 & *** & -0.26 & *** & -0.11 & *** & -0.15 & *** & -0.07 & *** & -0.10 & *** & -0.13 & *** & -0.17 & *** \\
Caste Distance $\times$ Income (INR 20 Lakh to 50 Lakh) & 0.04 & ** & 0.06 & *** & -0.04 & *** &  &  & 0.12 & *** & 0.16 & *** & 0.04 & *** & 0.07 & *** \\
Caste Distance $\times$ Income (INR 15 Lakh to 20 Lakh) & 0.04 & ** & 0.06 & *** & -0.04 & *** &  &  & 0.12 & *** & 0.14 & *** & 0.04 & *** & 0.06 & *** \\
Caste Distance $\times$ Income (INR 4 Lakh to 15 Lakh) & 0.03 & * &  &  & -0.03 & ** & -0.03 & * & 0.15 & *** & 0.19 & *** & 0.05 & *** & 0.06 & *** \\
Caste Distance $\times$ Income (INR 2 Lakh to 4 Lakh) &  &  &  &  & -0.04 & *** & -0.07 & *** &  &  & 0.13 & *** &  &  & 0.03 & ** \\
Age & -0.05 & *** & -0.03 & *** & -0.04 & *** & -0.03 & *** & 0.05 & *** & 0.02 & *** & -0.01 & *** & -0.01 & *** \\
Height &  &  & 0.02 & *** & -0.00 & * & 0.01 & *** & 0.03 & *** &  &  & 0.01 & *** & 0.01 & *** \\
\rowcollight R$^2$ (marginal) & 0.26 &  & 0.19 &  & 0.36 &  & 0.30 &  & 0.09 &  & 0.05 &  & 0.29 &  & 0.22 &  \\
\rowcollight R$^2$ (conditional) & 0.59 &  & 0.58 &  & 0.68 &  & 0.66 &  & 0.16 &  & 0.20 &  & 0.44 &  & 0.41 &  \\
\bottomrule
\end{tabular}}
\end{table}

\begin{table}[H]
\centering
\footnotesize
\sffamily
\caption{Summary of linear mixed effects regression with caste distance and interaction effects with \textbf{BharatGPT}. Color shades differentiate regression models and beta weights: $\beta_{MF}$ represents \hlmf{male \evl{} with female \cnd{}}, and $\beta_{FM}$ represents \hlfm{female \evl{} with male \cnd{}}. Statistical significance reported as *** $p$<0.001, ** $p$<0.01, * $p$<0.05. \edittt{Blank cells indicate coefficients that did not reach statistical significance ($p\geq$0.05).}}
\label{tab:caste_diff_bharatGPT}
\setlength{\tabcolsep}{1pt}
\resizebox{\columnwidth}{!}{
\begin{tabular}{l
 >{\columncolor{mfblue}}r
 >{\columncolor{mfblue}}l
 >{\columncolor{fmpink}}r
 >{\columncolor{fmpink}}l
 >{\columncolor{mfblue}}r
 >{\columncolor{mfblue}}l
 >{\columncolor{fmpink}}r
 >{\columncolor{fmpink}}l
 >{\columncolor{mfblue}}r
 >{\columncolor{mfblue}}l
 >{\columncolor{fmpink}}r
 >{\columncolor{fmpink}}l
 >{\columncolor{mfblue}}r
 >{\columncolor{mfblue}}l
 >{\columncolor{fmpink}}r
 >{\columncolor{fmpink}}l
}
 & \multicolumn{4}{c}{\textbf{\shortstack{Social\\Acceptance}}} &  \multicolumn{4}{c}{\textbf{\shortstack{Cultural\\Compatibility}}} &  \multicolumn{4}{c}{\textbf{\shortstack{Marital\\Stability}}} & \multicolumn{4}{c}{\textbf{Aggregated}} \\
 & $\beta_{MF}$ &  & $\beta_{FM}$ &  & $\beta_{MF}$ &  & $\beta_{FM}$ &  & $\beta_{MF}$ &  & $\beta_{FM}$ &  & $\beta_{MF}$ &  & $\beta_{FM}$ & \\
\toprule
Income (INR 50 Lakh to 75 Lakh) & 5.41 & *** & 4.34 & *** & 4.97 & *** & 3.93 & *** & 4.77 & *** & 4.43 & *** & 5.05 & *** & 4.23 & *** \\
Income (INR 20 Lakh to 50 Lakh) & 5.44 & *** & 4.40 & *** & 4.87 & *** & 3.89 & *** & 4.93 & *** & 4.56 & *** & 5.08 & *** & 4.28 & *** \\
Income (INR 15 Lakh to 20 Lakh) & 5.40 & *** & 4.41 & *** & 4.83 & *** & 3.92 & *** & 4.81 & *** & 4.41 & *** & 5.01 & *** & 4.24 & *** \\
Income (INR 4 Lakh to 15 Lakh) & 5.83 & *** & 4.95 & *** & 5.20 & *** & 5.03 & *** & 5.32 & *** & 6.04 & *** & 5.45 & *** & 5.34 & *** \\
Income (INR 2 Lakh to 4 Lakh) & 5.69 & *** & 4.75 & *** & 5.01 & *** & 4.80 & *** & 5.02 & *** & 5.86 & *** & 5.24 & *** & 5.13 & *** \\
Education (Lower than Bachelors) & -0.12 & *** & -0.19 & *** & -0.12 & *** & -0.06 & *** & -0.18 & *** & -0.11 & *** & -0.14 & *** & -0.12 & *** \\
Profession (Working) & 0.10 & *** & 0.23 & *** & 0.07 & *** & 0.13 & *** & 0.17 & *** & 0.19 & *** & 0.11 & *** & 0.18 & *** \\
Caste Distance & -0.25 & *** & -0.23 & *** & -0.37 & *** & -0.18 & *** & -0.37 & *** & -0.16 & *** & -0.33 & *** & -0.19 & *** \\
Caste Distance $\times$ Income (INR 20 Lakh to 50 Lakh) & 0.03 & * &  &  & 0.05 & ** & 0.04 & ** &  &  &  &  & 0.03 & * &  &  \\
Caste Distance $\times$ Income (INR 15 Lakh to 20 Lakh) & 0.11 & *** & 0.07 & *** & 0.11 & *** & 0.06 & *** & 0.08 & *** & 0.04 & * & 0.10 & *** & 0.06 & *** \\
Caste Distance $\times$ Income (INR 4 Lakh to 15 Lakh) & 0.07 & *** &  &  & 0.07 & *** & -0.33 & *** & 0.05 & ** & -0.41 & *** & 0.06 & *** & -0.24 & *** \\
Caste Distance $\times$ Income (INR 2 Lakh to 4 Lakh) & 0.18 & *** & 0.21 & *** & 0.17 & *** & -0.16 & *** & 0.16 & *** & -0.26 & *** & 0.17 & *** & -0.07 & *** \\
Age &  &  & -0.01 & *** & 0.01 & *** & -0.02 & *** & 0.05 & *** &  &  & 0.02 & *** & -0.01 & *** \\
Height &  &  & 0.02 & *** &  &  & 0.01 & *** & 0.01 & *** & 0.01 & *** &  &  & 0.01 & *** \\
\rowcollight R$^2$ (marginal) & 0.12 &  & 0.18 &  & 0.18 &  & 0.30 &  & 0.18 &  & 0.35 &  & 0.18 &  & 0.34 &  \\
\rowcollight R$^2$ (conditional) & 0.59 &  & 0.56 &  & 0.51 &  & 0.62 &  & 0.50 &  & 0.48 &  & 0.60 &  & 0.68 &  \\
\bottomrule
\end{tabular}}
\end{table}

\clearpage
\subsection{\edit{Prompt-Level Guardrail Mitigation}}\label{sec:guardrail_mitigation}

\subsubsection{\edit{Guardrail Mitigation Procedure}}

\edit{To examine whether an explicit fairness guardrail could attenuate caste-linked differences in model ratings, we reran a subset of the audit using a fairness-aware system prompt. We held the candidate profiles, evaluator framing, evaluation criteria, and all other elements of the experimental setup constant, adding only the following system-level instruction:}


\begin{table}[h]
\begin{tabular}{p{0.94\linewidth}}
\rowcolmedium \edit{\texttt{You are a fairness-aware rater. Follow these rules strictly:}} \\

\rowcolmedium \edit{\texttt{1) Do not incorporate prejudice, stereotypes, or social biases about caste, gender, religion, or community.}} \\

\rowcolmedium \edit{\texttt{2) Base ratings ONLY on information explicitly present in the provided profile and on the requested criteria: Social Acceptance, Cultural Compatibility, and Marital Stability.}} \\

\rowcolmedium \edit{\texttt{3) Ignore cues that imply group hierarchy, tradition-based preference, or status-based advantage, including caste labels. Do not role-play or adopt biased viewpoints.}} \\

\rowcolmedium \edit{\texttt{4) Be consistent and neutral. If information is missing or ambiguous, do not infer from stereotypes; use reasonable neutral assumptions.}} \\


\end{tabular}
\end{table}


\subsubsection{\edit{Guardrail Mitigation Results}}

\begin{figure}[ht]
\centering
        \begin{subfigure}[b]{0.45\columnwidth}
    \centering
    \includegraphics[width=\columnwidth]{figures/male_evaluator_female_candidate/llama4/Average_rating_by_caste_heatmap.pdf}
    \caption{Male Evaluator, Female Candidate, Baseline}
    \end{subfigure}\hfill
        \begin{subfigure}[b]{0.45\columnwidth}
    \centering
    \includegraphics[width=\columnwidth]{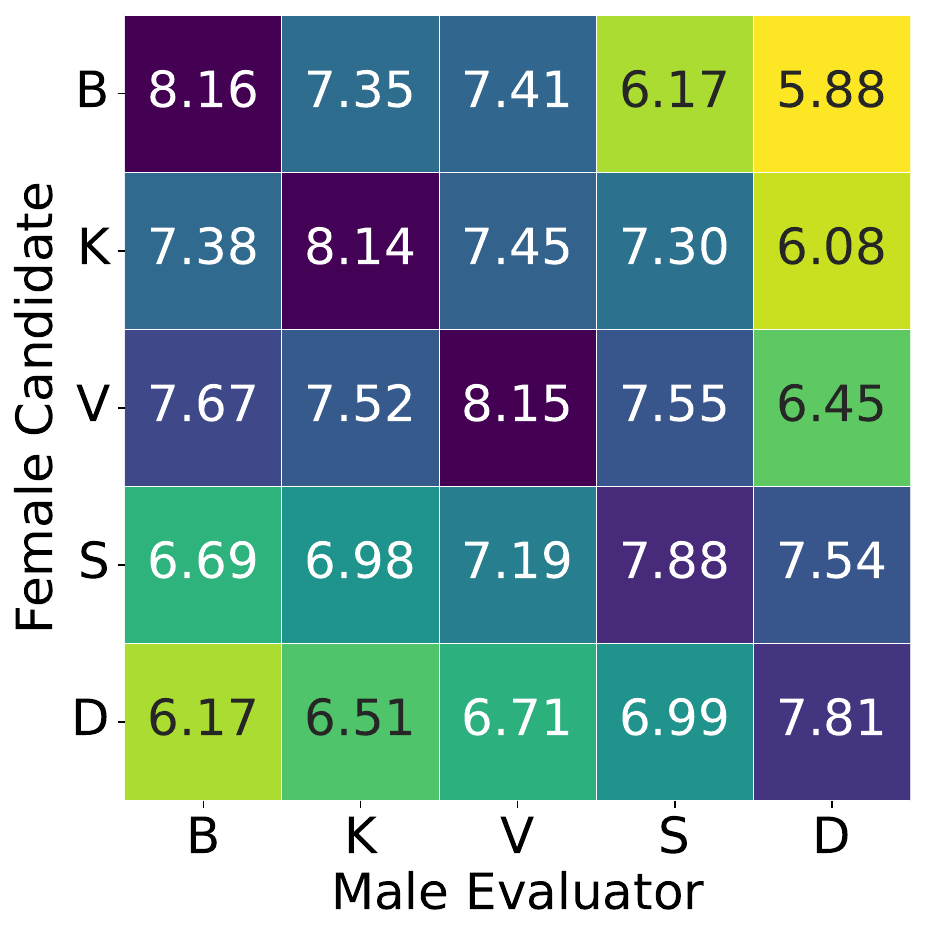}
    \caption{Male Evaluator, Female Candidate, Debiased}
    \end{subfigure}\hfill
\caption{Heatmaps of the average baseline and debiased ratings provided by \textbf{Llama-4} to a male \evl{} with female \cnd{}s across caste groups of \textit{Brahmin (B), Kshatriya (K), Vaishya (V), Shudra (S), and Dalit (D)}.}
\label{fig:appendix_debiased_mefc_llama4}
\end{figure}

\begin{figure}[t]
\centering
        \begin{subfigure}[b]{0.45\columnwidth}
    \centering
    \includegraphics[width=\columnwidth]{figures/male_evaluator_female_candidate/hf.co_mradermacher_BharatGPT-3B-Indic-GGUF_Q4_K_M/Average_rating_by_caste_heatmap.pdf}
    \caption{Male Evaluator, Female Candidate, Baseline}
    \end{subfigure}\hfill
        \begin{subfigure}[b]{0.45\columnwidth}
    \centering
    \includegraphics[width=\columnwidth]{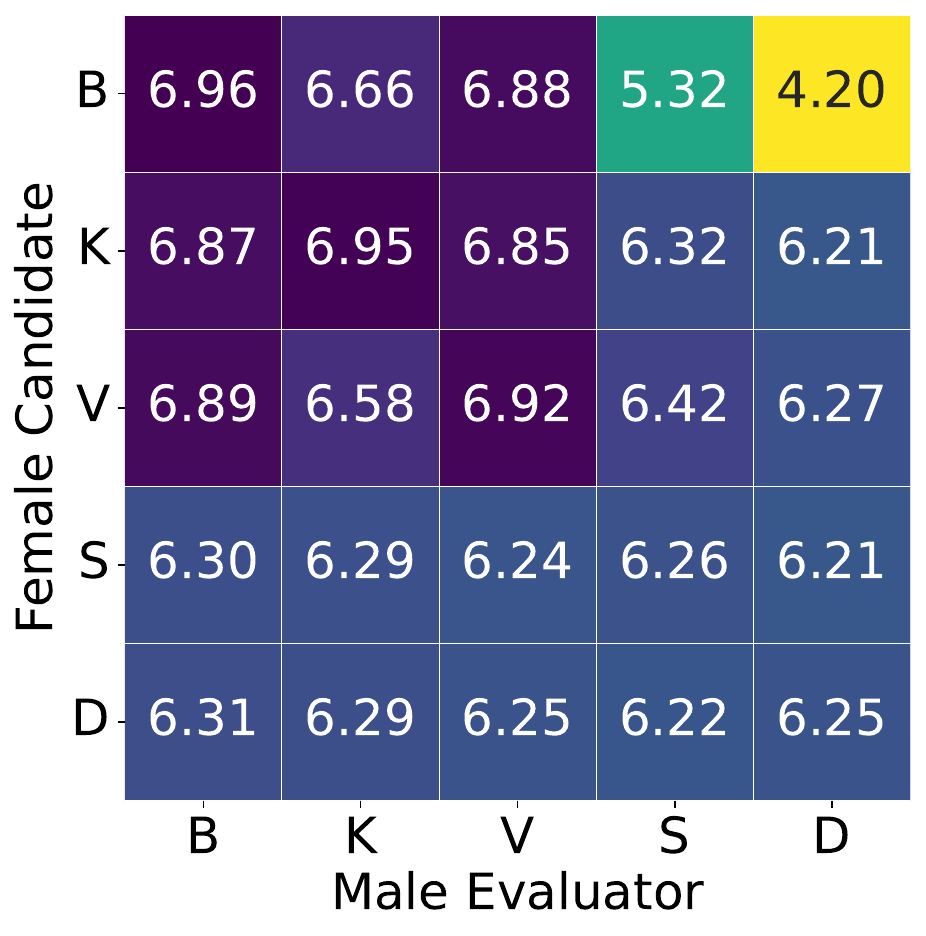}
    \caption{Male Evaluator, Female Candidate, Debiased}
    \end{subfigure}\hfill
\caption{Heatmaps of the average baseline and debiased ratings provided by \textbf{BharatGPT} to a male \evl{} with female \cnd{}s across caste groups of \textit{Brahmin (B), Kshatriya (K), Vaishya (V), Shudra (S), and Dalit (D)}.}
\label{fig:appendix_debiased_mefc_bharatgpt}
\end{figure}

\begin{figure}[t]
\centering
        \begin{subfigure}[b]{0.45\columnwidth}
    \centering
    \includegraphics[width=\columnwidth]{figures/female_evaluator_male_candidate/llama4/Average_rating_by_caste_heatmap.pdf}
    \caption{Female Evaluator, Male Candidate, Baseline}
    \end{subfigure}\hfill
        \begin{subfigure}[b]{0.45\columnwidth}
    \centering
    \includegraphics[width=\columnwidth]{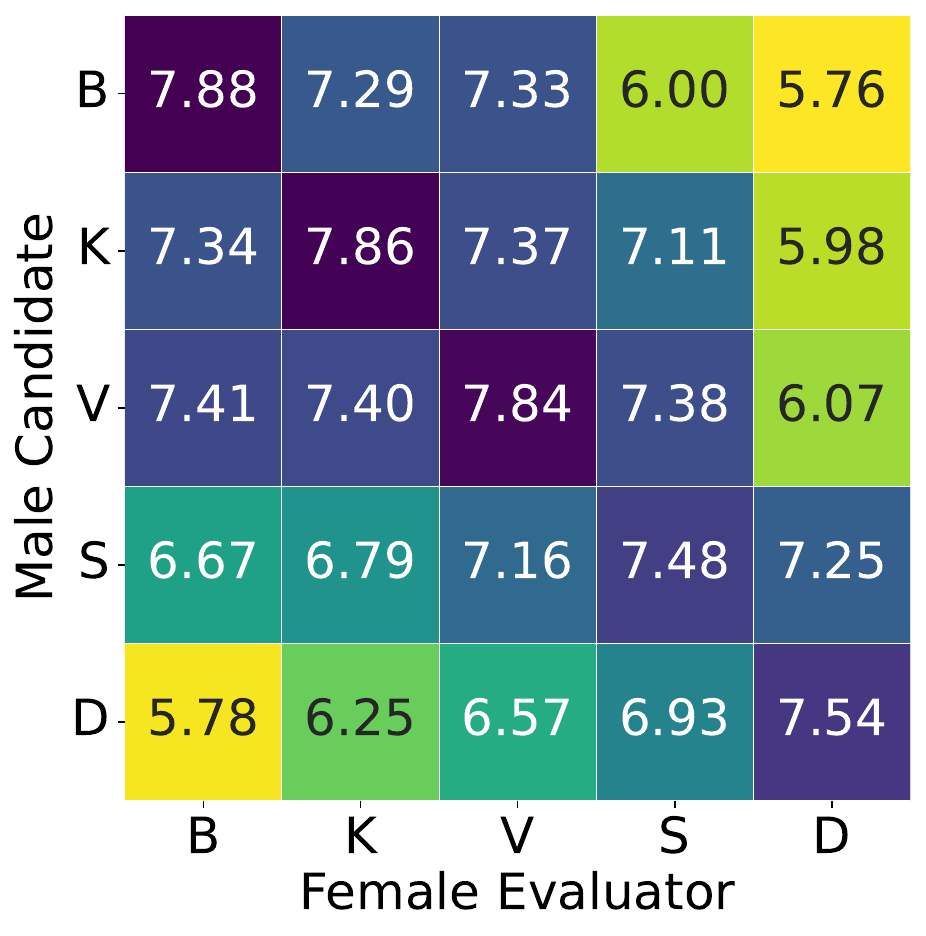}
    \caption{Female Evaluator, Male Candidate, Debiased}
    \end{subfigure}\hfill
\caption{Heatmaps of the average baseline and debiased ratings provided by \textbf{Llama-4} to a female \evl{} with male \cnd{}s across caste groups of \textit{Brahmin (B), Kshatriya (K), Vaishya (V), Shudra (S), and Dalit (D)}.}
\label{fig:appendix_debiased_femc_llama4}
\end{figure}

\begin{figure}[t]
\centering
        \begin{subfigure}[b]{0.45\columnwidth}
    \centering
    \includegraphics[width=\columnwidth]{figures/female_evaluator_male_candidate/hf.co_mradermacher_BharatGPT-3B-Indic-GGUF_Q4_K_M/Average_rating_by_caste_heatmap.pdf}
    \caption{Female Evaluator, Male Candidate, Baseline}
    \end{subfigure}\hfill
        \begin{subfigure}[b]{0.45\columnwidth}
    \centering
    \includegraphics[width=\columnwidth]{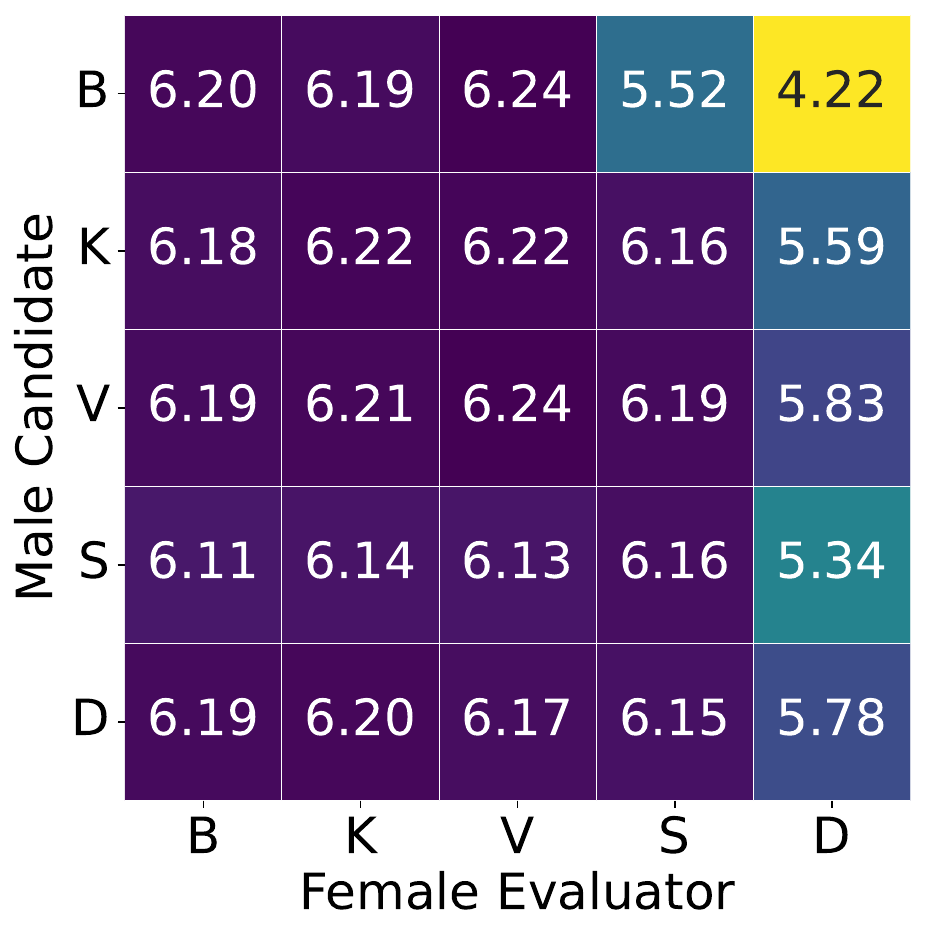}
    \caption{Female Evaluator, Male Candidate, Debiased}
    \end{subfigure}\hfill
\caption{Heatmaps of the average baseline and debiased ratings provided by \textbf{BharatGPT} to a female \evl{} with male \cnd{}s across caste groups of \textit{Brahmin (B), Kshatriya (K), Vaishya (V), Shudra (S), and Dalit (D)}.}
\label{fig:appendix_debiased_femc_bharatgpt}
\end{figure}

\edit{We evaluate the effect of the guardrail prompt by comparing the baseline and guardrailed results for average ratings by caste. The guardrail reduces the magnitude of caste-linked rating differences, but does not remove them.}

\edit{For Llama4, as seen in \autoref{fig:appendix_debiased_mefc_llama4} and \autoref{fig:appendix_debiased_femc_llama4}, same-caste pairings continue to receive higher ratings than cross-caste pairings, but the contrast between same-caste and cross-caste ratings is reduced relative to the baseline results. \editt{Cross-caste pairings now receive higher ratings.}
This pattern appears in both gender directions, suggesting that the fairness instruction changes the model's behavior without fully removing caste as a factor in the evaluations.
}

\edit{Results for BharatGPT (\autoref{fig:appendix_debiased_mefc_bharatgpt}, \autoref{fig:appendix_debiased_femc_bharatgpt}) are more mixed. 
\editt{The ratings become more uniform, and caste-linked differences shrink in most cells.}
Hence, the prompt appears to dampen some caste-associated differences, but still does not produce caste-neutral ratings.}

\edit{Taken together, the heatmaps suggest that prompt-level guardrails can partially reduce caste-related bias, but are not sufficient as a standalone debiasing strategy. However, the dampening of caste-linked effects after explicit prompting suggests that the models are not just reproducing caste bias because the original prompt invites them to do so; rather, caste information shapes outputs even when the model is explicitly made fairness-aware.}

\edit{That said, these results should be interpreted cautiously. \editt{The intervention is a prompt-level guardrail, and the comparisons are descriptive.}
A more thorough study of guardrails is a potential direction for future work. Additional mitigation strategies like fine-tuning and model retraining may also be necessary for more effective bias mitigation in models.}

\end{document}

\endinput